\documentclass{aa}  

\usepackage{graphicx}
\usepackage{txfonts}
\usepackage{multirow}
\usepackage{float}
\usepackage{natbib}
\bibpunct{(}{)}{;}{a}{}{,}
\usepackage{placeins}
\usepackage{stfloats}
\DeclareMathOperator\erf{erf}

\begin{document}

   \title{Convolutional neural networks as an alternative to Bayesian retrievals for interpreting exoplanet transmission spectra}

   %\subtitle{I. Overviewing the $\kappa$-mechanism}
    
    \titlerunning{Convolutional neural networks as an alternative to Bayesian retrievals}
   \author{F. Ard\'evol Mart\'inez
          \inst{1,2,3,4}
          \and
          M. Min\inst{2}
          \and
          I. Kamp\inst{1}
          \and
          P. I. Palmer\inst{3,4}
          }

   \institute{Kapteyn Astronomical Institute, University of Groningen, Groningen, The Netherlands\\
              \email{ardevol@astro.rug.nl}
         \and
             Netherlands Space Research Institute (SRON), Leiden, The Netherlands
        \and
            Centre for Exoplanet Science, University of Edinburgh, Edinburgh, UK
        \and
            School of GeoSciences, University of Edinburgh, Edinburgh, UK
             }

   \date{Received XXX; accepted YYY}

% \abstract{}{}{}{}{} 
% 5 {} token are mandatory
 
  \abstract
   {Exoplanet observations are currently analysed with Bayesian retrieval techniques to constrain physical and chemical properties of their atmospheres. Due to the computational load of the models used to analyse said observations, a compromise is usually needed between model complexity and computing time. Analysis of observational data from future facilities, e.g. JWST (James Webb Space Telescope), will need more complex models which will increase the computational load of retrievals, prompting the search for a faster approach for interpreting exoplanet observations. }
   {{The goal is to} compare machine learning retrievals of exoplanet transmission spectra with nested sampling (Bayesian retrieval), and understand if machine learning can be as reliable as a Bayesian retrieval {for a statistically significant sample of spectra while being orders of magnitude faster}.}
   {We generated grids of synthetic transmission spectra and their corresponding planetary and atmospheric parameters, one using free chemistry models, and the other using equilibrium chemistry models. Each grid was subsequently rebinned to simulate both HST (Hubble Space Telescope) WFC3 (Wide Field Camera 3) and JWST NIRSpec (Near-Infrared Spectrograph) observations, yielding four datasets in total. Convolutional neural networks (CNNs) were trained with each of the datasets. We performed retrievals for a set of 1,000  simulated observations for each combination of model type and instrument with nested sampling and machine learning. We also used both methods to perform retrievals for real HST/WFC3 transmission spectra of 48 exoplanets. Additionally, we carried out experiments to test how robust machine learning and nested sampling are against incorrect assumptions in our models.}
   {CNNs reach a lower coefficient of determination between predicted and true values of the parameters. Neither CNNs nor nested sampling systematically reach a lower bias for all parameters. Nested sampling underestimates the uncertainty in $\sim 8\%$ of retrievals, whereas CNNs correctly estimate the uncertainties. When performing retrievals for real HST/WFC3 observations, nested sampling and machine learning agree within $2\sigma$ for $\sim 86\%$ of spectra. When doing retrievals with incorrect assumptions, nested sampling underestimates the uncertainty in $\sim12\%$ to $\sim41\%$ of cases, whereas for the CNNs this fraction always remains below $\sim10\%$.}
   {}

   \keywords{planets and satellites: atmospheres --
                planets and satellites: gaseous planets --
                planets and satellites: composition
               }

   \maketitle
%
%-------------------------------------------------------------------

\section{Introduction}

The number of confirmed exoplanets has grown to over {4,900} in recent years\footnote{NASA Exoplanet Archive, \url{exoplanetarchive.ipac.caltech.edu}}. Transiting exoplanets provide  a golden opportunity for characterisation. The wavelength dependent transit depth contains information about the chemical and physical structure of the exoplanet's atmosphere. Yet retrieving this information is not an easy task. Traditional Bayesian retrievals involve computing tens to hundreds of thousands of forward models and comparing them to  observations to derive probability distributions for different physical and chemical parameters. Nested sampling \citep{Skilling2006NestedComputation}, and in particular the MultiNest (MN) implementation \citep{Feroz2009MultiNest:Physics} is currently the preferred Bayesian sampling technique.
The high computational load of Bayesian retrievals often makes it necessary to simplify the forward models by parameterizing different aspects of the atmosphere, such as the temperature structure, the chemistry and/or the clouds.
More complex models that can compute these self-consistently paired with data with higher SNR  and larger wavelength coverage, such as what will be provided by future facilities like JWST \citep[James Webb Space Telescope, ][]{Gardner2006TheTelescope} or ARIEL \citep[Atmospheric Remote-sensing Exoplanet Large-survey, ][]{2018a_tinetti}, will only exacerbate this problem.

Machine learning approaches have recently started being considered as means to reduce the computational load of atmospheric retrievals \citep[e.g., ][]{Waldmann2016DREAMINGATMOSPHERES, Zingales2018ExoGAN:Networks, Marquez-Neila2018SupervisedAtmospheres, Cobb_2019, nixon2020ml}. In principle, once a supervised learning algorithm has been trained with pairs of forward models and their corresponding parameters, it should be possible to perform retrievals in seconds. The feasibility of this approach was first demonstrated by \citet{Waldmann2016DREAMINGATMOSPHERES}, who used a Deep Belief Network (DBN) to recognise molecular signatures in exoplanetary emission spectra. DBNs are generative models, which means that they learn to replicate the inputs (in this case the emission spectra) in an unsupervised manner. Subsequently, they can be trained with supervision to perform classification. Indeed, this first method did not do a full retrieval and could only predict whether different molecules were present in the spectrum. Another generative model was used by \citet{Zingales2018ExoGAN:Networks}, who trained a Deep Convolutional Generative Adversarial Network (DCGAN). In their case, the DCGAN was trained to generate a 2D array containing both the spectra and the parameters. This is accomplished by pitting two convolutional neural networks against each other, one trying to generate fake inputs and the other trying to discriminate the real inputs from the fake ones. Training concludes when the discriminator can no longer tell which is which. The DCGAN can then be used to reconstruct incomplete inputs, in this case a 2D array missing the parameters. Having to train two neural networks puts this method at a disadvantage, as more fine tuning is necessary to optimise the architectures of both networks. Lastly, \citet{Zingales2018ExoGAN:Networks} find that a very large training set of $\sim 5\cdot10^5$ forward models is necessary for the DCGAN to learn, while the methods presented below can be trained with $\sim 5\cdot10^4$ forward models, a factor ten fewer.

Aside from generative models, random forests and neural networks have also been used to retrieve atmospheric parameters from transmission spectra of exoplanets. \citet{Marquez-Neila2018SupervisedAtmospheres} trained a random forest to do atmospheric retrievals of transmission spectra, obtaining a posterior distribution for WASP-12b consistent with a Bayesian retrieval. Random forests are ensemble predictors composed of multiple regression trees, each trying to find which features and thresholds best split the data into categories which best match the labels. They are simple, easy to interpret, and fast to train (although they can require large amounts of RAM). However, due to their simplicity, they are outperformed by neural networks in complex tasks, such as the ones we will be using in this work (see Section \ref{sec:discussion}).
\citet{Cobb_2019} trained a Dense Neural Network able to predict both the means and covariance matrix of the parameters, showing again good agreement with nested sampling for WASP-12b. Additionally, machine learning has also been used to interpret albedo spectra \citep{Johnsen2020ASpectra}, as well as planetary interiors \citep{Baumeister2020Machine-learningExoplanets} and high resolution ground-based transmission spectra \citep{Fisher2020InterpretingLearning}. More recently, \citet{Yip2020PeekingRetrievals} investigated the interpretability (that is, understanding what these algorithms are `looking at' when making their predictions) of these deep learning approaches, showing that the decision process of the neural networks is largely aligned with our intuitions.

These previous studies were limited to retrieving isothermal atmospheres, free chemistry (meaning the abundances of chemical species are free parameters) with only a few molecules, and grey clouds.  Retrieving the chemistry self consistently is necessary to constrain macroscopic parameters such as the C/O or the metallicity, which are crucial to understand the formation history of the planet \citep[e.g., ][]{Madhusudhan2016ExoplanetaryHabitability, Cridland2020ConnectingAstrochemistry}. Additionally, the previous literature only contains comparisons between machine learning and Bayesian retrievals for a handful of spectra. More extensive testing is needed to establish machine learning retrieval frameworks as valid alternatives to nested sampling.

We therefore trained convolutional neural networks (CNN) to do free and equilibrium chemistry retrievals of JWST/{NIRSpec} and HST/WFC3 transmission spectra of exoplanets. We also compare CNN and Multinest retrievals for large samples of spectra.

In this paper, we first introduce the generation of the data used to train the CNNs and its pre-processing in section \ref{sec:data}. In Section \ref{sec:training} a brief explanation of how CNNs work is presented, followed by a description of our implementation and the evaluation techniques employed. Our results are presented in section \ref{sec:results}. In Section \ref{sec:uncomfortable} we test the robustness of both retrieval methods against incorrect assumptions in our forward models. In Section \ref{sec:discussion} the results presented previously are discussed. Finally, in Section \ref{sec:conclusion} we present our conclusions.

\section{Generation of the training data}\label{sec:data}

Training a supervised learning algorithm requires training examples, containing features (inputs) and labels (outputs).
In the case of inferring physical and chemical parameters from transmission spectra of exoplanets, the features are the transit depths at each wavelength and the labels are the planetary and atmospheric parameters. 
We generated two grids of transmission spectra using two different types of models with ARCiS \citep[ARtful modelling of Cloudy exoplanet atmosphereS, ][]{Min2020TheAtmospheres}, each containing 100,000 spectrum-parameters pairs.

ARCiS is an atmospheric modelling code attempting to strike a balance between fully self-consistent and parametric codes. This balance allows it to predict observations from physical parameters while limiting the computational load, making it suitable for nested sampling retrievals.

ARCiS is a flexible framework that allows for computing models of different complexity. In this work, we use two different types of models, which we will hereinafter refer to as `type 1' and `type 2'. The difference between the two lies in the chemistry. Type 1 models are computed with free chemistry, meaning the abundances of the chosen molecules (in our case, H$_2$O, CO, CO$_2$, CH$_4$ and NH$_3$) are free parameters. In the type 2 models, the thermochemical equilibrium abundances are computed  using GGchem \citep{Woitke2018EquilibriumRatio} from the carbon to oxygen ratio and the metallicity of the atmosphere\footnote{For supersolar C/O ratios ARCiS adjusts the oxygen abundance while keeping the carbon at solar abundance, and for subsolar C/O ratios ARCiS adjusts the carbon abundance while oxygen remains at a solar abundance. All other elements are then scaled so that the solar Si/O is matched. The metallicity is then adjusted  by scaling the H and He abundances}. The following species are considered in the chemistry as well as in the opacity sources: H$_2$O, CO, CO$_2$, CH$_4$, NH$_3$, HCN, TiO, VO, AlO, FeH, OH, C$_2$H$_2$, CrH, H$_2$S, MgO, H$^-$, Na, K.  

In both cases, the atmospheres are characterized by an isothermal temperature structure. We also include grey clouds parameterised by the cloud top pressure $P_{\rm{cloud}}$; in this parameterisation, the atmosphere is transparent for $P<P_{\rm{cloud}}$ and opaque otherwise. Finally, hazes are also included, which are modelled by adding a grey opacity $\kappa_{\rm{haze}}$ to the atmosphere. 

We generated two grids of transmission spectra, one for each type of model. For more details regarding the forward models, the ARCiS input files used to generate them have been provided in the replication package\footnote{All data and codes necessary to reproduce the results of this work can be found here: \url{https://gitlab.astro.rug.nl/ardevol/exocnn}}.

Tables \ref{table:level_1} and \ref{table:level_2} detail the parameters that were allowed to vary for both grids, along with the ranges from which they were randomly sampled.

\begin{table}[h]
    \centering
    \begin{tabular}{r | c}
        \hline
         Parameter &  Range\\
         \hline\hline
         $T$ (K) & (100, 5000)\\
         $\log {\rm H_2O}$ & (-12, 0)\\
         $\log {\rm CO}$ & (-12, 0)\\
         $\log {\rm CO_2}$ & (-12, 0)\\
         $\log {\rm CH_4}$ & (-12, 0)\\
         $\log {\rm NH_3}$ & (-12, 0)\\
         $\log {\kappa_{\rm{haze}}}$ & (-8, 2)\\
         $\log {P_{\rm{cloud}}}$ & (-3, 3)\\
         $R_{\rm P}$ ($R_{\rm J}$) & (0.1, 1.9)\\
         $\log {g}$ & (2.3, 4.3)\\
         \hline
    \end{tabular}
    \caption{Parameters and corresponding ranges for the type 1 grid.}
    \label{table:level_1}
\end{table}
\quad
\begin{table}[h]
\centering
    \begin{tabular}{r | c}
        \hline
         Parameter &  Range\\
         \hline\hline
         $T$ (K) & (100, 5000)\\
         $\rm C/\rm O$ & (0.1, 2.0)\\
         $\log {Z}$ & (-3, 3)\\
         $\log {\kappa_{\rm{haze}}}$ & (-8, 2)\\
         $\log {P_{\rm{cloud}}}$ & (-3, 3)\\
         $R_{\rm P}$ ($R_{\rm J}$) & (0.1, 1.9)\\
         $\log {g}$ & (2.3, 4.3)\\
         \hline
    \end{tabular}
    \caption{Parameters and corresponding ranges for the type 2 grid.}
    \label{table:level_2}
\end{table}

All of the forward models were generated for {NIRSpec}'s wavelength range and spectral resolution obtained from a Pandexo \citep{Batalha2017Pandexo:HST} simulation, and then rebinned to WFC3's spectral range and resolution. The exact wavelength bins were extracted from real WFC3 observations reduced by the Exoplanets-A project \citep{Pye2019ExoplanetExoplanets}.

\subsection{Noise}

To make the retrievals as realistic as possible, and to facilitate comparison to Bayesian retrievals, we trained the algorithms with noisy spectra. To this end we generated a number of noisy copies of each model, drawing from normal distributions with standard deviations described below.  This process is called `data augmentation' and allows us to make the training sets larger without computing more forward models, which would be significantly more time consuming. Section \ref{number_examples} contains a discussion on the number of models and noisy copies necessary to train the machine learning algorithms.

\begin{figure}
    \centering
    \includegraphics[width=0.49\textwidth]{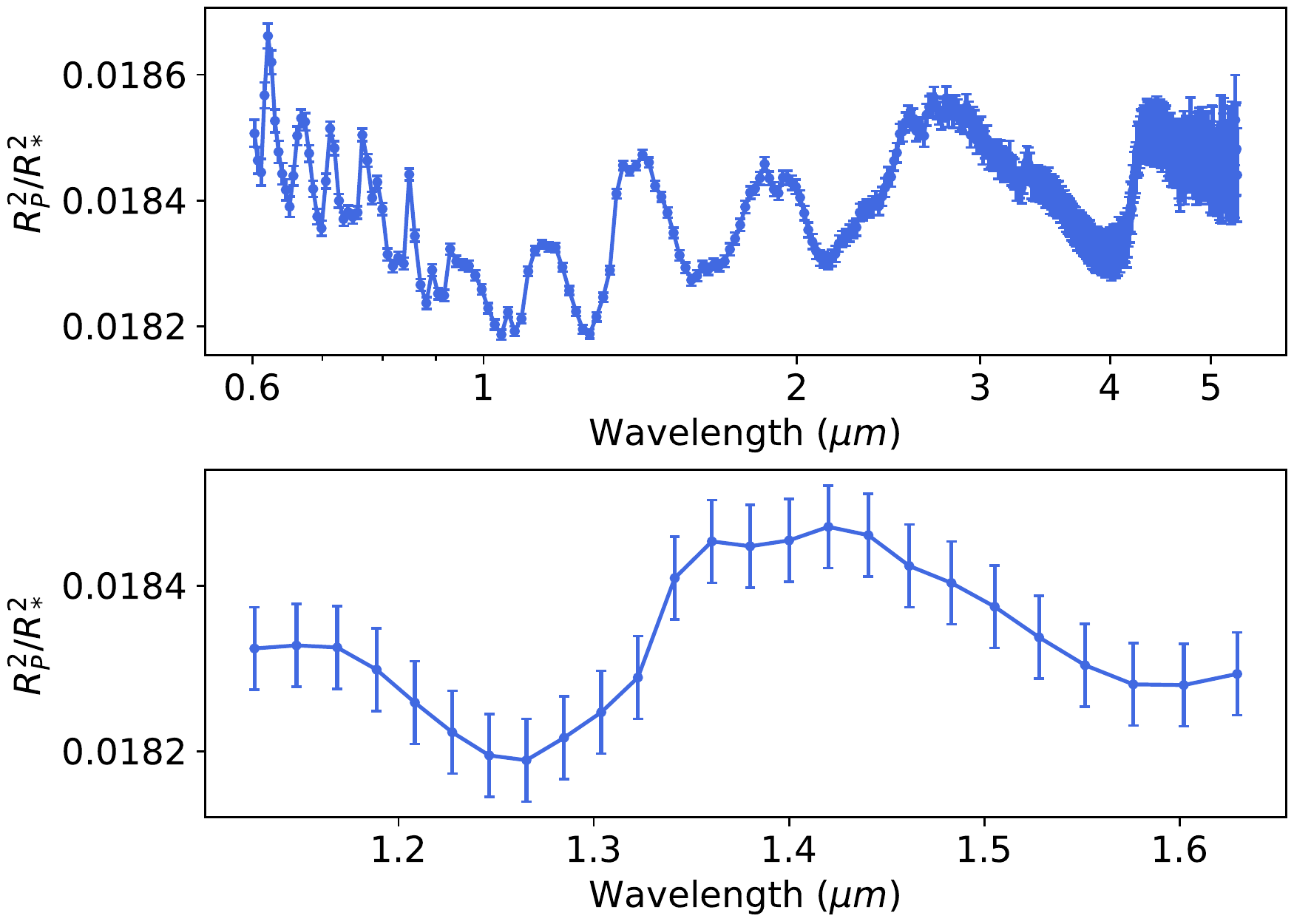}
    \caption{Examples of {NIRSpec} (Top) and WFC3 (Bottom) transmission spectra of a cloudless hot Jupiter ($T=1000\,K$, $R_{\rm P} = 1\,R_{\rm J}$, $M_{\rm P} = 1\,M_{\rm J}$) with solar C/O and $\log Z$.}
    \label{fig:ex_spectra}
\end{figure}

For the HST/WFC3 forward models we introduced constant noise throughout the wavelength range with $\sigma=50$ ppm for consistency with the previous literature \citep{Marquez-Neila2018SupervisedAtmospheres}. For JWST/NIRSpec forward models we calculated the noise with PandExo, setting the noise floor to 10 ppm. We used the stellar parameters of HD209458, which is an optimistic case due to its brightness ($m_J = 6.591$), but realistic nevertheless. HD209458 is a solar type  host to HD209458b, one of the best studied hot Jupiters and  target of several accepted JWST observing programs. In addition, we limited the noise calculation to one transit. In any case, scaling the noise for a different number of transits or a star of a different magnitude is straightforward, and a new simulation would only be required for a star of a different spectral type.

\subsection{Pre-conditioning of the training data}

\begin{figure}
    \centering
    \includegraphics[width=0.49\textwidth]{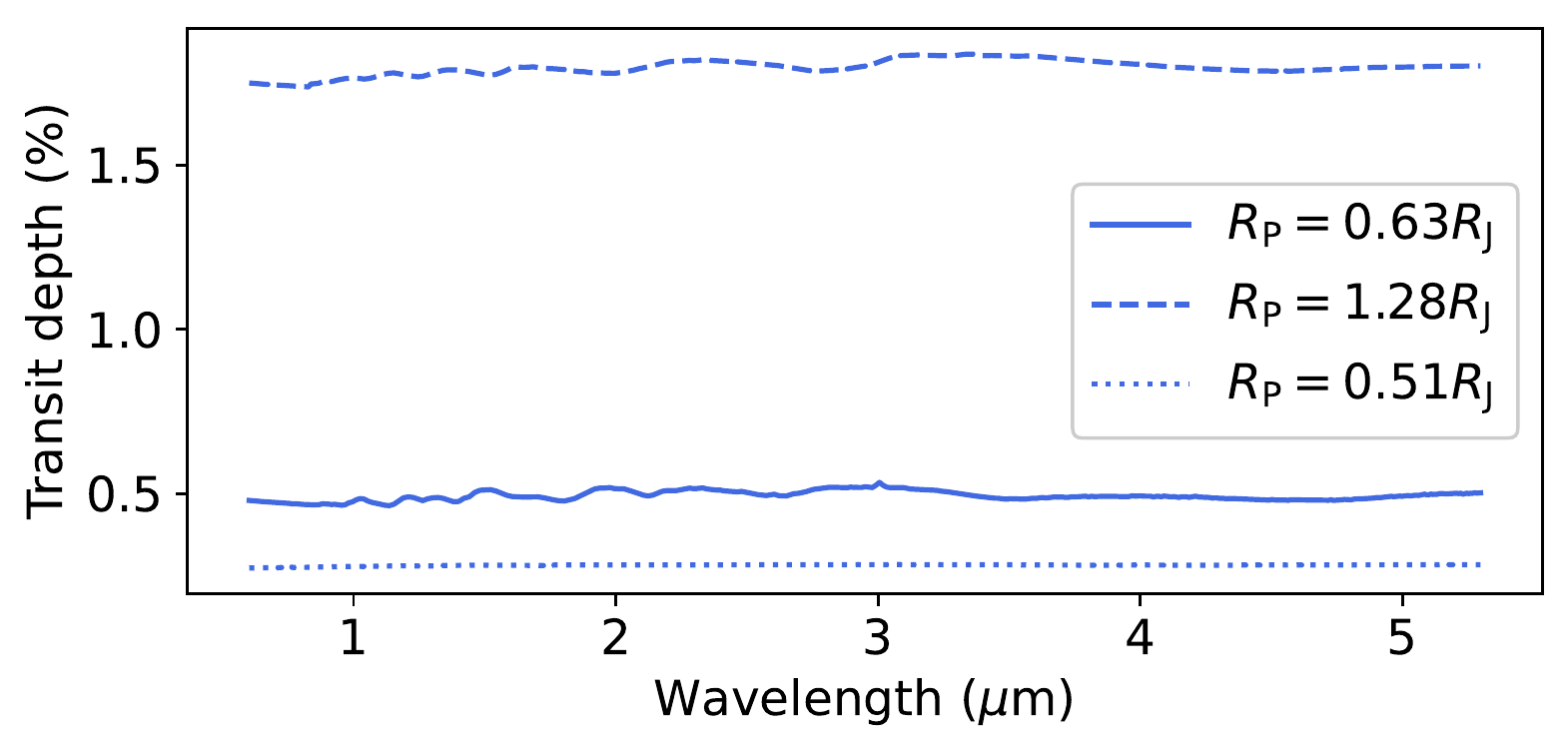}
    \includegraphics[width=0.49\textwidth]{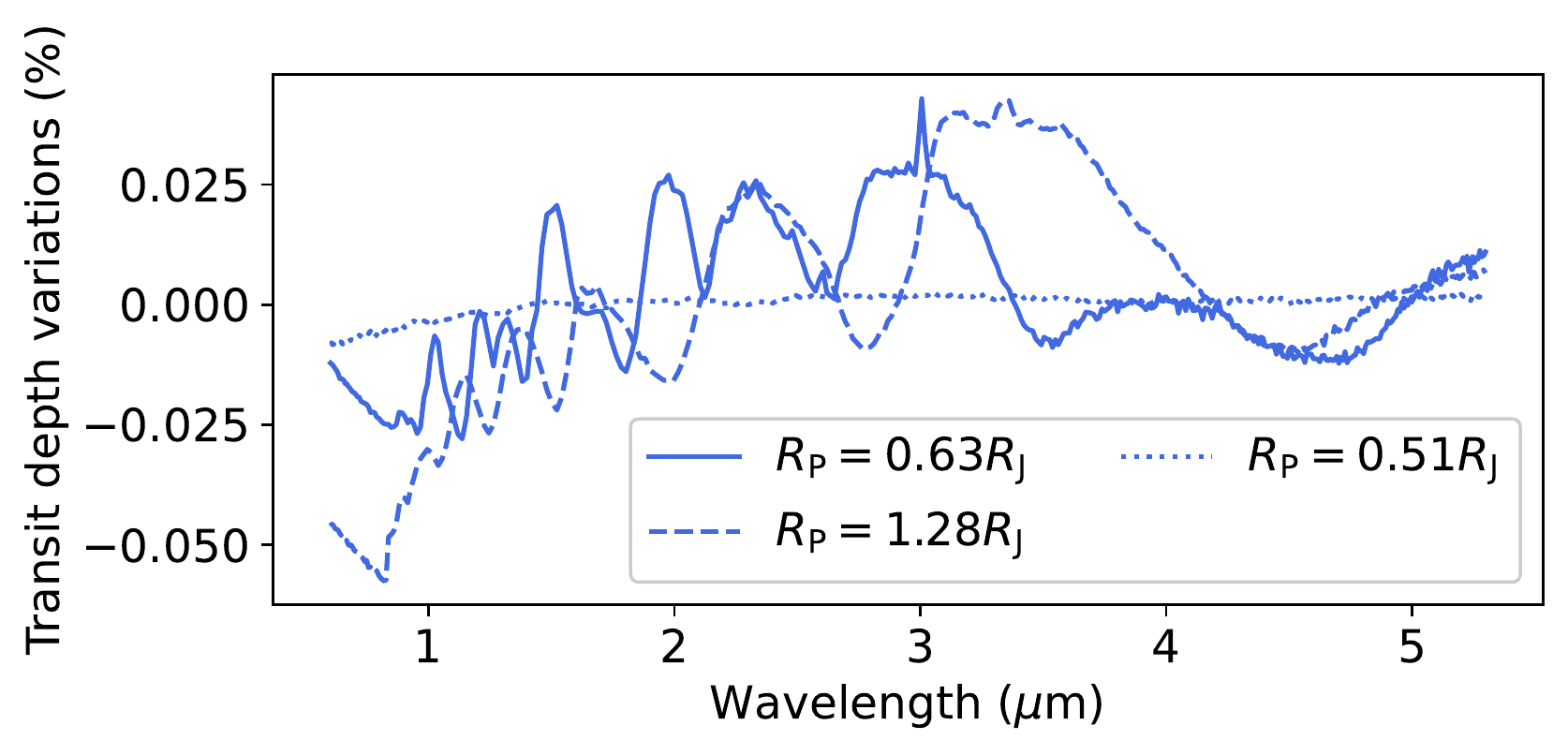}
    \caption{(Top) Clean transmission spectra of planets with different radii. (Bottom) Same spectra as above but with their respective means subtracted.}
    \label{fig:normalize_spectra}
\end{figure}

The bulk of the transit depth is due to the planetary radii, with small variations due to the atmospheric features. As a result, spectra of planets with different radii can look essentially flat when compared to each other. This is illustrated in Fig. \ref{fig:normalize_spectra} (Top). 
If no pre-processing is applied, the CNNs will need to learn how the different features look at every radius and will learn to predict the radius virtually perfectly, but their performance for the rest of the parameters will lag behind. 
After testing different approaches, we found that simply subtracting from each individual spectrum its own mean was the best way to increase the performance of the CNNs. As can be seen in Fig. \ref{fig:normalize_spectra} (Bottom), by removing the contribution from the planet itself, the atmospheric features become easily comparable between different spectra. This way, the CNNs can learn how spectral features look for all radii simultaneously. In order not to discard information, both the original and the normalized spectra were given as inputs as a two column array. Finally, the spectra were converted from $R_P^2/R_*^2$ to $R_P^2$ units. In this way, the trained algorithms do not depend on the host star properties.

The spectra (features) are not the only thing that needs to be normalized, but also the parameters (labels). Due to the parameters having different ranges, the CNNs can reduce the loss by focusing on predicting well the parameters with greater absolute values, like the temperature $T$ in our case. Because this parameter can be between 500 and 5,000 K, improving the temperature predictions will have a higher impact in minimizing the loss than improving the $\rm C/\rm O$, which only varies between 0.1 and 2. In order to avoid this we normalise the parameters to all have consistent ranges so that the CNN does not favour one over the rest. We used the \verb|MinMaxScaler| included in \textit{scikit-learn} to linearly transform all parameters to be between 0 and 1.

Some  combinations of parameters (namely high $T$ and low $\log g$) can cause unphysically large transit depths. To avoid this, we imposed the condition $\overline{R^2_P(\lambda)}<1.5 R^2_{P,0}$, where $ \overline{R^2_P(\lambda)/R^2_*}$ is the average transit depth across the whole wavelength range and $R_{P,0}$ is the planetary radius at a pressure $P=10$ bar. This left us with 68,000 and 64,000 training examples for the type 1 and type 2 retrievals respectively. In both cases 8,000 spectra were used for validation, and 1,000 were reserved for testing.

\section{Training and evaluation of the networks}\label{sec:training}

We used the datasets described in the previous section to train convolutional neural networks (CNNs). The choice of using convolutional layers for our neural networks was informed by previous results in the literature, which found that CNNs were the best performers amongst different machine learning algorithms \citep[e.g. ][]{Soboczenski2018BayesianRetrieval, Yip2020PeekingRetrievals}, while at the same time having a lower number of trainable weights than fully connected neural network. A brief description of what CNNs are, along with our implementation and how they were evaluated is presented below.

\subsection{Convolutional neural networks}

A {dense neural network} is a succession of linear transformations followed by non-linear activations. Given an input $\vec{x}$ with $N$ features, the activations $a_i$ of a dense layer with $M$ neurons can be written as follows:
\begin{equation}
    a_i = g\left(\sum^N_{j=0}w_{ij}x_j\right)\, \; (0\leq i\leq M)
\end{equation}
where $w_{ij}$ are the network weights and $g$ is a non-linear function referred to as an activation function. These activations are subsequently used as the inputs for the next layer. 

\begin{figure}
    \centering
    \includegraphics[width=0.49\textwidth]{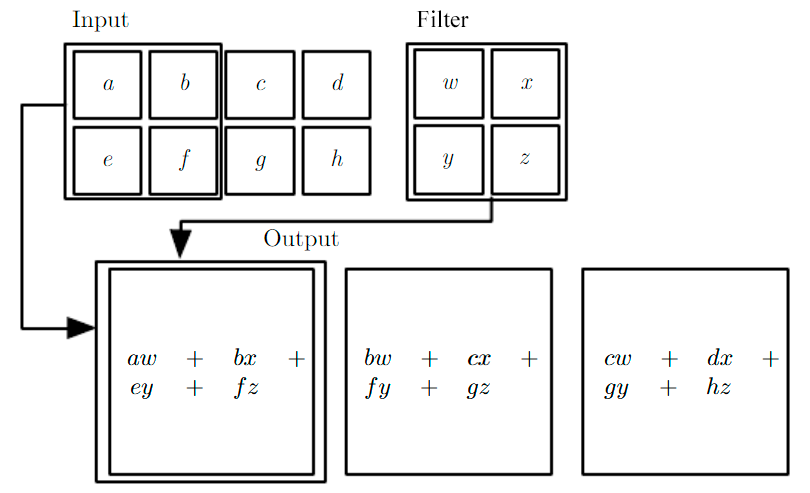}
    \caption{Example of a 1D convolutional layer with a single filter. Adapted from \citet{IanGoodfellowYoshuaBengio2015DeepBook}.}
    \label{fig:cnn_diagram}
\end{figure}

A one-dimensional {convolutional neural network} (CNN) substitutes some of the dense layers by 1D convolutional layers. A 1D convolutional layer is defined by a number of filters of a certain size. Given an input $\vec{x}$ of size $N\times L$, the activations of a 1D convolutional layer with $K$ filters $\vec{f}^k$ of size $M\times L$ ($M<N$) can be written as:
\begin{equation}
    a_{ik} = g\left(\sum^L_{l=0}\sum^M_{j=0}f_{jl}^kx_{i+j,l}\right)\; (0\leq i\leq N-M).
    \label{eq:cnn}
\end{equation}
In our case, $N$ is the number of wavelength bins, and since we feed both the original and pre-processed spectra as a 2-column array, the input size is $N\times2$. After applying $K$ convolutional filters, the output has size $N\times K$.
A graphical representation of the inner workings of a CNN can be seen in Fig. \ref{fig:cnn_diagram}.
The filter `slides' along the input, using Eq. \ref{eq:cnn} to obtain the output. Because the same filter is applied throughout the input, the number of trainable weights is reduced considerably when compared to fully connected layers. In addition, because the filter is applied to a group of inputs, CNNs are particularly suited for feature detection, and are commonly used for tasks involving images.

Training consists of finding the $w_{ij}$ or $f_{jl}^k$ that minimise a loss function. For a more detailed explanation see \citet{IanGoodfellowYoshuaBengio2015DeepBook}.  

We used the Python package \textit{Tensorflow} \citep{tensorflow2015-whitepaper} to implement our CNN. The network architecture is shown in Table \ref{table:cnn_arch}. To select the architecture, a grid search in hyperparameter space was conducted, choosing the smallest network possible without sacrificing performance. The total number of trainable weights for each combination of instrument and model type is shown in Table \ref{table:trainable_weights}

\begin{table}
    \label{table:cnn_arch}
    \centering
    \begin{tabular}{c c}
    \hline\hline 
         Layer Type & Hyperparameters  \\
         \hline
         Conv& $K=16$, $M=17$, $L=2$\\
         Conv& $K=32$, $M=9$, $L=16$\\
         Conv& $K=64$, $M=7$, $L=32$\\
         Dense& $M=128$\\
         \hline
    \end{tabular}
    \caption{Architecture of our CNN. We denote the number of filters and the filter size in convolutional layers by `f', and the number of neurons in the dense layer by `h'.}
\end{table}

\begin{table}
    \centering
    \begin{tabular}{r | c | c }
        \hline
         Instrument &  Type & Trainable weights\\
         \hline\hline
         \multirow{2}{4em}{WFC3} & 1 & 52,689\\
          & 2 & 48,819\\    
          \hline
         \multirow{2}{4em}{NIRSpec} & 1 & 437,713\\
          & 2 & 433,843\\
         \hline 
    \end{tabular}
    \caption{Number of trainable weights of the four CNNs.}
    \label{table:trainable_weights}
\end{table}

Max-pooling was used after each convolutional layer. Max-pooling reduces the size of an input by taking the maximum value over a group of inputs. We chose a pooling size of 2, downsizing an input of size $N\times L$  to size $(N/2+1)\times L$ or $[(N+1)/2]\times L$ depending on whether $N$ is odd or even.  The activation function was \textit{ReLU} for every layer except for the last one, for which we used a sigmoid activation function. \textit{ReLU} stands for \textit{Rectified Linear Unit}, and is defined as:
\begin{equation}
    f(x) = \max(0,\, x)
\end{equation}
The sigmoid activation function returns a value between 0 and 1 and is defined as:
\begin{equation}
    f(x) = \frac{1}{1+e^{-x}}
\end{equation}
{Although a linear activation function in the output layer was found to reach a virtually identical performance, ultimately the sigmoid function was chosen to ensure that the predictions are within the selected parameter ranges.}

The optimisation algorithm used to update the network's weights during training was the Adam optimizer \citep{adam-optimizer}. 

\begin{figure}
    \centering
    \includegraphics[width=0.495\textwidth]{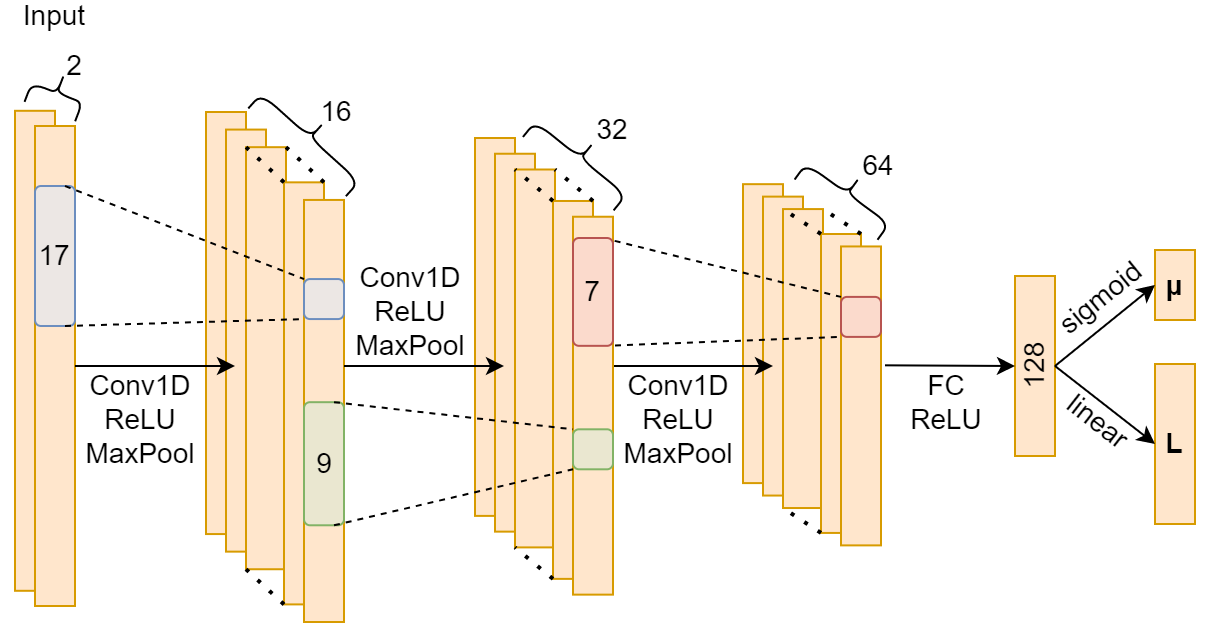}
    \caption{Schematic representation of the architecture of the CNNs used in this paper.}
    \label{fig:cnn_schem}
\end{figure}

Fig. \ref{fig:cnn_schem} shows a schematic representation of the architecture. The blue, green, and red boxes represent the convolutional filters. Although in the diagram they are only drawn in the first column, the filters are 2D and are applied to all columns. The height of the columns decreases after each MaxPooling operation and depends on the input size, and is therefore different for WFC3 and {NIRSpec} spectra.

\begin{figure}
    \centering
    \includegraphics[width=0.49\textwidth]{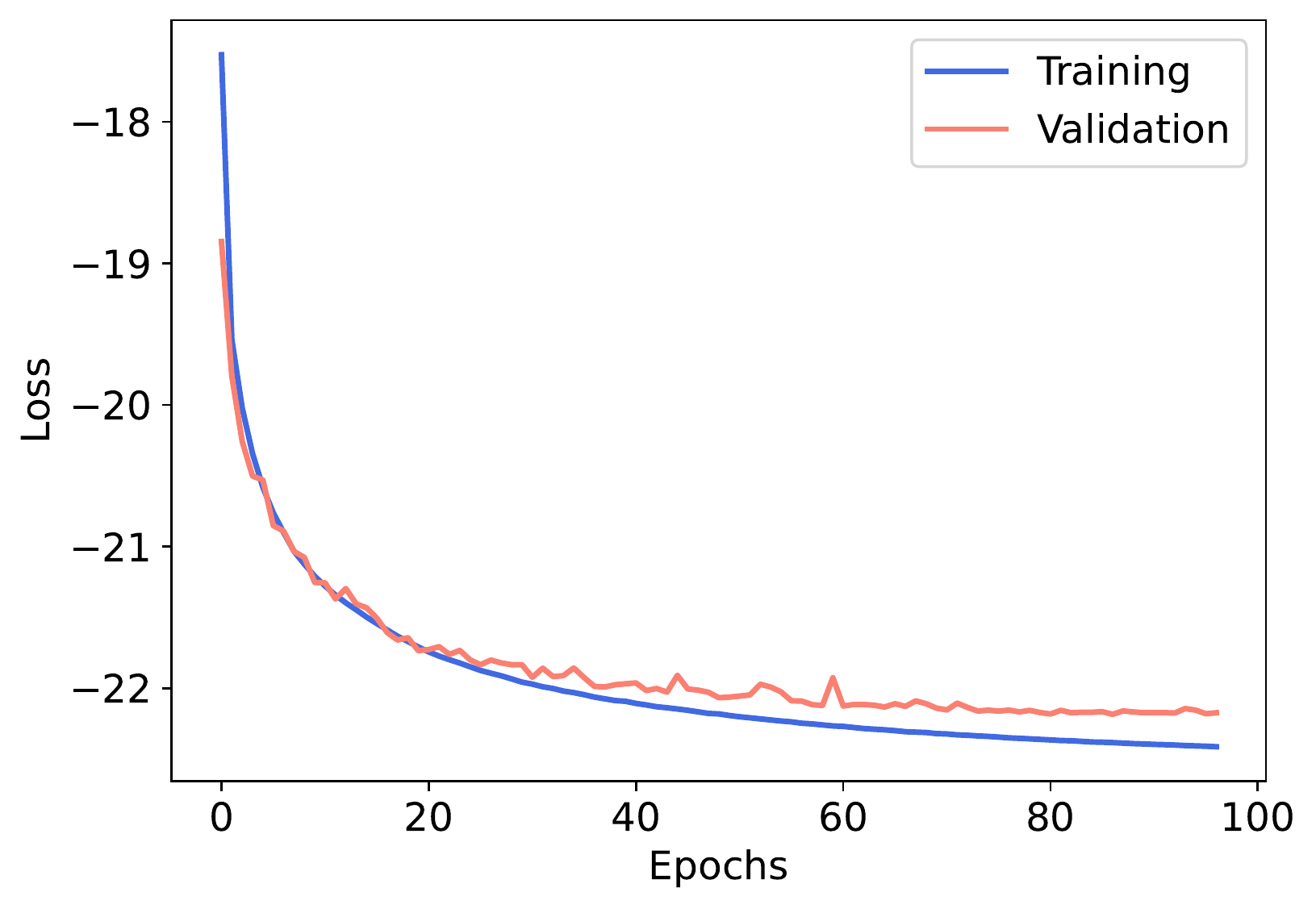}
    \caption{{Training and validation losses as a function of training epoch for HST type 2.}}
    \label{fig:train_vs_val}
\end{figure}

To obtain a probability distribution, our CNN does not predict only a single value for the parameters given a spectrum, but the means and the covariance matrix of a multivariate Gaussian. The latter is predicted via its Cholesky decomposition $\Sigma = (LL^T)^{-1}$. The loss function used to achieve this was the negative log-likelihood of the multivariate Gaussian, as presented in \citet{Cobb_2019}:
\begin{equation}
    \mathcal{L} = -2 \sum^D_{d=1}\log (l_{dd}) + (\vec{y}-\vec{\mu})^TLL^T(\vec{y}-\vec{\mu}),
\end{equation}
where $D$ is the number of dimensions, $l_{dd}$ the diagonal elements of $L$, $\vec{y}$ the true values and $\vec{\mu}$ the predictions. {Figure \ref{fig:train_vs_val} shows how the training and validation losses decrease as the training progresses. We stop the training if the validation loss has not decreased for 10 epochs.}

Once trained, to account for the observational error during a retrieval, we generate 1,000 noise realisations and do 1,000 forward passes of the CNN, one for each noise realisation. Finally, we randomly pick one point from each of the 1,000 normal distributions returned, which collectively form the probability distribution of the parameters. 

\subsection{Evaluating the performance of the CNNs}\label{sec:method_performance}

To evaluate the machine learning methods, as well as to facilitate the comparison with nested sampling, we generated sets of 1,000 simulated observations for each combination of type of model (1 and 2) and instrument (WFC3 and {NIRSpec}), yielding four sets in total. We then performed retrievals for these simulated observations with our CNNs and Multinest. All the Multinest retrievals had flat priors for all of the parameters.

The first metric we take into account is the coefficient of determination $R^2$ between the predicted and true values of the parameters. We consider the predicted value to be the median of the posterior distribution. In particular, we used the \textit{scikit-learn} \citep{scikit-learn} implementation of the $R^2$, defined as follows:
\begin{equation}
    R^2_{y,\, \mu} = 1-\frac{\sum_{i=1}^N(y_i-\mu_i)^2}{\sum_{i=1}^N(y_i-\bar{y})^2},
\end{equation}
where $y_i$ are the true values of a given parameter, $\bar{y}$ the average of the $y_i$, $\mu_i$ the predicted value, and $N$ the number of spectra. The coefficient of determination is measuring the normalised sum of squared differences between predicted and true values, therefore being a joint proxy for the bias and the variance of the model. In order to disentangle both, we also calculate the mean bias:
\begin{equation}
MB = \frac{1}{N}\sum_{i=1}^N(\mu_i - y_i).
\end{equation}
Both a high $R^2$ and low mean bias are desired. Computing both metrics is useful to determine whether a lower $R^2$ is due to a higher bias or a lower variance, and vice versa.

Additionally, we also compare the predicted values with the true values of the parameters. This is helpful to identify trends and compare the performance of the different retrieval methods in different regions of parameter space.

Retrieving atmospheric parameters from transmission spectra is a degenerate problem \citep[e.g. ][]{Brown2001TransmissionAtmospheres,Fortney2005TheSpectroscopy,Griffith2014DisentanglingExoplanets, Fisher2018RetrievalDegeneracy}. As such, the retrieved values are often incorrect. When evaluating the performance of a retrieval method it is crucial that we also consider the uncertainties of the retrieved values. To do so for our large samples of simulated observations we compute the difference between the predicted and true values in units of standard deviation. Because the posterior probability distributions are typically not Gaussian, a true standard deviation cannot be calculated. Instead we determine the percentile of the posterior distribution that the true value of the parameter falls in, or in other words, the fraction of points $p_-$ in the posterior distribution that are smaller than the true value. This can be converted into an equivalent standard deviation:
\begin{equation}
    \sigma_{eq} = \sqrt{2}\erf^{-1}(1-2p_-),
\end{equation}
{where $\erf$ is the error function and is defined as:
\begin{equation}
    \erf{z} = \frac{2}{\sqrt{\pi}}\int_0^z e^{-t^2} dt.
\end{equation}}
We can then evaluate the histogram of these differences in $\sigma_{eq}$ to see whether the retrieval method is overconfident, underconfident, or predicting accurate uncertainties.

This metric is arguably more important than the $R^2$ or the $MB$, at least for our goal of characterizing exoplanetary atmospheres. Predicted values that are further from the `truth' but have uncertainties that account for this distance will be preferred over predictions which are closer to the `truth', but due to too small uncertainties, are statistically incompatible with the true value.

\subsection{On the number of training examples} \label{number_examples}

We investigated the number of training examples that were needed to train our CNNs. This is an important factor to take into account when considering using machine learning, as its use may not be advantageous enough if too many forward model computations are needed to train it. We trained our CNNs using training sets of varying sizes and measured {how the loss decreased with increasing training set size}. 

\begin{figure}
    \centering
    \includegraphics[width=0.49\textwidth]{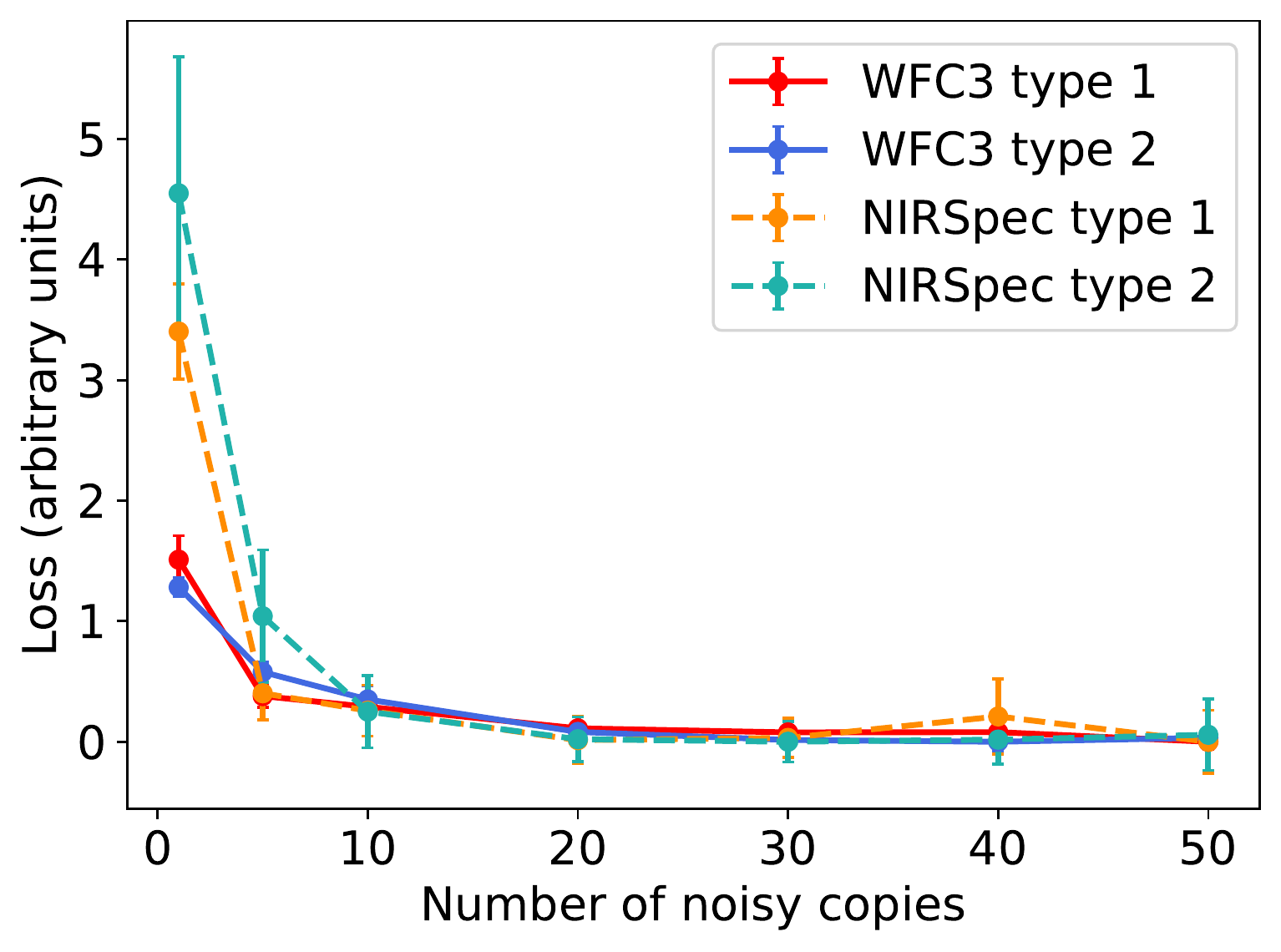}
    \includegraphics[width=0.49\textwidth]{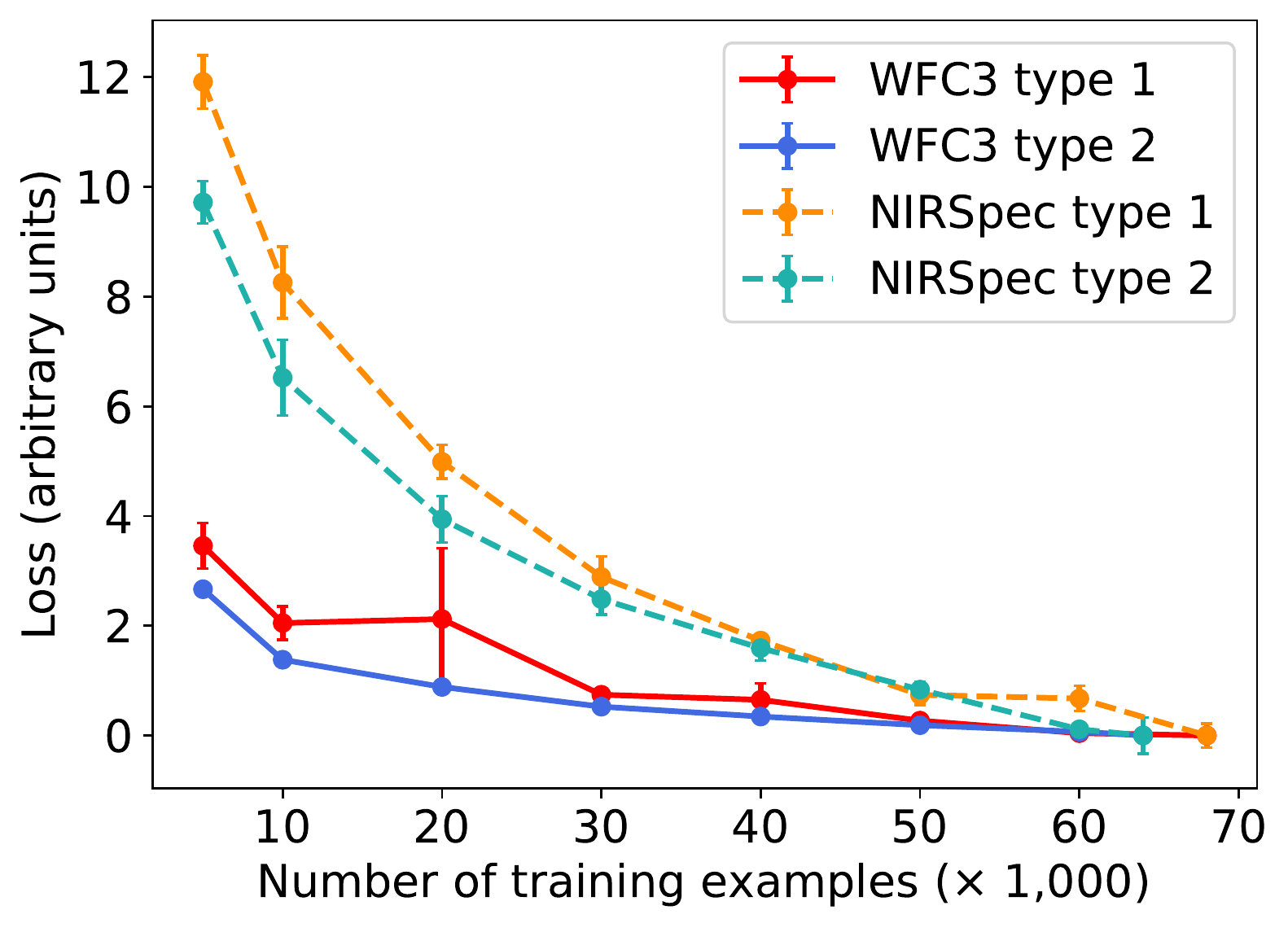}
    \caption{(Top) {Loss reached by the CNN for all combinations of instrument and model complexity versus the number of noisy copies}. (Bottom) {Loss reached by the CNN for all combinations of instrument and model complexity versus the number of forward models in the training set.} {The curves have been shifted vertically so they reach a value of zero}.}
    \label{fig:r2_vs_nc}
\end{figure}

First of all, we modified the number of noisy copies made of every forward model computed with ARCiS. Fig. \ref{fig:r2_vs_nc} (Top) shows how the {loss decreases with increasing number of noisy copies}, although there is little improvement beyond 20 noisy copies, if at all. {To facilitate comparison, all the curves were shifted so they reached a loss of zero.}

If we instead modify the number of forward models used to train the CNNs while keeping the number of noisy copies constant (20), we see a much more drastic decrease in performance with low number of forward models (see Fig. \ref{fig:r2_vs_nc} {(Bottom)}). This is an intuitive result as the CNN sees fewer combinations of parameters during training; in the previous case, it still sees many parameter combinations, but it does not see as many noise realisations. {When increasing the number of
training examples, the loss improves much more dramatically for NIRSpec than for WFC3, with the latter showing very little improvement beyond 30,000 examples. Since the NIRSpec spectra contain more information, it was to be expected that more examples would be needed to train the CNNs.}

\subsection{Limitations of the current approach}

The machine learning retrieval framework presented in this paper has some limitations, some related to our current implementation, and others inherent to supervised machine learning approaches. In our current implementation, the noise level is not an input but instead all training examples are sampled from the same noise distribution. This results in sub-optimal results when performing retrievals for observations with different noise levels. It is possible to retrain the CNNs with the observational noise, which we do in Section \ref{sec:wfc3}, but it is far from ideal. Additionally, we are training the CNNs to predict the means and covariance matrix of a multivariate Gaussian, constraining the shape of the posterior distributions. This makes it impossible to predict multimodal or uniform distributions. The latter in particular are very common as often only an upper or lower bound for a parameter can be retrieved.

The limitations related to the supervised learning approach itself are the fixed wavelength grid and the fixed model type. The training examples contain only the transit depth, no wavelength axis is provided. Therefore to perform a retrieval for a given observation, its wavelength axis must match exactly the one used during training. This is a lesser inconvenience, as observations from the same instruments will share the same wavelength axis and so one could simply train networks for specific instruments. Lastly, this approach lacks flexibility regarding the kinds of models used in the retrieval. Here we have trained CNNs for two different kinds of models, and if we wanted to analyse an observation with a different model (say, for example, with a different list of chemical species), it would be necessary to generate a completely new training set and train a CNN. Of course, since eventually the different kinds of models would likely be used to analyse more than one observation, using machine learning would probably still be computationally advantageous in the long run.

\section{Results}\label{sec:results}

In order to test the performance of the CNNs and have a benchmark to compare it against, we generated four sets of 1,000 simulated observations (one for each combination of instrument and model type), which we then retrieved with the CNNs and Multinest. Below we present the results for the different evaluation metrics discussed in Section \ref{sec:training}. A summary of these results can be found in Table \ref{table:summary_performance}.

\begin{table}
    \centering
    \begin{tabular}{r | c | c | c | c | c }
        \hline
        & & \multicolumn{2}{|c|}{$R^2$}  & \multicolumn{2}{|c}{\% $>3\sigma$}\\
        \hline
         Instrument &  Type & CNN & MN  & CNN & MN\\
         \hline\hline
         \multirow{2}{4em}{WFC3} & 1 & 0.30 & 0.30  & 0.0 & 8.5\\
          & 2 & 0.36 & 0.29 & 0.1 & 6.2\\    
          \hline
         \multirow{2}{4em}{NIRSpec} & 1 & 0.67 & 0.76  & 0.1 & 10.6\\
          & 2 & 0.71 & 0.76 &  0.2 & 8.2\\
         \hline
         \multicolumn{6}{c}{} 
    \end{tabular}
    \caption{Summary statistics of the CNN and Multinest bulk retrievals.}
    \label{table:summary_performance}
\end{table}

\subsection{Predicted vs true values}

The predicted vs true values plots for all parameters, instrument, and model type can be seen in  {Appendix \ref{app:pred_vs_true}}. 
For WFC3 type 1 retrievals the CNN and Multinest reach the same $R^2$, whereas for WFC3 type 2, the CNN actually outperforms Multinest in this metric. The CNNs also achieve a lower $MB$ (except for the $\log \kappa_{\rm{haze}}$ in the type 2 retrievals), although neither method shows large biases.

For {NIRSpec} retrievals, Multinest reaches a higher average $R^2$ in both cases. Yet if we focus on the type 2 retrievals, the $R^2$ of all the parameters is actually very close between both methods, with the CNN only doing significantly worse for the $\rm C/\rm O$ and the $P_{\rm{cloud}}$. Regarding the bias, there is no clear winner, and no method reaches systematically lower biases for all parameters. However, as for the WFC3 spectra, the bias is low for both methods.

\begin{figure}
    \centering
    \includegraphics[width=0.24\textwidth]{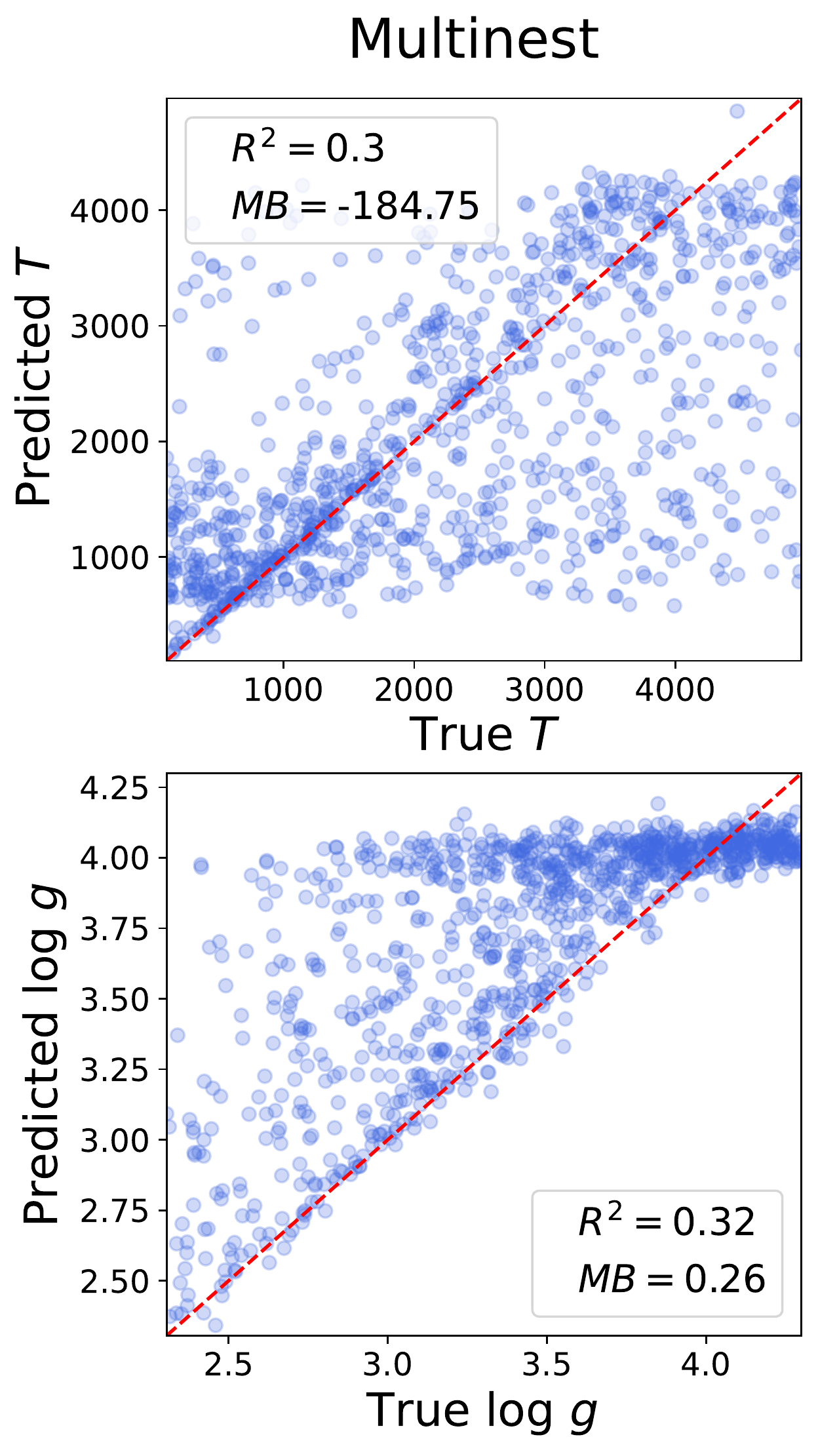}
    \includegraphics[width=0.24\textwidth]{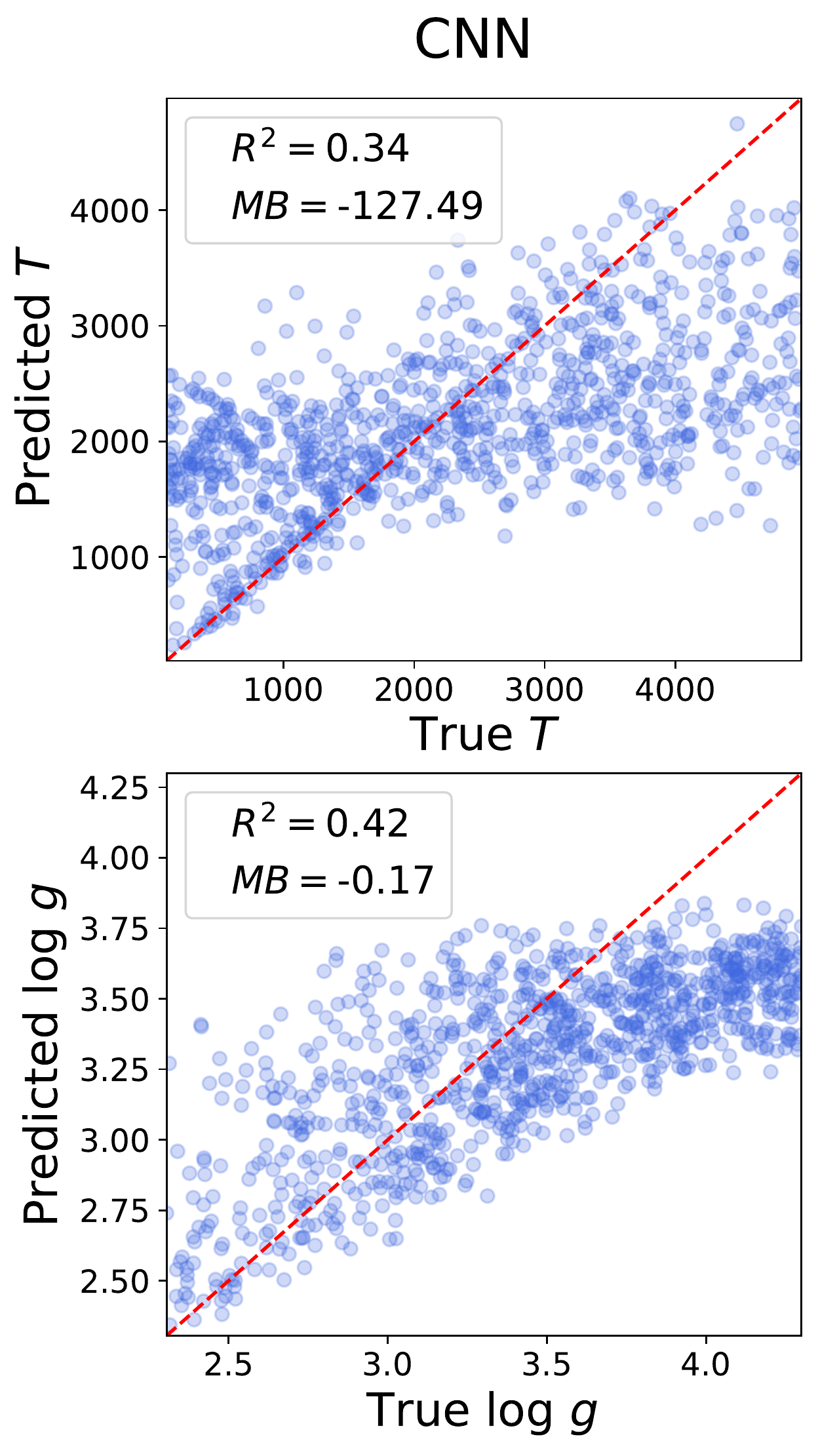}
    \caption{Predicted vs true values of the temperature and the $\log g$ for type 1 retrievals performed for simulated WFC3 transmission spectra. {(Left)} Multinest. {(Right)} CNN.}
    \label{fig:t_logg_wfc3_l1}
\end{figure}

More importantly, we observe similar trends in the predictions vs truths for both methods, which is an indication that the lack of predictive power in some regions of parameter space is mostly due to the lack of information in the data itself. 
There are however a few differences worth mentioning. Firstly, for WFC3 data (for both type 1 and 2 retrievals), the predictions vs truth plots for the temperature $T$ and the $\log g$ are markedly different, as can be seen in Fig. \ref{fig:t_logg_wfc3_l1} (only illustrated for type 1 retrievals, type 2 can be found in {Appendix \ref{app:pred_vs_true}}). In both cases the CNN achieves a higher $R^2$ and lower $MB$.

\begin{figure}
    \centering
    \includegraphics[width=0.24\textwidth]{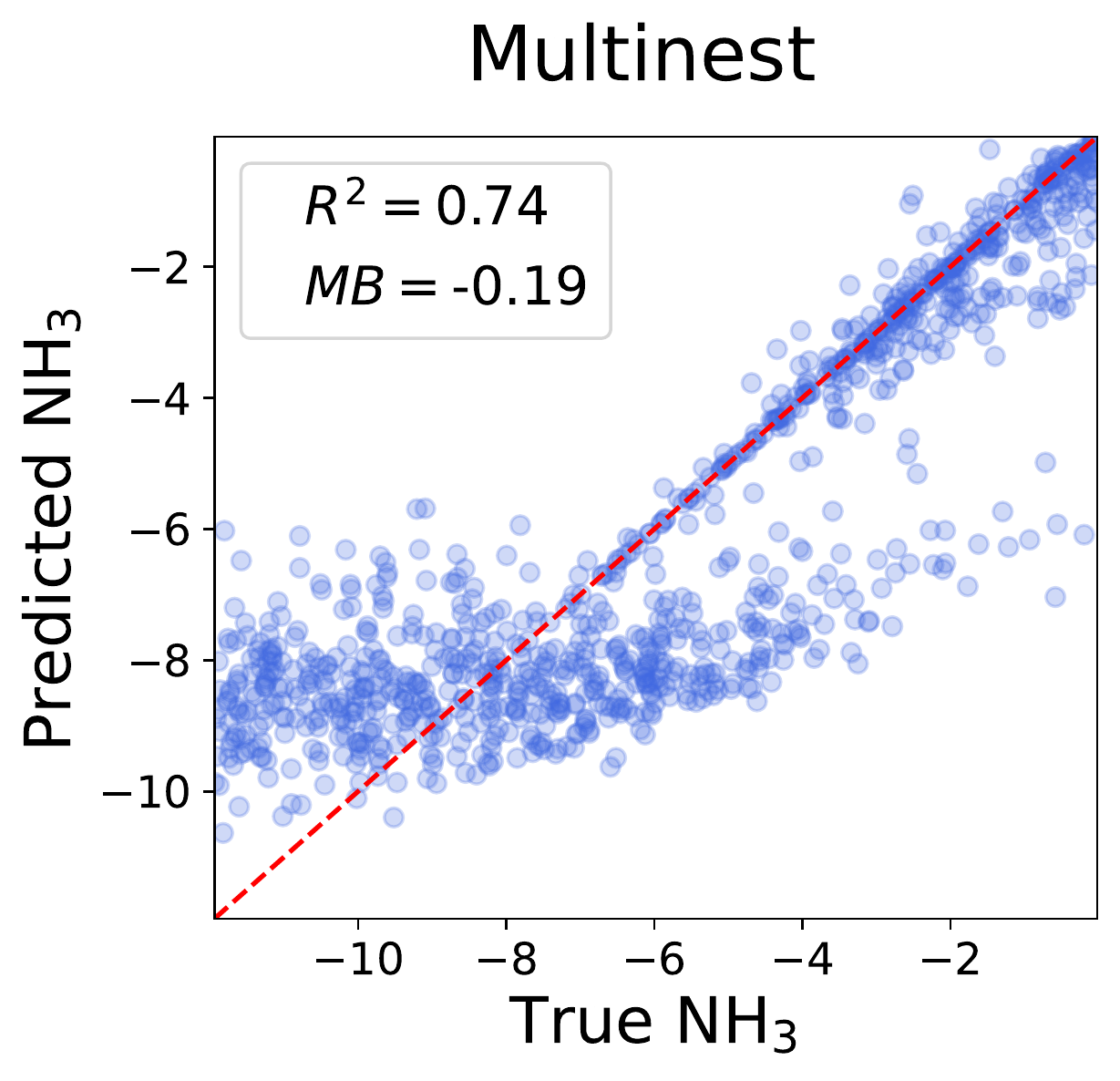}
    \includegraphics[width=0.24\textwidth]{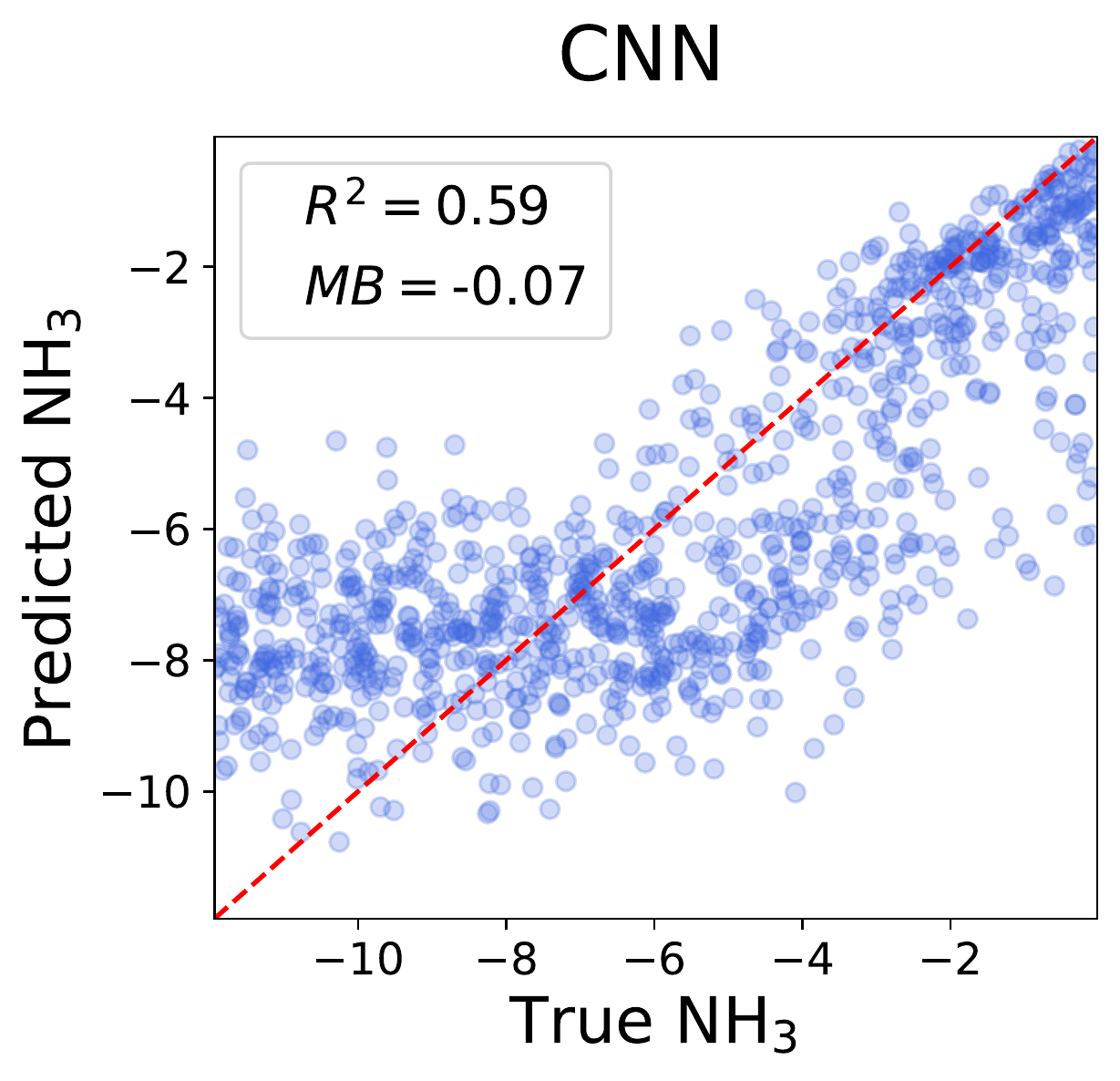}
    \caption{Predicted vs true values of the NH$_3$ abundance for type 1 retrievals performed for simulated {NIRSpec} transmission spectra. {(Left)} Multinest. {(Right)} CNN.}
    \label{fig:nh3_jwst_l1}
\end{figure}

The other main differences between both methods can be seen for parameters where the predicted vs true values clump along two `tails'. This can be interpreted as a degeneracy in the data for which Multinest predicts one of two possible values. In these cases, we observe that the CNN predictions are distributed between both `tails' without clumping. This is very evident in the predictions of the molecular abundances in type 1 retrievals. Fig. \ref{fig:nh3_jwst_l1} illustrates this for the retrieved values of the NH$_3$ abundance from {NIRSpec} spectra.

\subsection{Uncertainty estimates}

The coefficients of determination between predicted and true values discussed above are, of course, only half of the picture. As we have seen, there are many regions in parameter space for which the correct values of the parameters are not retrieved. Whether the retrieved parameters agree with the `ground truth' then depends on the entire posterior distribution. Following the method detailed in section \ref{sec:training}, we quantify whether the different methods get closer to the true values of the parameters.

\begin{figure*}
    \centering
    \includegraphics[width=0.49\textwidth]{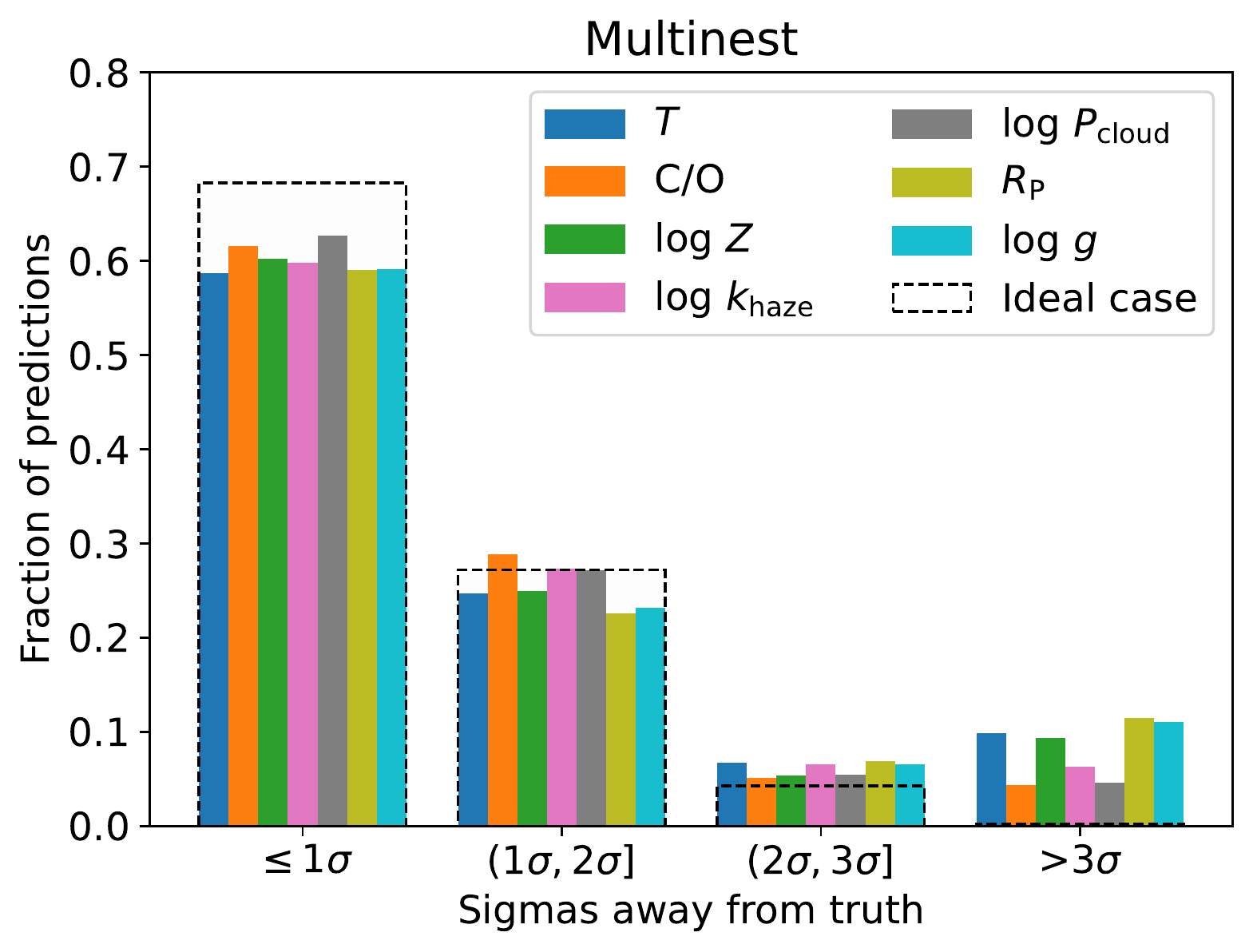}
    \includegraphics[width=0.49\textwidth]{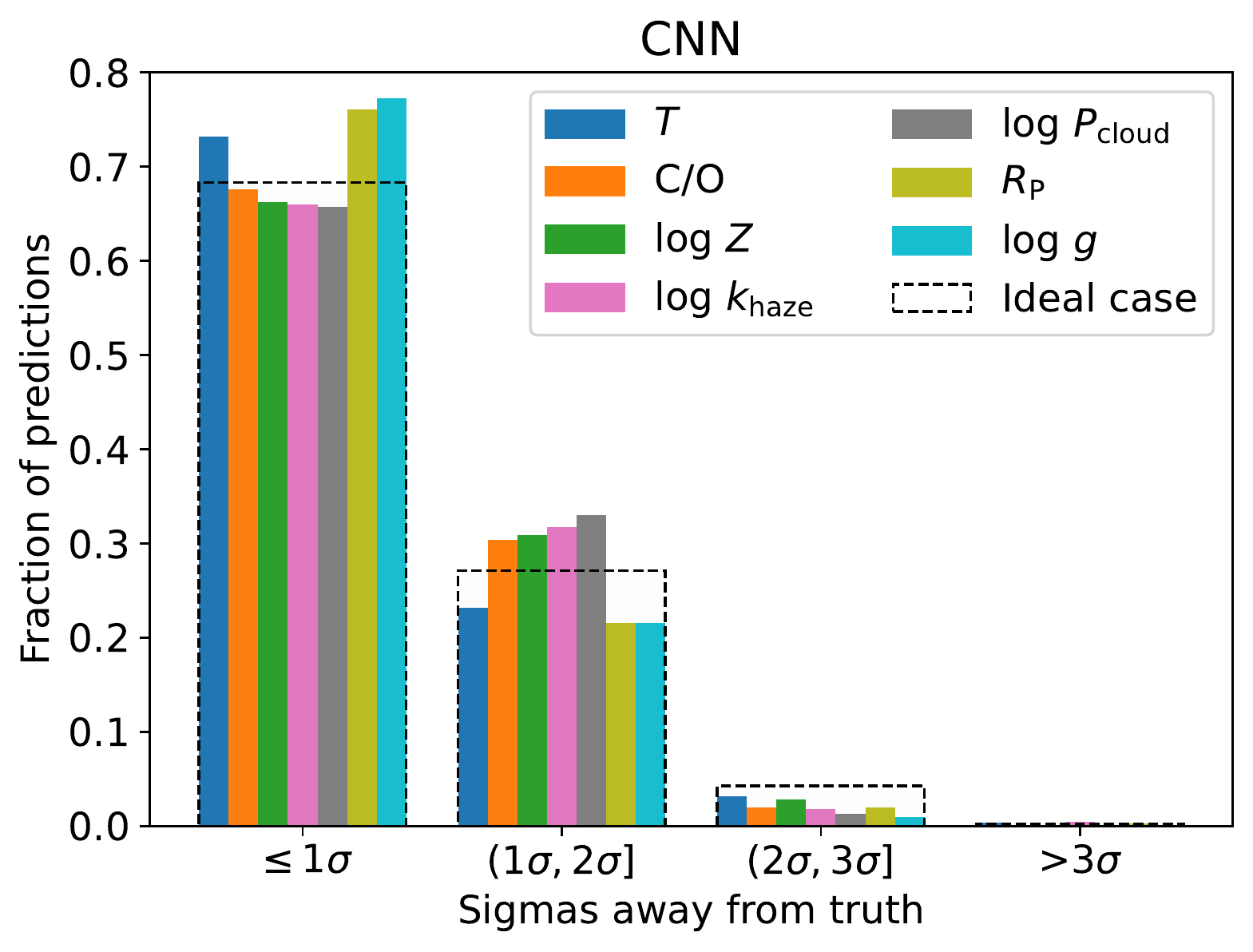}
    \caption{Distance in $\sigma$ between predictions and ground truths for type 2 retrievals performed for simulated NIRSpec observations. {(Left)} Multinest. {(Right)} CNN.}
    \label{fig:NIRSpec_level2_unc}
\end{figure*}

For type 2 retrievals of {NIRSpec} spectra, the histogram of the distances (in units of standard deviation) between predictions and true values for both methods can be seen in Fig. \ref{fig:NIRSpec_level2_unc}. {The `ideal case' represents our statistical expectation that $\sim68\%$ of predictions should be within $1\sigma$ of the ground truth, $\sim27\%$ within $2\sigma$ and so on}. For all other combinations of model type and instrument, the figures can be found in {Appendix \ref{app:sigma_diff}}. Here our CNNs follow closely the expected values, doing better than Multinest. For the CNNs the fraction of predictions lying more than $3\sigma$ away from the true value is always below $0.3\%$. While this remains true for the retrievals of synthetic WFC3 spectra, in this case we find an excess of predictions between $1\sigma$ and $2\sigma$ from the true value and a lack of predictions within $1\sigma$ and between $2\sigma$ and $3\sigma$ of the ground truth. Multinest on the other hand can be off by more than $3\sigma$ in between $\sim 2\%$ and $\sim20\%$ of retrievals, depending on the parameter, the model type, and the instrument. This means that there is a non-negligible fraction of spectra for which Multinest underestimates the uncertainties of the parameters it retrieves. This goes against our expectations as it is typically assumed that Multinest finds the `true' posterior probability distribution. Although finding the reasons behind this would be extremely valuable to the community, our goal in this paper is simply to compare the results obtained with Multinest and our CNNs.

\subsection{Real HST observations}\label{sec:wfc3}

Until now the CNNs have only been tested with simulated observations. 
The Exoplanets-A database of consistently reduced HST/WFC3 transmission spectra of 48 exoplanets \citep{Pye2019ExoplanetExoplanets} provide us with a unique opportunity to test our CNNs on a relatively large sample of real data. With this dataset we can study the real world performance of our CNNs, and compare it with Multinest. Retrievals of these spectra were ran with both our CNNs and Multinest, and the results were compared. In order to apply our machine learning technique to the real dataset, for each of the planets the CNNs were retrained with the observational noise from the real data applied to the training sets (see Section \ref{sec:data}). This is not ideal and work is ongoing to avoid it in the future, but for now it allows us to test our framework with real observations. Examples of the corner plots for retrievals for WASP-12b can be seen in Appendix \ref{app:corner-plots}.

\begin{figure*}
    \centering
    \includegraphics[width=0.49\textwidth]{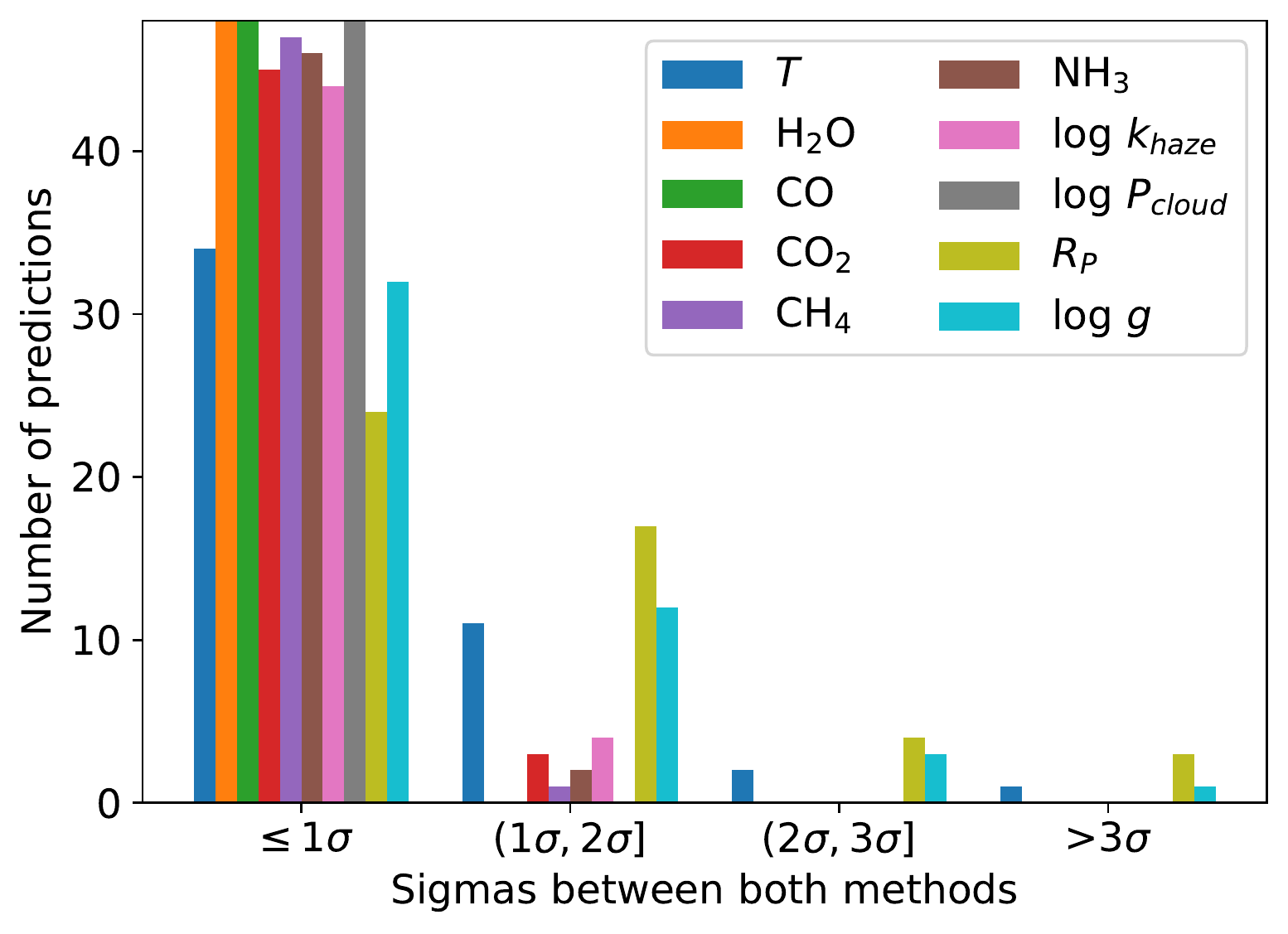}
    \includegraphics[width=0.49\textwidth]{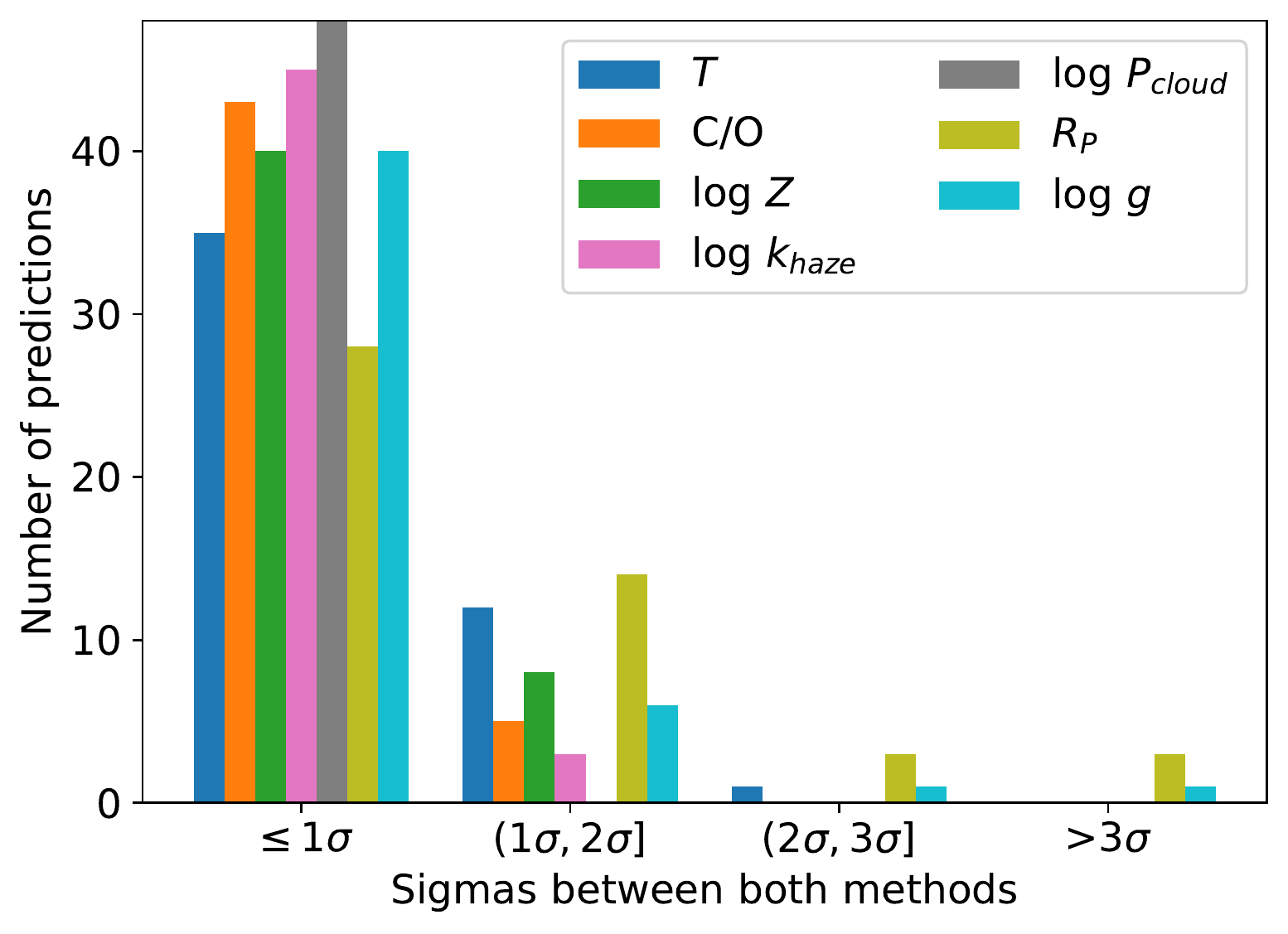}
    \caption{Differences between the parameters retrieved by Multinest and our CNNs for the 48 exoplanets in the Exoplanets-A database. Although these figures look similar to previous ones, here we are showing the difference between the predictions of both methods, and not the differences between the predictions and the truths. Therefore ideally the predictions for all spectra would be within $1\sigma$. {(Left)} Type 1 retrievals. {(Right)} Type 2 retrievals.}
    \label{fig:exoa_mn-cnn}
\end{figure*}

 The differences (measured in $\sigma$) between the parameters retrieved by the two methods can be seen in Fig. \ref{fig:exoa_mn-cnn}. These were calculated by assuming, for the method that predicted a lower value, a normal distribution with mean its median and standard deviation the 84th minus the 50th percentile; and similarly for the other method but with the 50th minus 16th percentile as the standard deviation. For most planets, both methods agree within $1\sigma$, with very few cases in which they disagree by more than $2\sigma$. These differences are discussed in Sect. \ref{sec:discussion}.

\section{Uncomfortable retrievals}\label{sec:uncomfortable}

The forward models we use to analyse exoplanet observations will virtually always be an incomplete description of the real atmosphere. On top of this, the assumptions made in the models can be wrong, causing our models to never be a one-to-one match of the actual observation. In machine learning jargon, these are generally referred to as `adversarial examples'. It is therefore important to know beforehand what to expect in these cases. Will our retrieval framework still provide good results? Or will it mislead us into a wrong characterisation of the exoplanet? Or will it break, letting us know that this is not something it was prepared for? To figure out how our machine learning retrieval framework responds in these scenarios, and to compare it with nested sampling, we modified our synthetic observations in different ways while keeping the retrieval frameworks identical. In particular we ran three experiments, namely adding an extra chemical species to our type 1 models, removing species from our type 2 models, and simulating the effect of stellar spots. 

\subsection{Type 1 retrievals of synthetic observations with ${\rm AlO}$}

\begin{figure}
    \centering
    \includegraphics[width=0.49\textwidth]{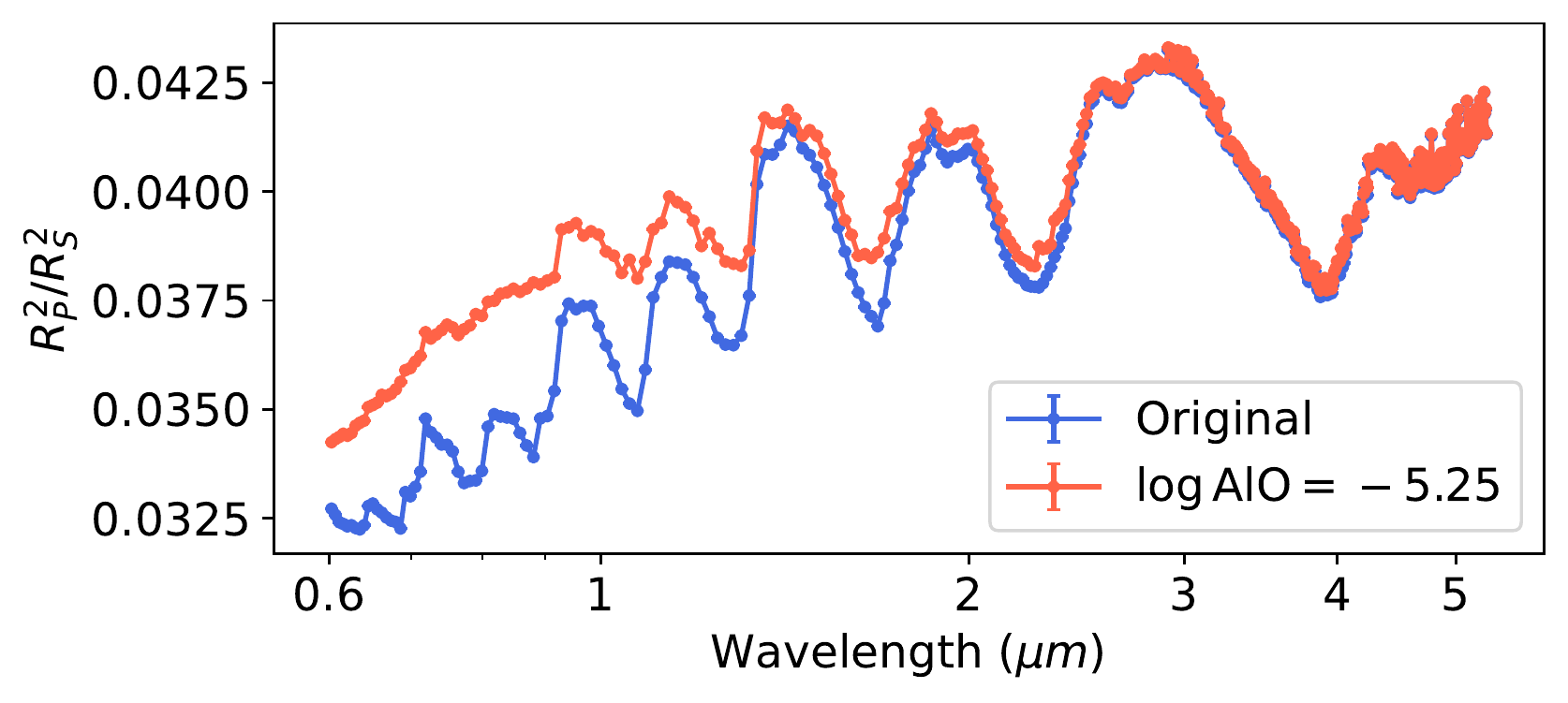}
    \caption{Comparison between the transmission spectra of an AlO-free atmosphere and an atmosphere with $\log \rm{AlO}=-5.25$. All other parameters remain identical.}
    \label{fig:alo-525}
\end{figure}

Firstly, we added AlO to our type 1 models with an abundance $\log {\rm AlO} = -5.25$. This choice was influenced by \cite{chubb_alo} reporting AlO on WASP-43b with this abundance. We generated 735 simulated {NIRSpec} observations (the original set was 1,000, but we filtered unphysical atmospheres as discussed in section 2), sampled randomly from the parameter space. These synthetic spectra were then retrieved using the same retrieval frameworks as in Section \ref{sec:results}, which assume that no AlO is present in the atmosphere. An example of how the addition of AlO at an abundance of $\log {\rm AlO} = -5.25$ modifies the transmission spectrum can be seen in Fig. \ref{fig:alo-525}.

\begin{figure*}
    \centering    
    \includegraphics[width=0.49\textwidth]{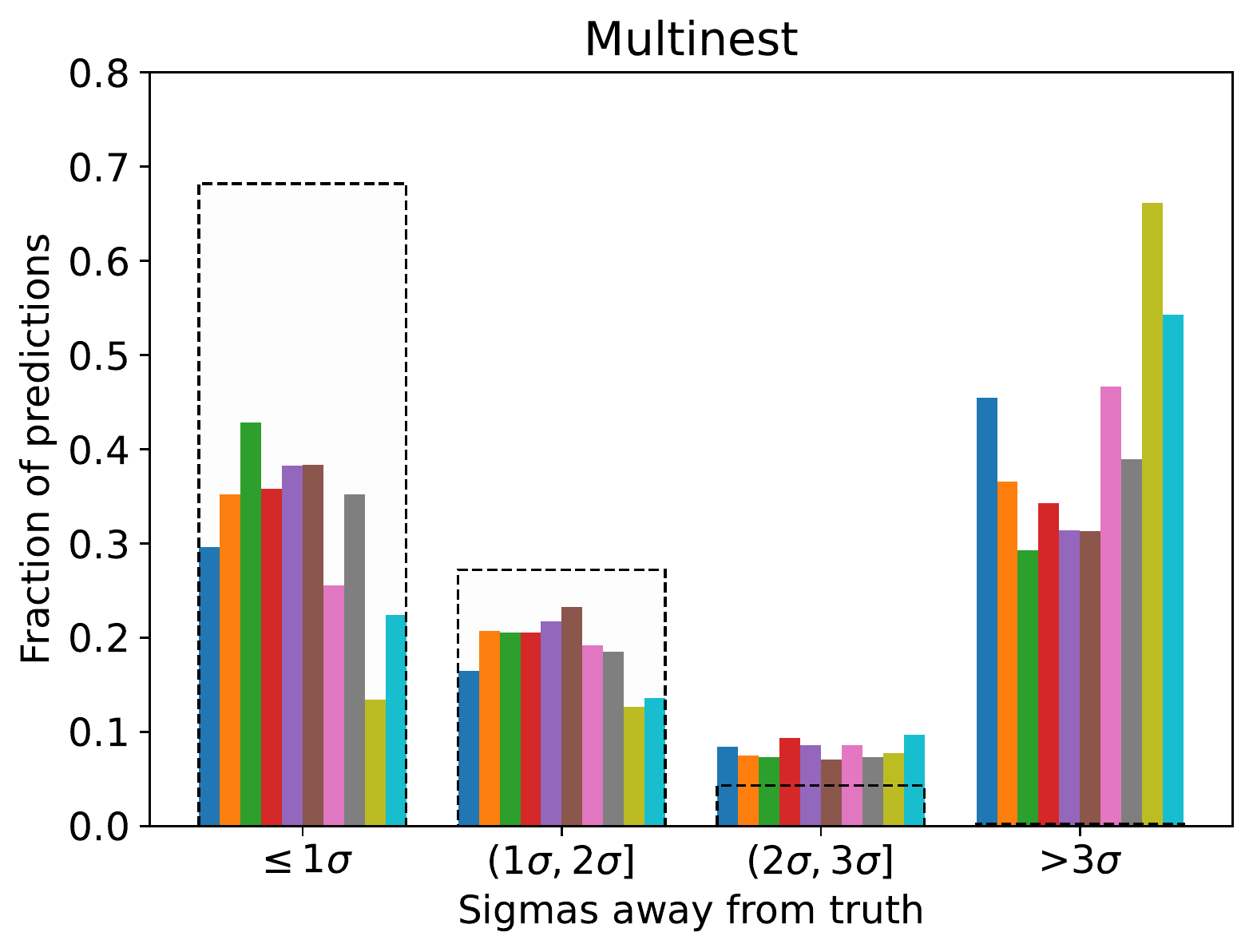}
    \includegraphics[width=0.49\textwidth]{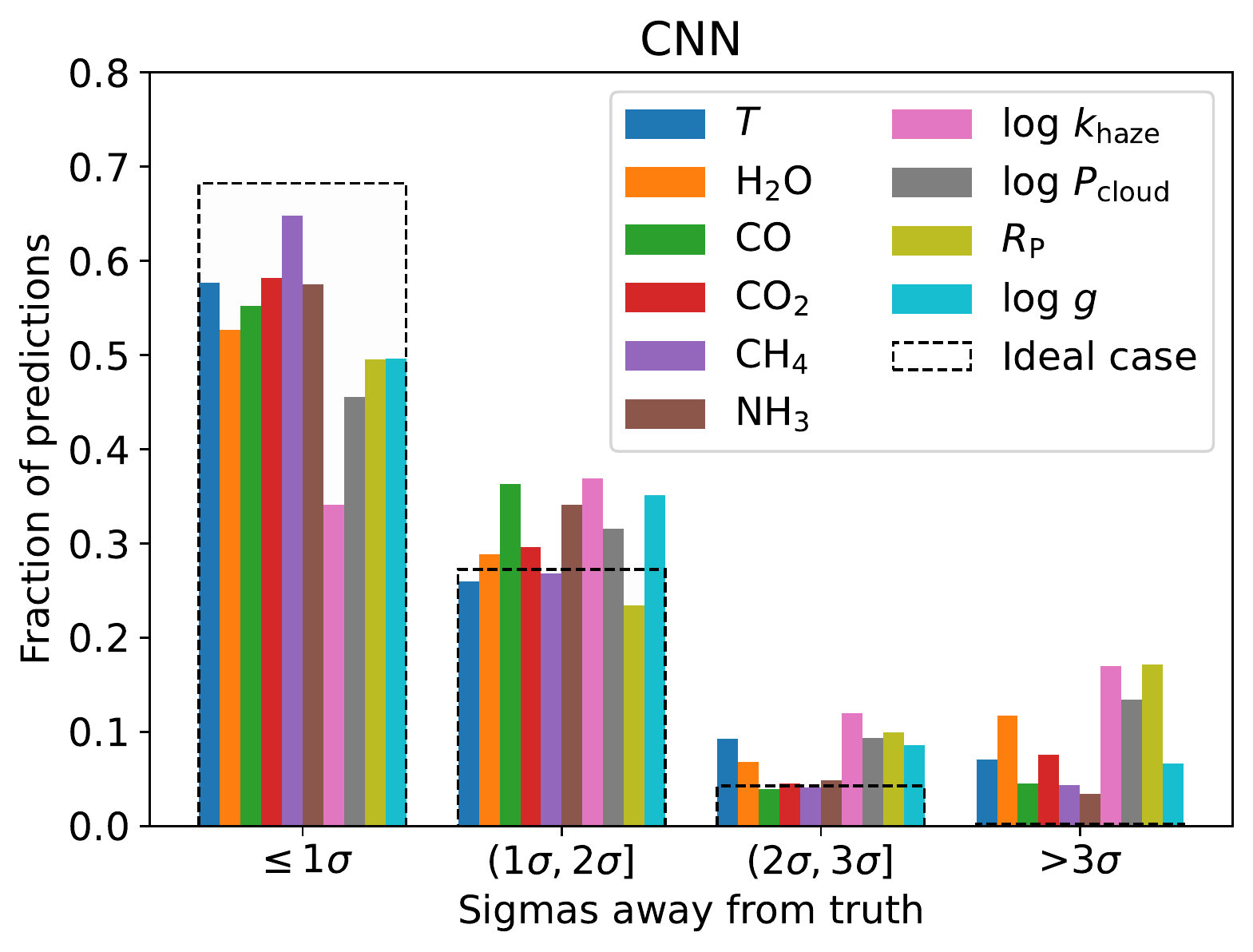}
    \caption{Distance in $\sigma$ between predictions and ground truths for level 1 retrievals of simulated NIRSpec observations of exoplanet atmospheres with $\log \rm AlO = -5.25$. {(Left)} Multinest. {(Right)} CNN.}
    \label{fig:sigmas_away_alo}
\end{figure*}

We analysed the retrievals as detailed in Section \ref{sec:training}, just as we did in Section \ref{sec:results}. As expected, both Multinest and the CNN achieved a lower $R^2$, with Multinest's $R^2$ still being better than that of the CNN ($R^2=0.39$ vs. $R^2=0.26$). Multinest also achieves a lower $MB$ for most parameters. However when we look at the whole picture and take the uncertainties into consideration (Fig. \ref{fig:sigmas_away_alo}), the CNN would still be the preferred method, with only $\sim 9\%$ of predictions further than $3\sigma$ from the truth compared to $\sim 41\%$ for Multinest.

\subsection{Type 2 retrievals of synthetic observations without $\rm TiO$ or $\rm VO$}

\begin{figure}
    \centering
    \includegraphics[width=0.49\textwidth]{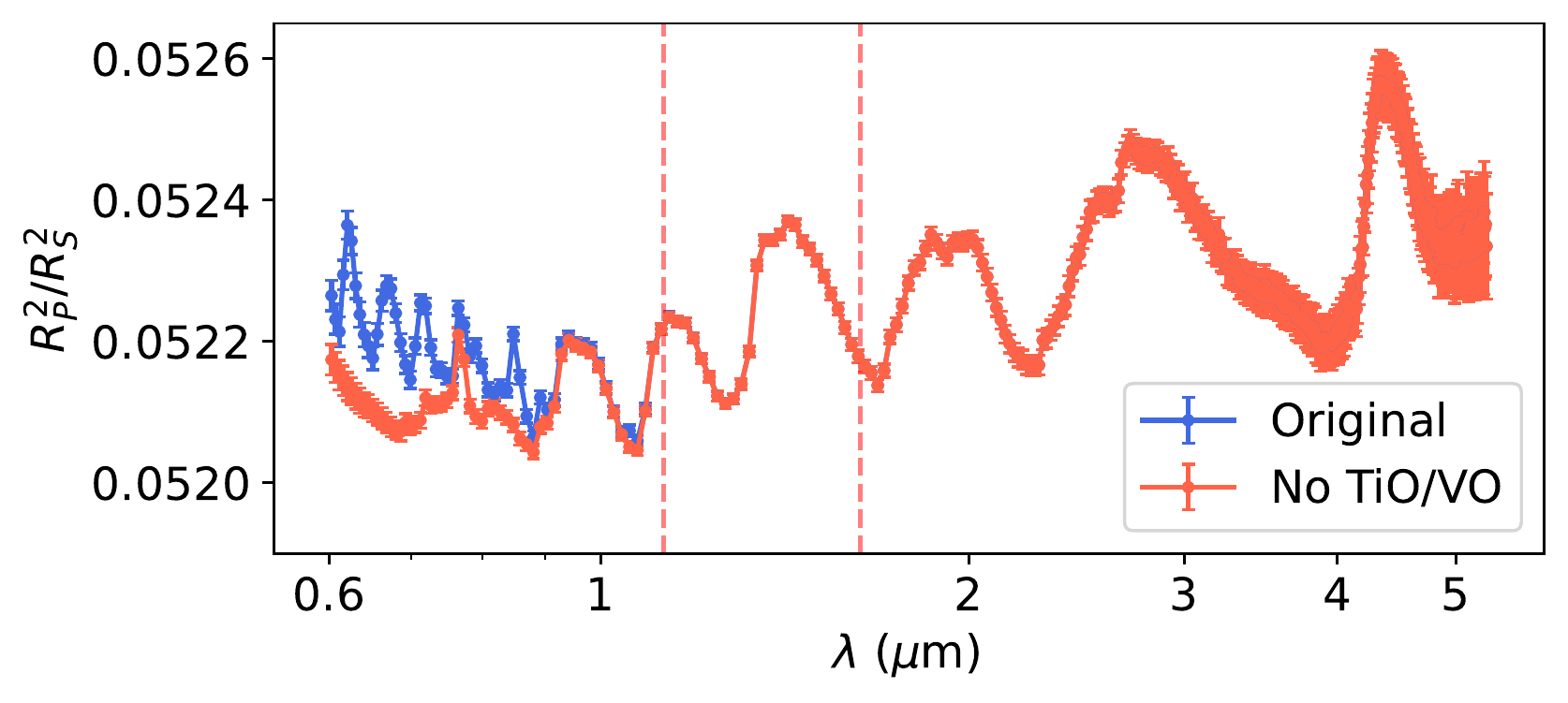}
    \caption{Comparison between transmission spectra of two identical atmospheres, except for the removal of TiO and VO in one of them. The red dashed lines indicate HST/WFC3's wavelength coverage, clearly showing its inability to detect this difference.}
    \label{fig:-tio-vo_spectra}
\end{figure}

To test the effects of an incorrect list of chemical species in type 2 retrievals, we removed TiO and VO and observed the effect it had on the retrieved parameters.
TiO and VO were expected to be major opacity sources in the atmospheres of hot Jupiters \citep[e.g.,][]{Hubeny2003APlanets, Fortney2008AAtmospheres}, however the lack of detection in multiple planets \citep[e.g.,][]{Helling2019AstronomyIonosphere, Merritt2020Non-detectionSpectroscopy, Hoeijmakers2020HotB} suggests that perhaps these species are condensing out of the gas phase at the day-night limb. {We therefore investigate how different assumptions regarding the presence of TiO and VO in the gas phase affect the results of the retrievals}, stressing again that our goal here is simply to compare how Multinest and our CNN behave in this scenario. To compare the behaviour of the CNN and Multinest under this scenario, we generated a set of 349 simulated observations (filtered down from 500 as discussed in Section \ref{sec:data} to get rid of extreme atmospheres) and retrieved them with our CNN and Multinest, in both cases assuming TiO and VO are part of the gas phase chemistry of the atmosphere. Removing these species changes only the optical part of the spectrum (see Fig. \ref{fig:-tio-vo_spectra}), and thus HST is unable to observe the differences. We therefore limit this experiment to simulated {NIRSpec} observations.

\begin{figure}
    \centering
    \includegraphics[width=0.24\textwidth]{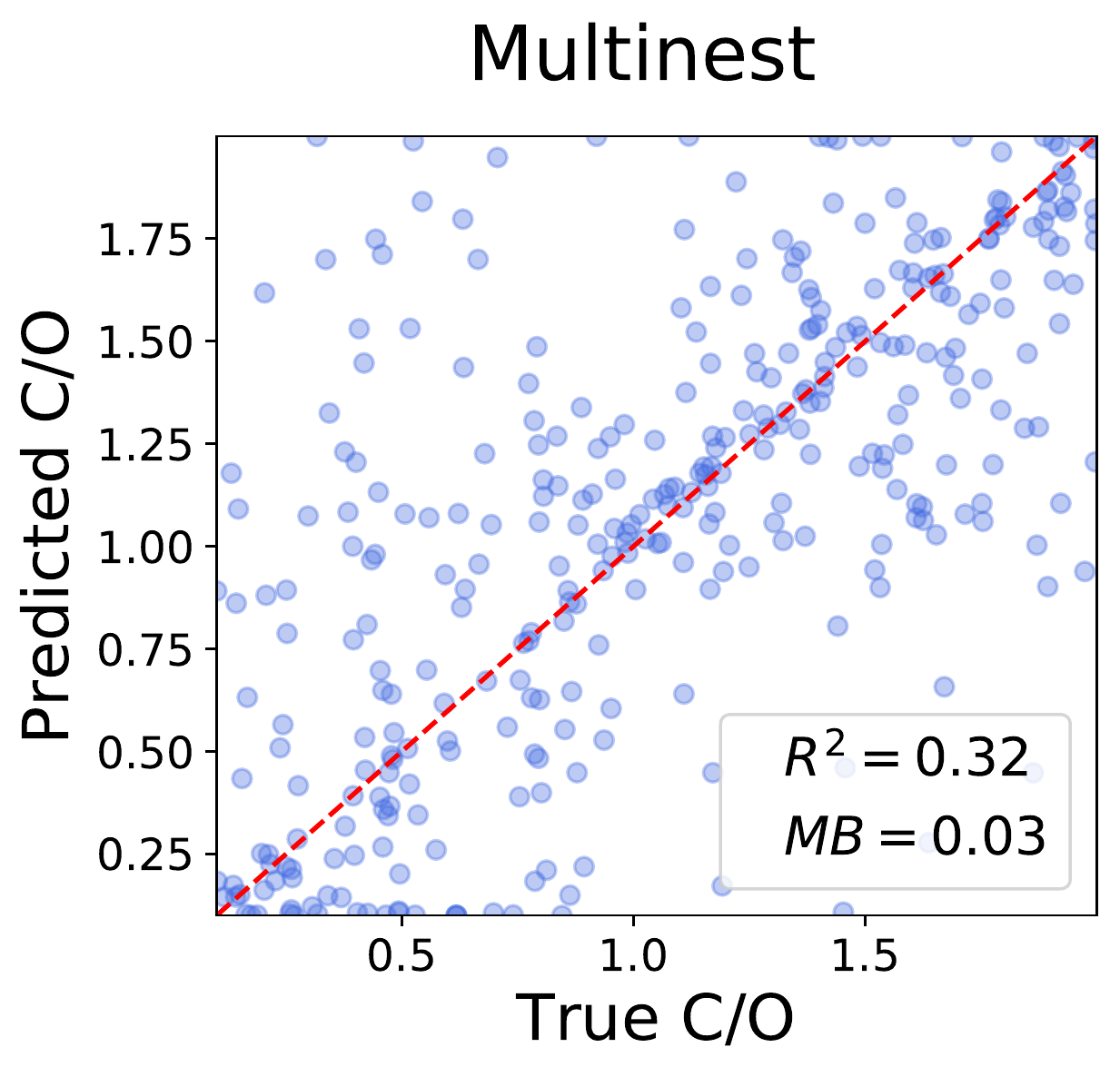}
    \includegraphics[width=0.24\textwidth]{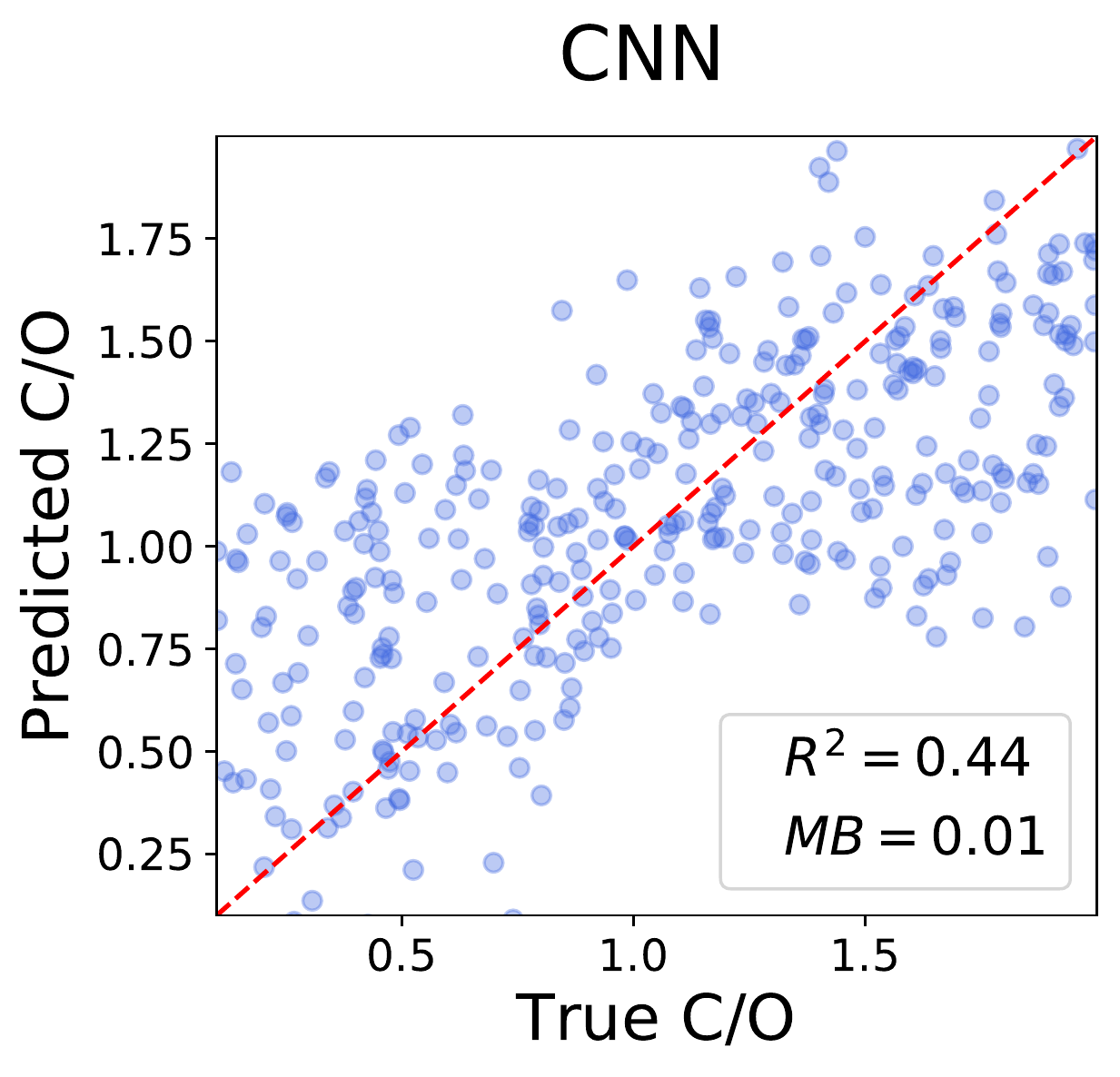}
    \caption{Predicted vs true values of the C/O of simulated spectra without TiO and VO. {(Left)} Multinest. {(Right)} CNN.}
    \label{fig:-tio_co}
\end{figure}

\begin{figure*}
    \centering
    \includegraphics[width=0.49\textwidth]{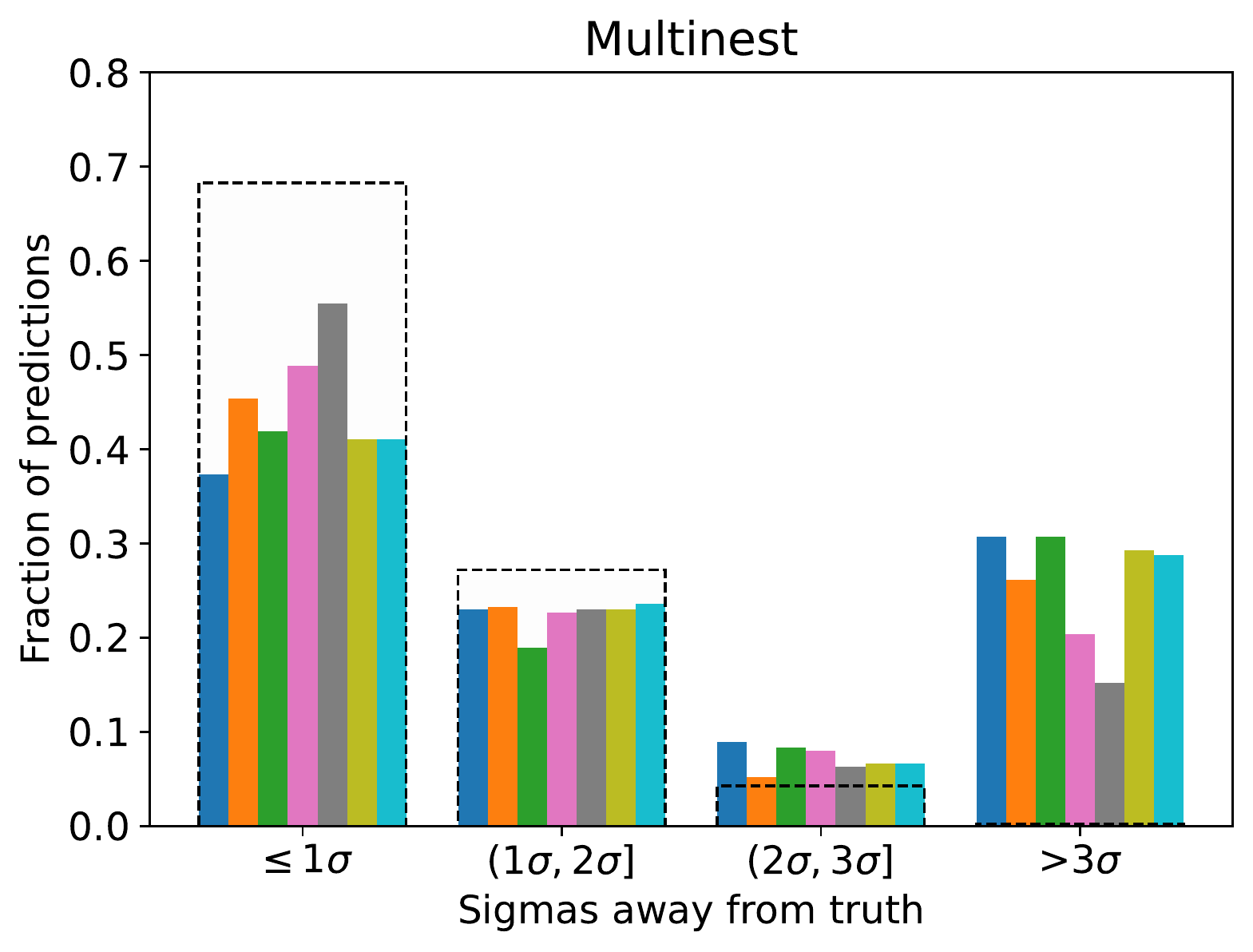}
    \includegraphics[width=0.49\textwidth]{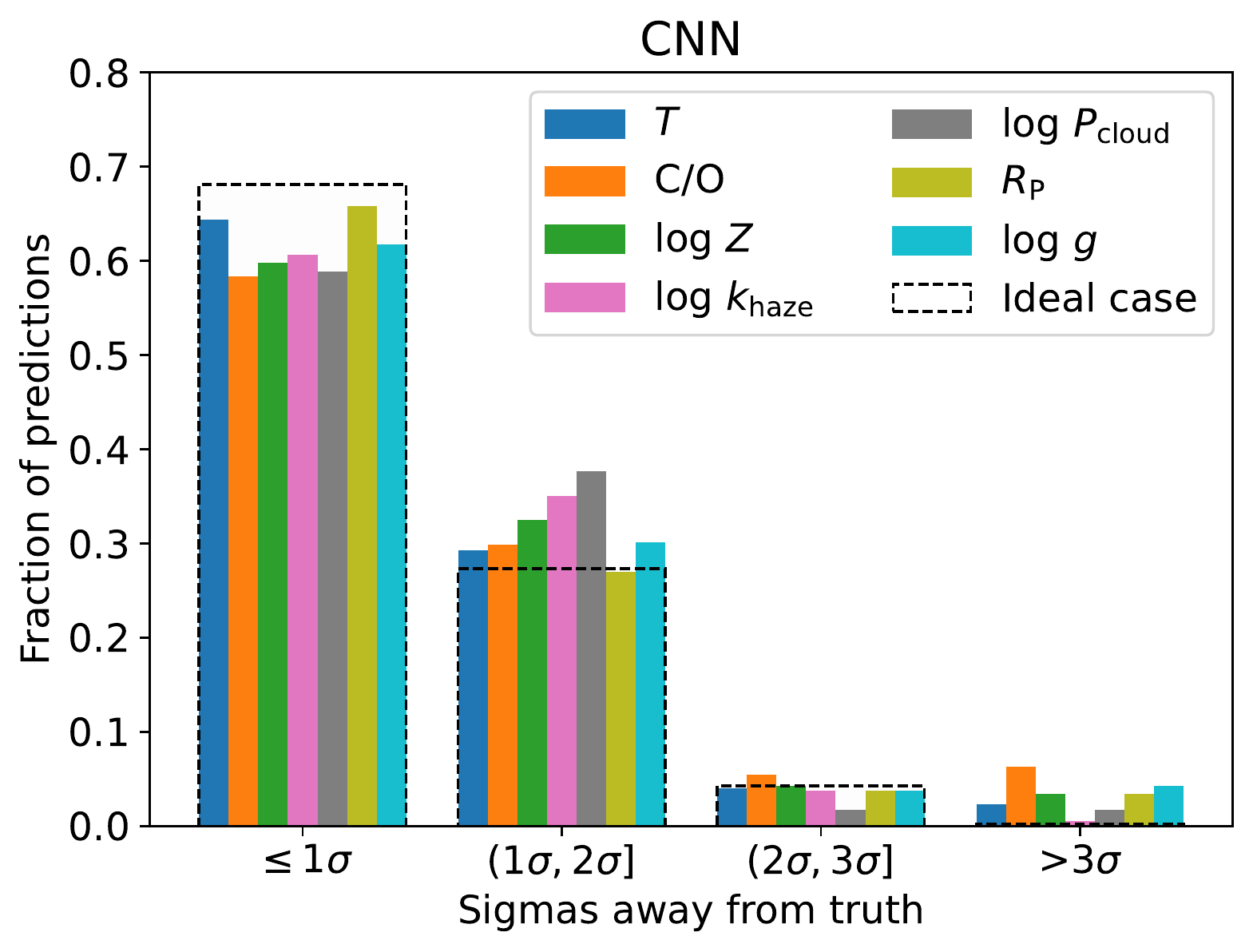}
    \caption{Distance in $\sigma$ between predictions and ground truths for type 2 retrievals of simulated {NIRSpec} observations of exoplanet atmospheres without TiO and VO in the gas phase. {(Left)} Multinest. {(Right)} CNN.}
    \label{fig:sigmas_away_-tio-vo}
\end{figure*}

 The CNN reaches a coefficient of determination between predictions and truths of $R^2=0.67$, falling from the regular value of $R^2=0.71$, while for Multinest it falls from $R^2=0.76$ to $R^2=0.63$. In particular the C/O predictions made by our CNN in this scenario are significantly better than those of Multinest, as can be seen in Fig. \ref{fig:-tio_co}. We find the bias to be small with both methods, with no method doing systematically better for all parameters. Most importantly however, we can see in Fig. \ref{fig:sigmas_away_-tio-vo} that Multinest is much more overconfident, with $\sim26\%$ of predictions more than $3\sigma$ away from the true value compared to only $\sim3\%$ for the CNN.

\subsection{Simulating unocculted star spots in the synthetic observations}

Finally, we also simulated the effect of unocculted star spots in the transmission spectra following the method described by \citet{ZellemForecastingSpectra}. If the spot coverage of the host star $f$ (defined as the fraction of the surface covered by spots) is heterogeneous and the planet transits in front of an area with no spots (or fewer than in the unocculted region), the planet will be blocking a hotter, bluer region than the overall stellar disk and thus the transit depth will be larger in shorter wavelengths. The transit depth modulated by the unocculted stellar activity can be calculated using
\begin{equation}\label{delta_active}
    \delta_{active,transit}=\left(\frac{R_P}{R_S}\right)^2 \left[1+f\left(\frac{B_{spots}}{B_{stars}}-1\right)\right]^{-1},
\end{equation}
where $\delta_{active,transit}$ is the transit depth and $B_{spots}$ and $B_{star}$ are the black body spectra of the spots and the star respectively.

\begin{figure}
    \centering
    \includegraphics[width=0.49\textwidth]{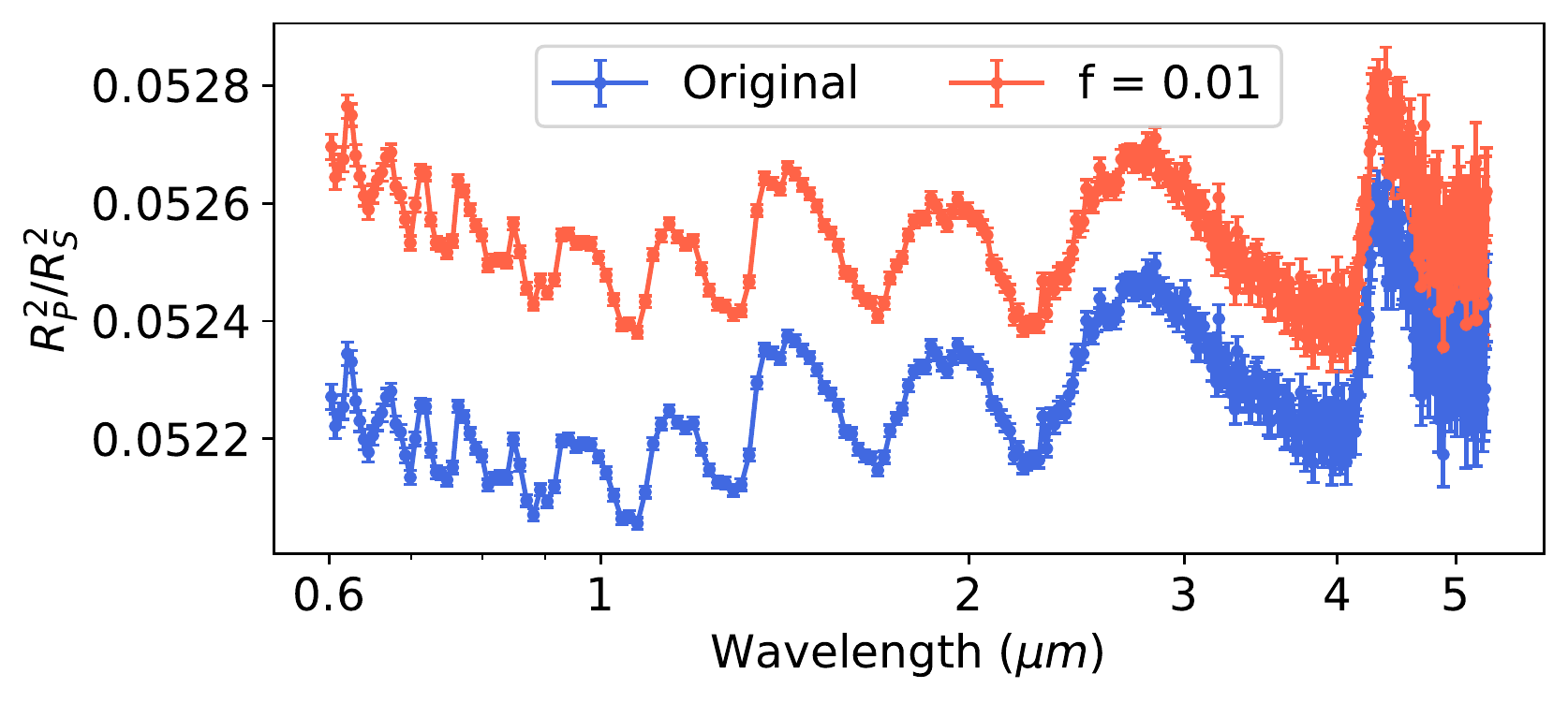}
    \caption{Comparison between simulated transmission spectra of an exoplanet orbiting a spot-less star and a planet orbiting a star with 1\% of its unocculted surface covered by spots. The transit depth increases more at shorter wavelengths.}
    \label{fig:spots_spectra}
\end{figure}

To test how this changes the retrieved parameters, we have assumed a star with $T_{star}=5500\,K$, $T_{spot}=4000K\,$ and a spot coverage of $f=0$ in the transiting region and $f=0.01$ in the non-transiting region. These values are consistent with each other and have been taken from \cite{Rackham2018THESTARS}. Fig. \ref{fig:spots_spectra} shows how these spots modify the observed transmission spectrum.

\begin{figure*}
    \centering
    \includegraphics[width=0.49\textwidth]{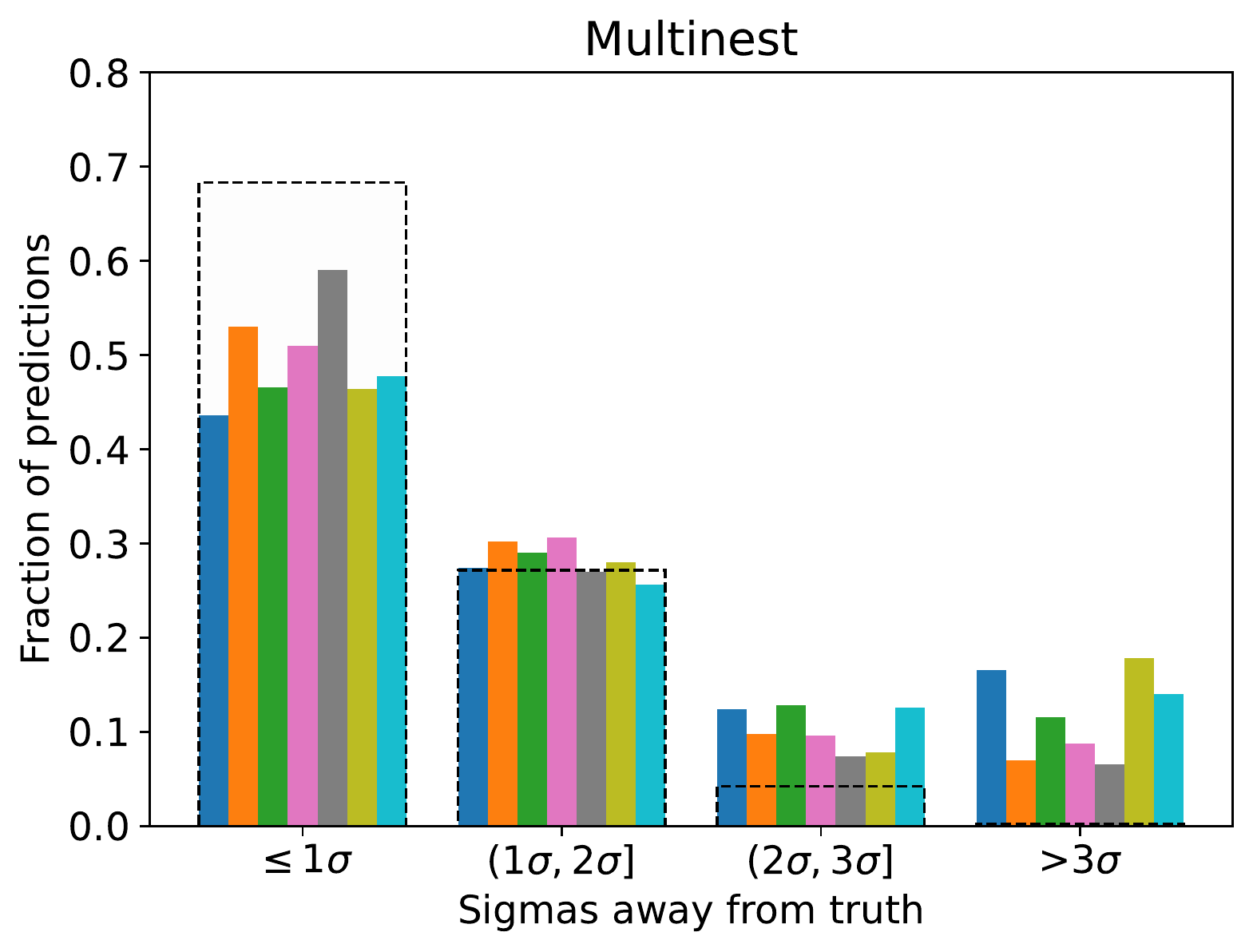}
    \includegraphics[width=0.49\textwidth]{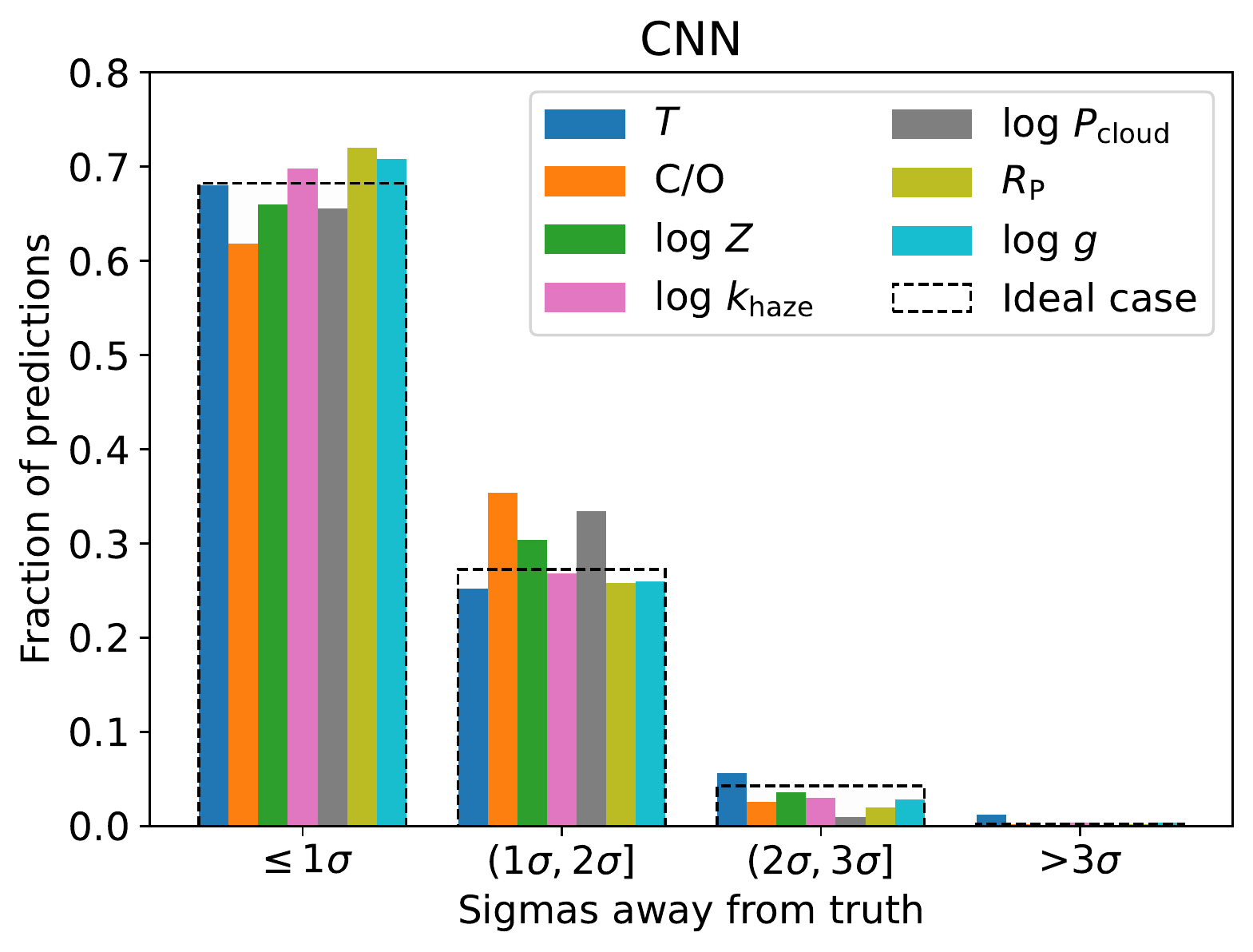}
    \caption{Distance in $\sigma$ between predictions and ground truths for type 2 retrievals of simulated {NIRSpec} observations of exoplanets orbiting around heterogeneously spotted host stars. {(Left)} Multinest. {(Right)} CNN.}
    \label{fig:sigmas_away_starspots}
\end{figure*}

Again, we generated a set of 500 simulated {NIRSpec} observations of exoplanets orbiting heterogeneously spotted host stars as described above (with type 2 model complexity) and retrieved them with our CNN and Multinest. Similar to previous experiments; we observe a reduction in $R^2$, although now Multinest still reaches a higher coefficient of determination ($R^2=0.70$) than our CNN ($R^2=0.64$). Despite this, the CNN retrievals would be preferred since Multinest is again quite overconfident as is made clear by Fig. \ref{fig:sigmas_away_starspots}. As for the previous experiments, the bias is low for all parameters with no method performing systematically better than the other for all parameters. Averaging over all parameters, $\sim12\%$ of Multinest predictions are more than $3\sigma$ away from the ground truth, whereas this fraction remains at the expected value of $\sim0.3\%$ for our CNN. 

\section{Discussion}\label{sec:discussion}

As presented in section \ref{sec:results}, our CNNs outperform Multinest when retrieving WFC3 simulated observations, achieving both a higher $R^2$, a lower bias, and a better estimate of the uncertainty. For {NIRSpec} though, the CNNs never reach a higher $R^2$ than Multinest. However this is not the case for the bias, which is low in all cases and is not systematically lower for any of the methods. Nevertheless when we also consider the uncertainty of the retrieved values, our CNN still performs better, with almost all predictions being within $3\sigma$ of the true value. 

It is also important to highlight the similar trends in the predicted values when compared with the ground truths for the CNNs and Multinest, particularly for {NIRSpec} data. This tells us that both methods are recovering similar information from the data, and both struggle in the same regions of parameter space. Interestingly, when both methods differ significantly (which is only for some parameters in retrievals of WFC3 simulated data), the CNN achieves a higher $R^2$ and lower $MB$.

Aside from the CNN retrievals and their comparison to Multinest, to the best of our knowledge, the exercise of retrieving a large sample of simulated data with Multinest had not been done previously. This is useful in and of itself, as we now have statistical information on how commonly Multinest returns too narrow posterior distributions, as well as being able to see how well different parameters can be retrieved in different regions of parameter space. The results of these bulk retrievals are worthy of a more in-depth analysis, which falls outside the scope of this paper.

One of the advantages of our CNNs is the ability to still provide a plausible answer even when the models with which they were trained did not match the observation to be retrieved, as shown in Section \ref{sec:uncomfortable}. Multinest did poorly in these experiments.
{\cite{Yip2020PeekingRetrievals} showed that CNNs focus solely on specific features to predict the parameters. On the other hand, Multinest tries to find the best fit to the whole spectrum, and is therefore more easily thrown off by modifications to the spectrum, even when the feature/s caused by the parameter we want to predict remain unchanged.}
This ability is crucial in the real world, as the models used will always be simplifications of the real atmospheres, and/or will be based on a set of limiting assumptions. 

On the other hand Multinest has the advantage that it calculates the bayesian evidence, making it possible to compare models with different assumptions and select the best one. This does not negate the advantage that our CNNs in these experiments, as running a model selection requires multiple retrievals, incurring in a higher computational cost.

The main limitation of our machine learning framework is that it can only predict Gaussian posteriors. This is a poor approximation of multimodal posteriors or uniform posteriors with lower and upper bounds. The latter is often the case for the molecular abundances in type 1 retrievals, for which an upper bound might be the only information that can be inferred from the spectrum; for the C/O, where it might only be possible to infer whether the atmosphere is carbon or oxygen rich; or for $k_{haze}$ and $P_{cloud}$, for which sometimes only an upper and lower bound, respectively, might be inferred.

However, when we tested likelihood-free machine learning methods like a random forest or a CNN with Montecarlo dropout \citep{Gal2015DropoutLearning} trained with the MSE loss (mean squared error) they performed worse than the CNNs trained with the negative log likelihood loss. A random forest reaches a lower $R^2$ and is slightly worse at estimating the uncertanties, while a likelihood-free CNN reaches a similar (even slightly higher) $R^2$ but is very overconfident and yields too narrow posterior distributions.

The other limitations of our CNNs are the fixed wavelength grid and noise, which really hinder its real world usability. For the purpose of retrieving real WFC3 transmission spectra, we retrained the CNNs with the observational noise of each observation, which is impractical. Future work will focus on adding flexibility in this regard, as well  as make it possible to retrieve non Gaussian posterior distributions.

Regarding the retrievals of the Exoplanets-A transmission spectra of 48 exoplanets, we want to highlight that although we are benchmarking our CNNs against Multinest, when both methods disagree, our previous analyses pointing the direction of the CNN's predictions being more trustworthy, even if less constraining.  Because  no `ground truth' exists for these spectra it is difficult to say which is actually correct.

Not surprisingly, where the CNNs excel is in the time they take to retrieve an observation. On a regular laptop (Intel i7-8550U with 16GB of RAM) it takes $\sim 1$ s with 1000 points in the posterior (meaning 1000 forward passes of the network each with a different noise realisation of the spectrum). With Multinest it would take from $\sim 10$ min to multiple hours, depending on whether it is a WFC3 or NIRSpec observation and the model complexity.

The CNNs were trained on a cluster with 80 Intel Xeon Gold 6148 CPUs and 753 GB of RAM. Table \ref{table:training_time} summarises the training times of the four CNNs. 

\begin{table}
    \centering
    \begin{tabular}{r | c | c }
        \hline
         Instrument &  Type & Training time\\
         \hline\hline
         \multirow{2}{4em}{WFC3} & 1 & $\sim 1$ h\\
          & 2 & $\sim 1$h\\    
          \hline
         \multirow{2}{4em}{NIRSpec} & 1 & $\sim 3$ h\\
          & 2 & $\sim 3$ h\\
         \hline 
    \end{tabular}
    \caption{Training times of the four CNNs.}
    \label{table:training_time}
\end{table}

This difference will only become larger when using higher model complexity or data from upcoming facilities with lower noise, higher spectral resolution and larger wavelength coverage. As an example, 3D nested sampling retrievals with self-consistent clouds and chemistry might not be feasible or practical with current techniques, and methods like machine learning might be the solution.

Because of their speed, machine learning retrievals are also specially suited for any application in which multiple retrievals need to be done. An example of this might be retrievability studies in which many spectra across parameter space are retrieved to see what information can be extracted and how different parameters affect the retrievability of other parameters. This information is of great relevance to inform the design of future instrumentation or observing programs, and the increased speed of  machine learning retrievals provides a huge advantage.

\section{Conclusions}\label{sec:conclusion}

In this paper we have presented extensive comparisons between machine learning and nested sampling retrievals of transmission spectra of exoplanetary atmospheres. This represents a step forward from the previous literature where comparisons had been limited to a few test cases. 

Machine learning methods can be a powerful alternative to Bayesian sampling techniques for performing retrievals on transmission spectra of exoplanetary atmospheres. Our work has shown that CNNs can achieve a similar performance and even behave more desirably than Multinest in certain situations. In particular, when taking into account the retrieved values and their uncertainties, the answers our CNN provides are virtually always correct, contrary to Multinest. For the latter, despite being the current standard for retrieving exoplanetary transmission spectra, we find that for a significant fraction of spectra ($\sim 8\%$), its answer will be off by more than $3\sigma$. 

When comparing both methods with real HST transmission spectra, we find them to agree within $1\sigma$ of each other for most cases. However in the cases where they disagree it is difficult to reach a verdict on which of the two is correct.

Finally, when retrieving spectra with different characteristics than those used in our retrieval frameworks (such as having different chemistry or being contaminated by stellar activity), the CNNs performed significantly better, with Multinest underestimating the uncertainty in $\sim 12\%$ to $\sim41\%$ of cases. 

\begin{acknowledgements}
      We wish to thank Daniela Huppenkothen for our fruitful discussions and her valuable comments on the manuscript.
      \\
      
      We would also like to thank the Center for Information Technology of the University of Groningen for their support and for providing access to the Peregrine high performance computing cluster.
      \\
      
      This project has received funding from the European Union’s Horizon 2020 research and innovation programme under the Marie Sklodowska-Curie grant agreement No. 860470.
      \\
\end{acknowledgements}

\bibliographystyle{aa} % style aa.bst
\bibliography{main.bib}% your references Yourfile.bib

\begin{appendix}

\clearpage

\section{Prediction versus truth plots}\label{app:pred_vs_true}

Here we present the full comparisons between predicted and true values of the parameters for all the different retrievals ran in this work. In these figures, each blue dot corresponds to an individual spectrum and the predicted value of the parameter is the median of the retrieved posterior. The red diagonal line represents perfect predictions.
\FloatBarrier
\begin{figure*}[b]
    \centering
    \includegraphics[width=0.49\textwidth]{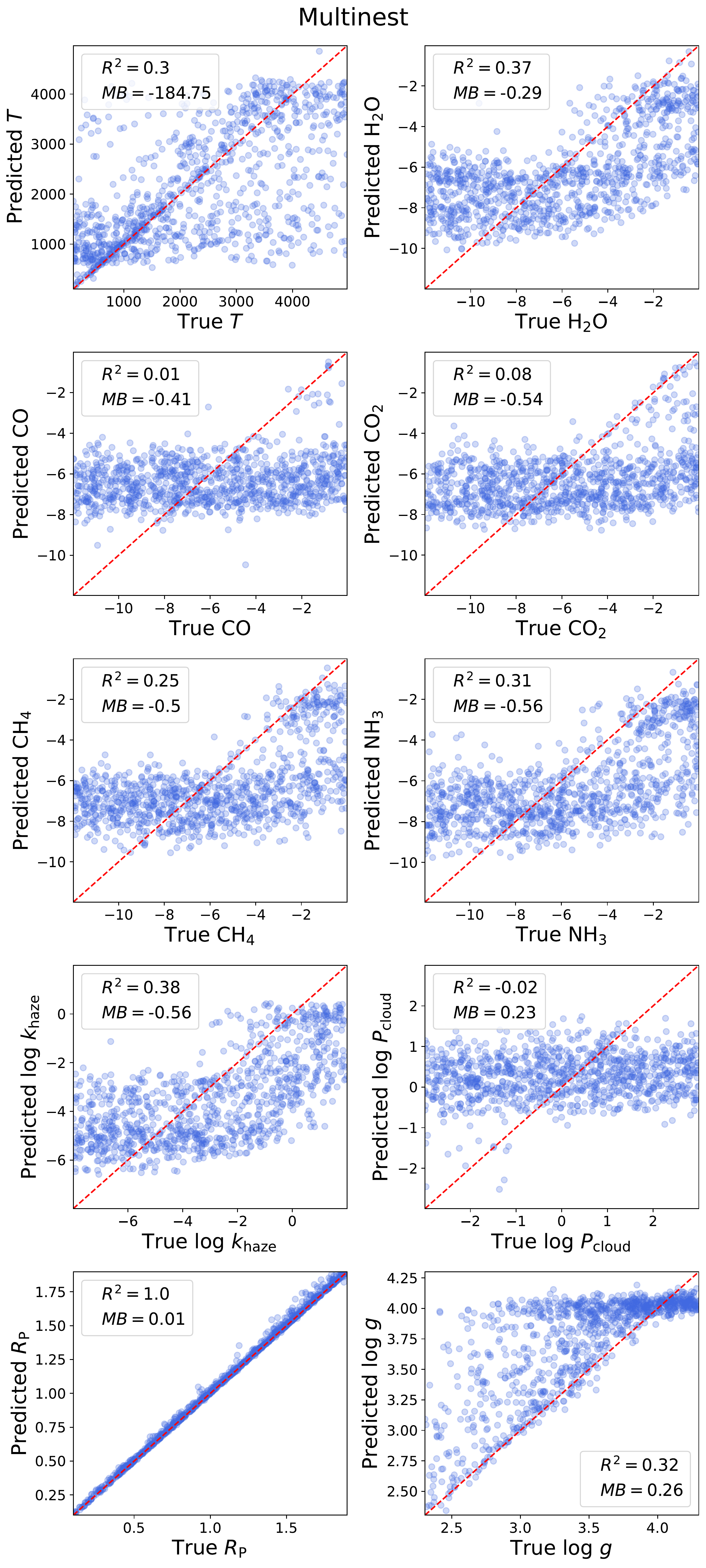}
    \includegraphics[width=0.49\textwidth]{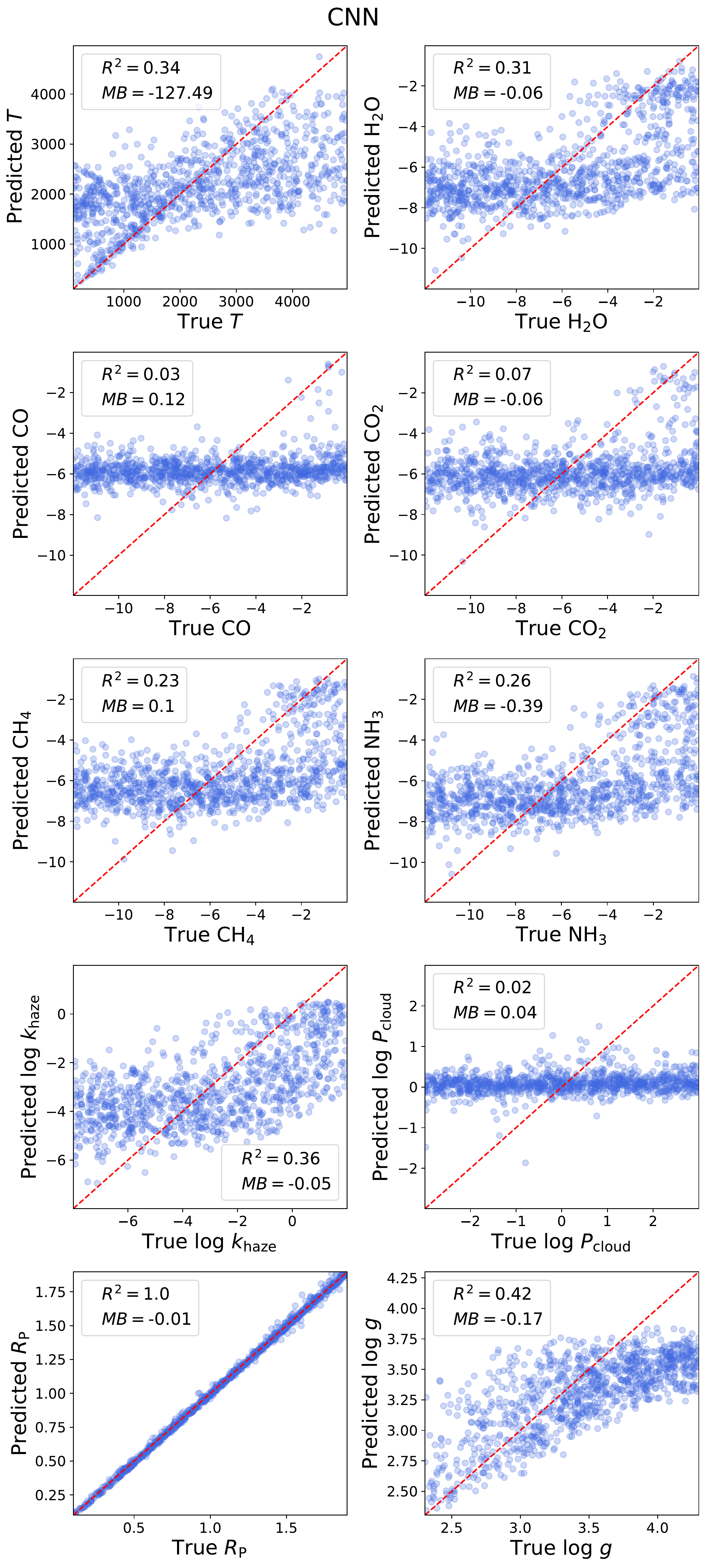}
    \caption{Predicted vs true values for type 1 retrievals performed for simulated WFC3 transmission spectra. {(Left)} Multinest. {(Right)} CNN.}
    \label{fig:wfc3_level1_corr}
\end{figure*}

\begin{figure*}
    \centering
    \includegraphics[width=0.49\textwidth]{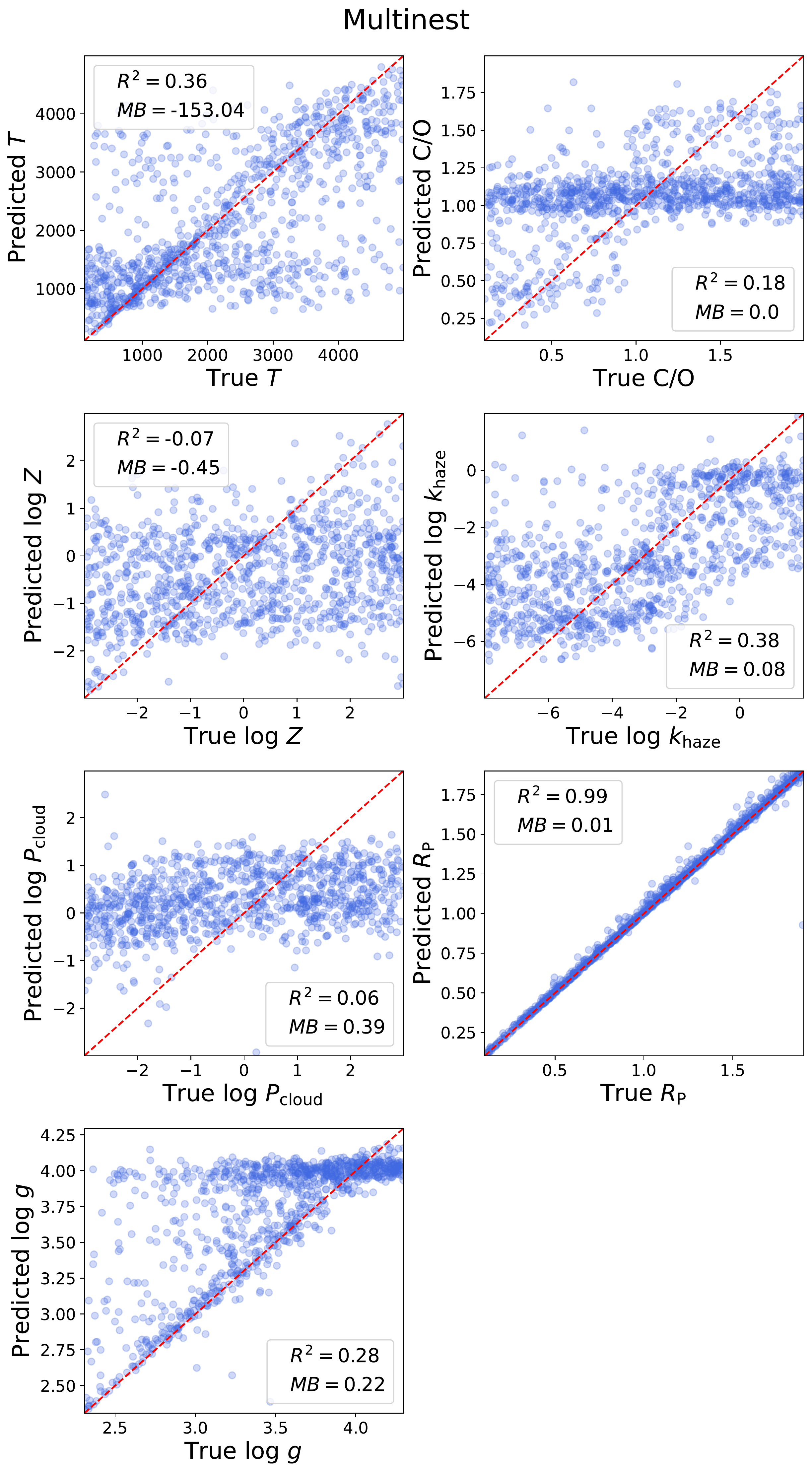}
    \includegraphics[width=0.49\textwidth]{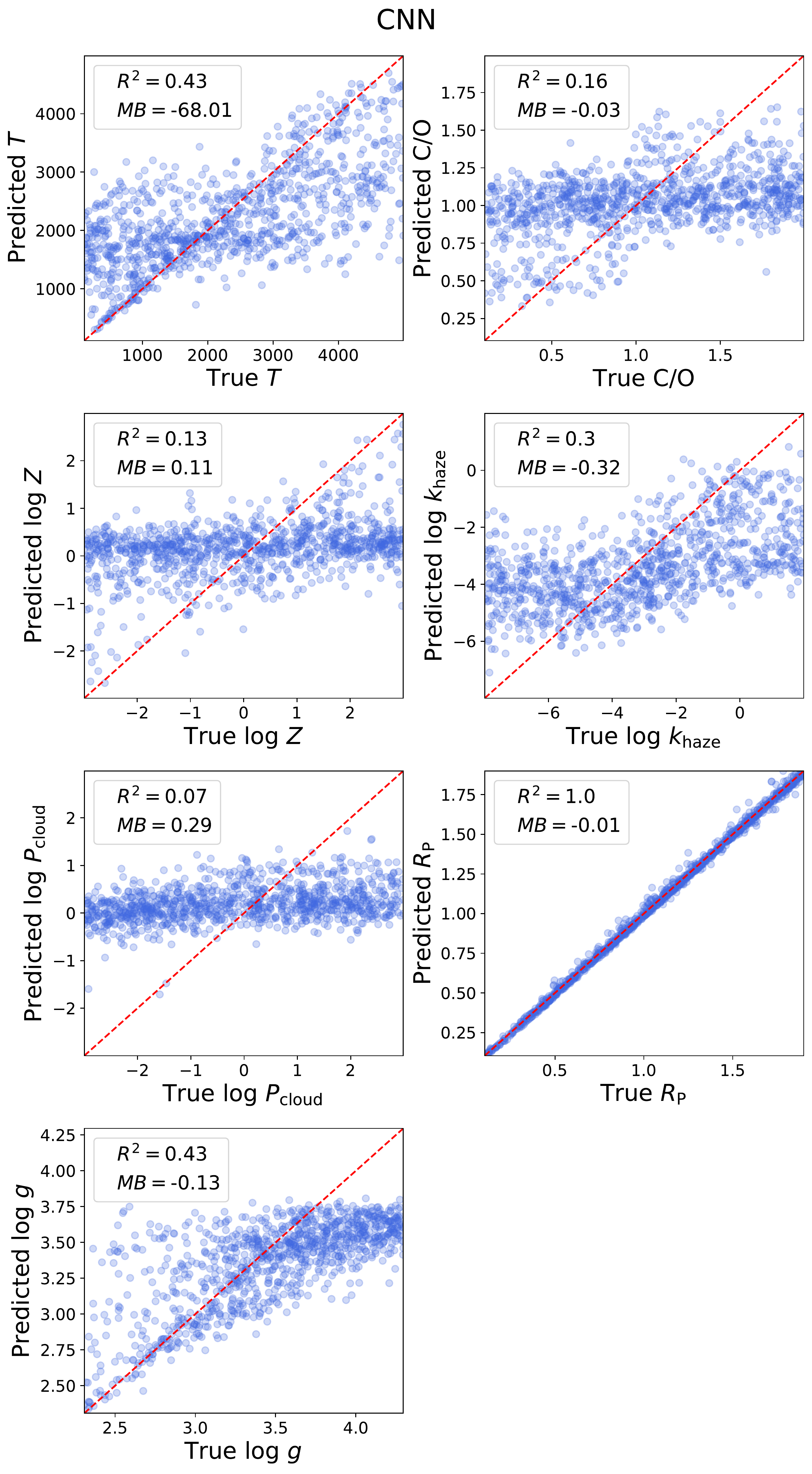}
    \caption{Predicted vs true values for type 2 retrievals performed for simulated WFC3 transmission spectra. {(Left)} Multinest. {(Right)} CNN.}
    \label{fig:wfc3_level2_corr}
\end{figure*}

\begin{figure*}
    \centering
    \includegraphics[width=0.49\textwidth]{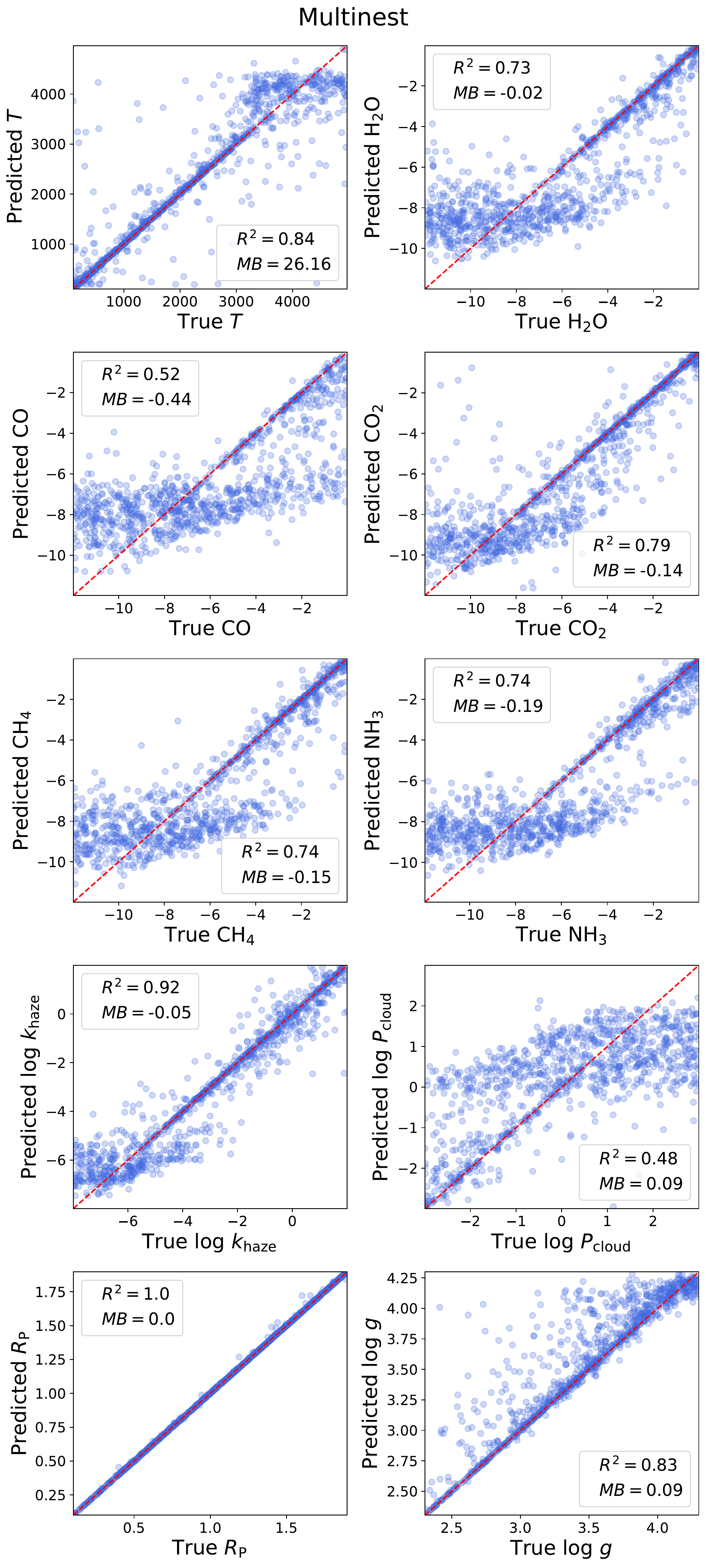}
    \includegraphics[width=0.49\textwidth]{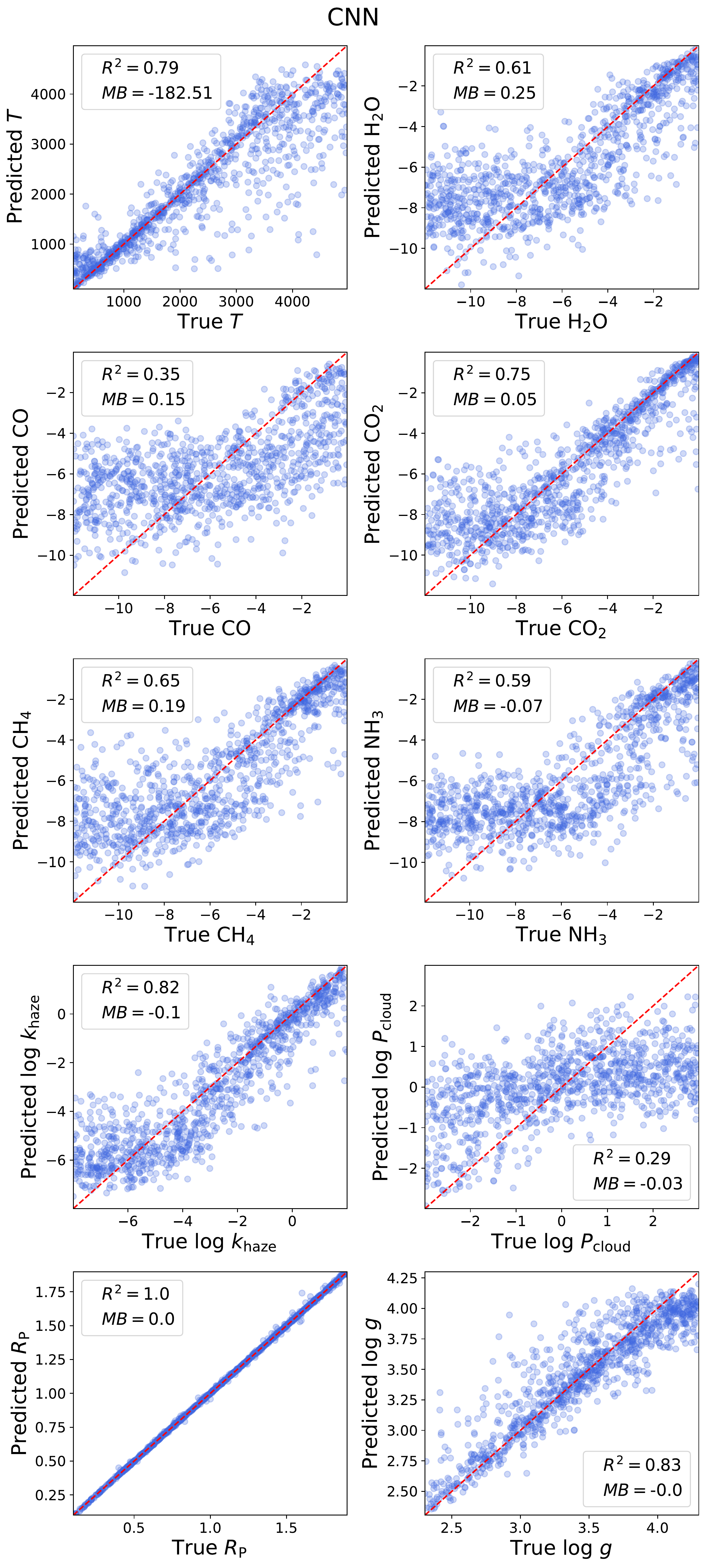}
    \caption{Predicted vs true values for type 1 retrievals performed for simulated NIRSpec transmission spectra. {(Left)} Multinest. {(Right)} CNN.}
    \label{fig:NIRSpec_level1_corr}
\end{figure*}

\begin{figure*}
    \centering
    \includegraphics[width=0.49\textwidth]{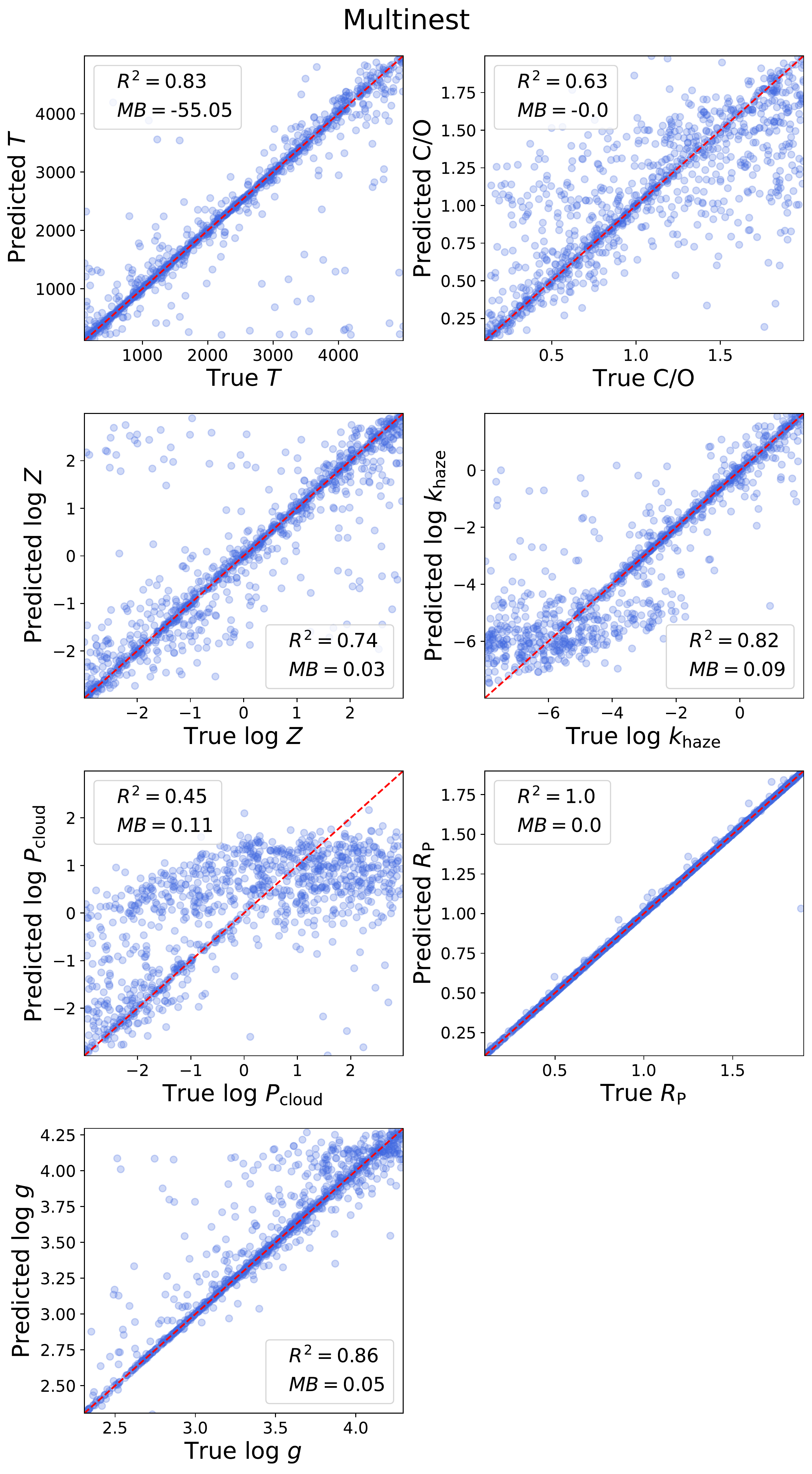}
    \includegraphics[width=0.49\textwidth]{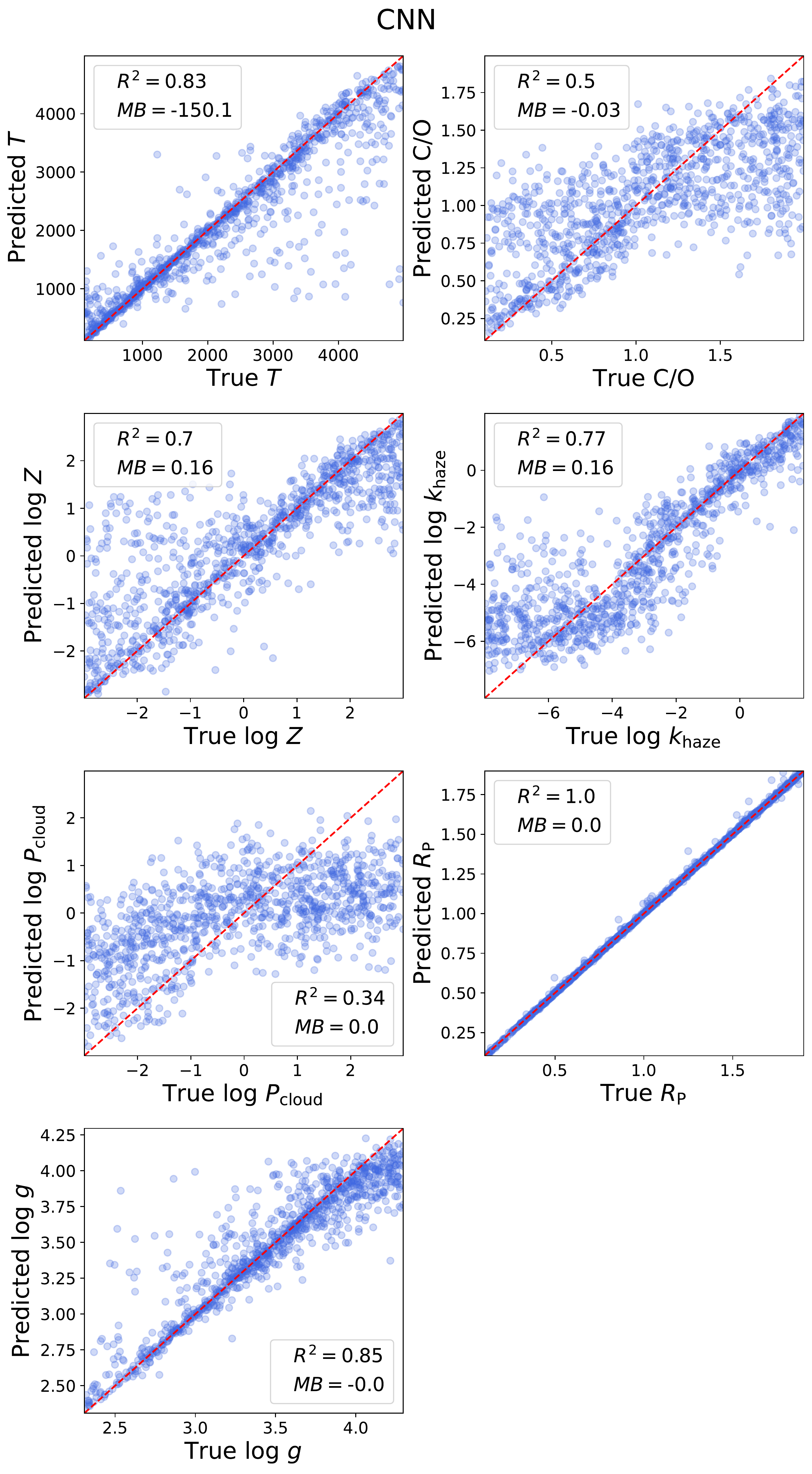}
    \caption{Predicted vs true values for type 2 retrievals performed for simulated NIRSpec transmission spectra. {(Left)} Multinest. {(Right)} CNN.}
    \label{fig:NIRSpec_level2_corr}
\end{figure*}

\begin{figure*}
    \centering
    \includegraphics[width=0.49\textwidth]{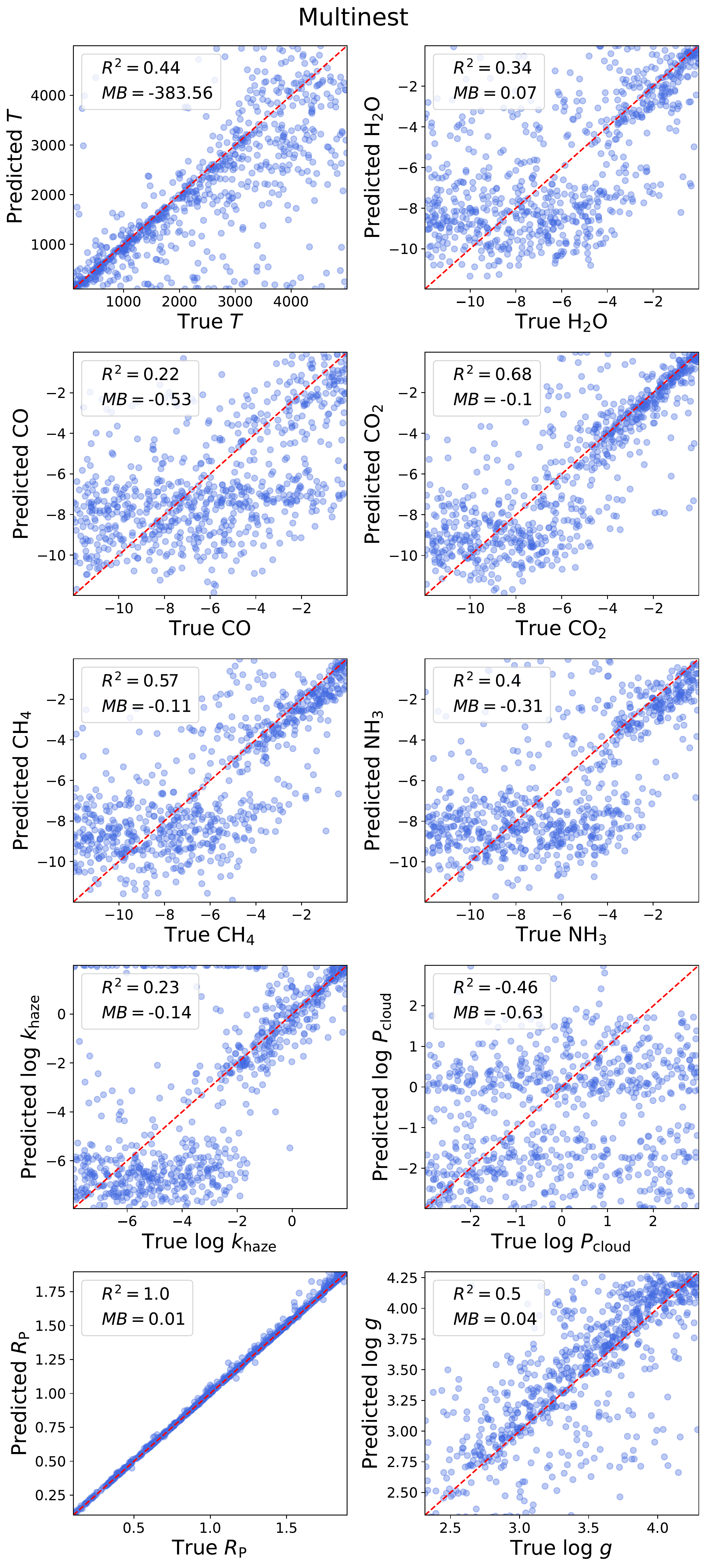}
    \includegraphics[width=0.49\textwidth]{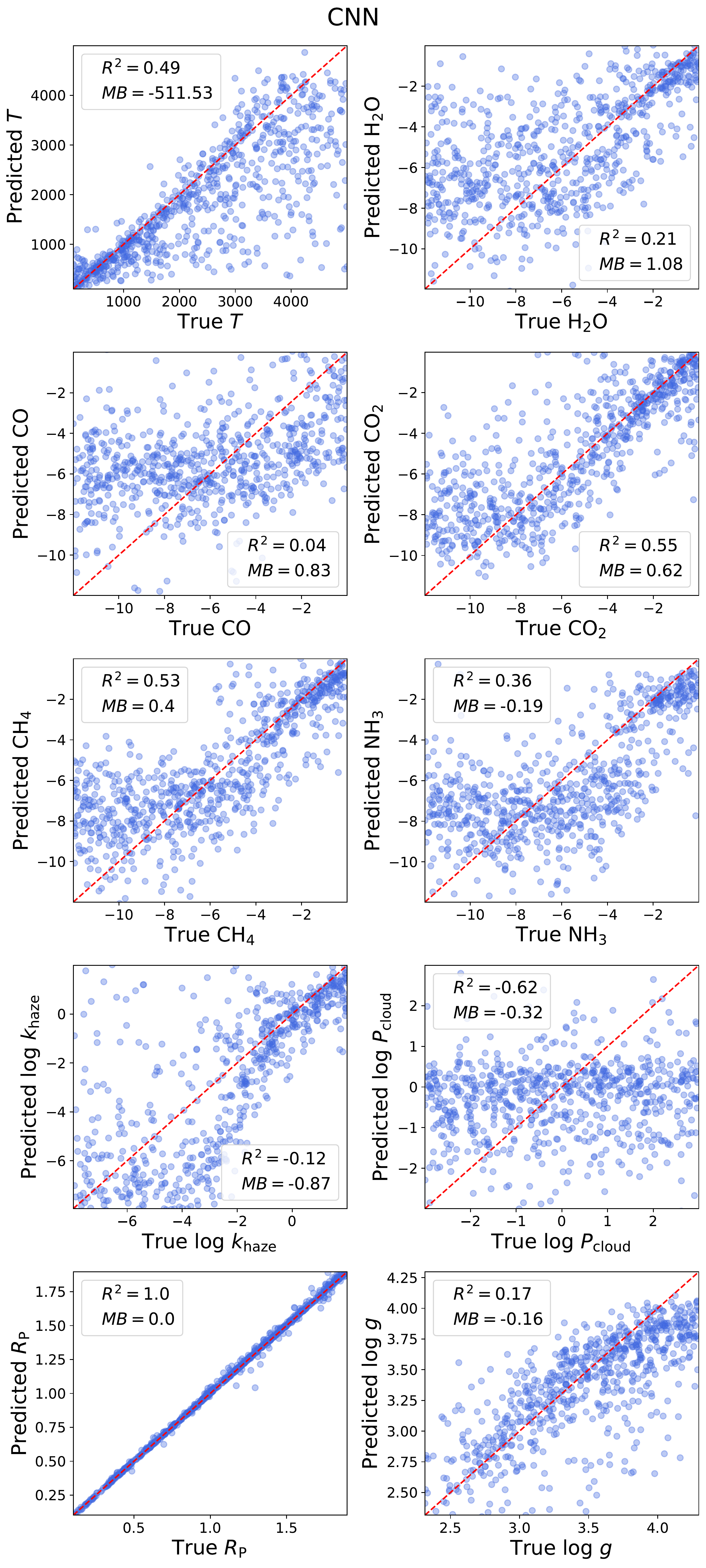}
    \caption{Predicted vs true values for type 1 retrievals performed for simulated NIRSpec transmission spectra with $\log AlO=-5.25$. {(Left)} Multinest. {(Right)} CNN.}
    \label{fig:alo_corr}
\end{figure*}

\begin{figure*}
    \centering
    \includegraphics[width=0.49\textwidth]{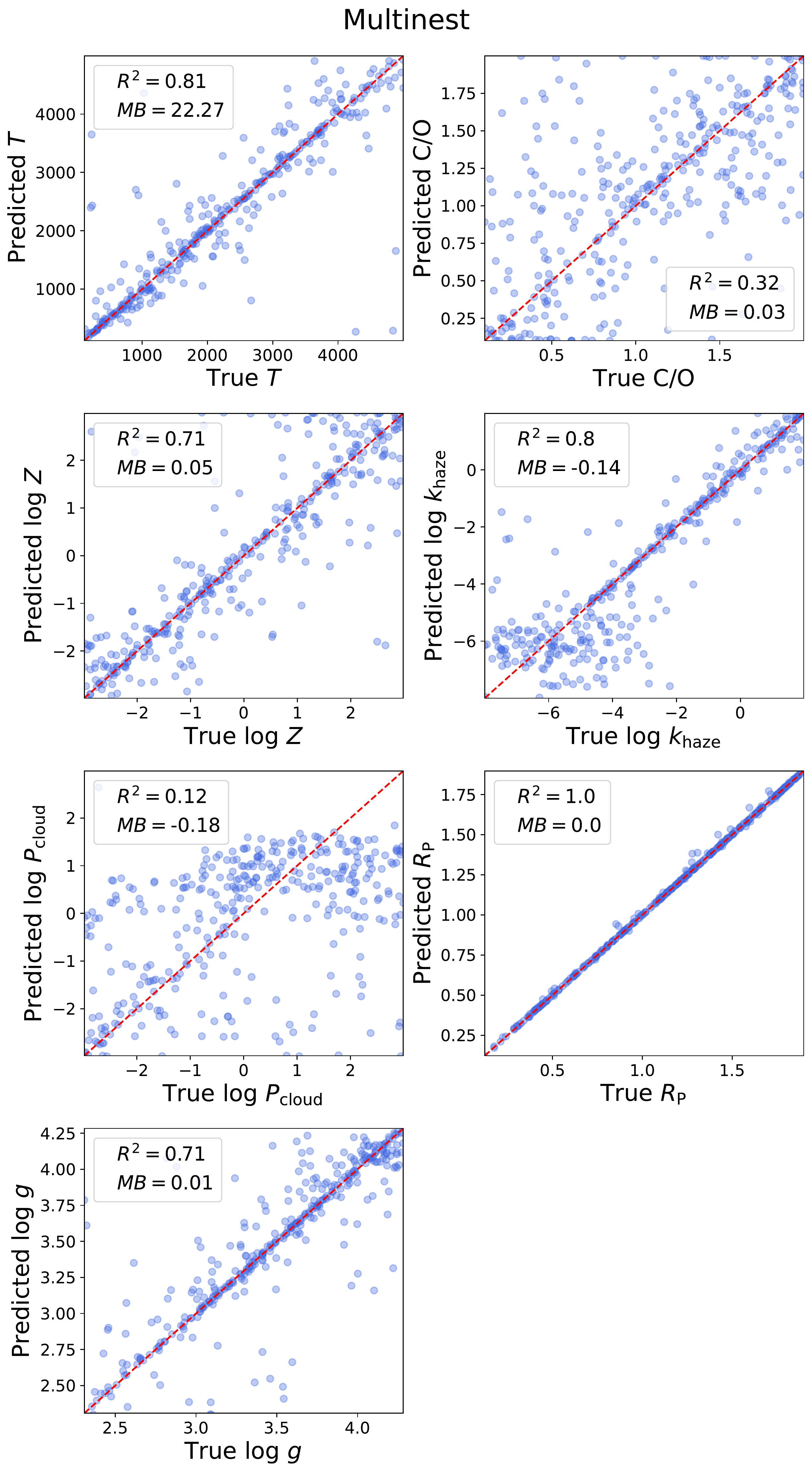}
    \includegraphics[width=0.49\textwidth]{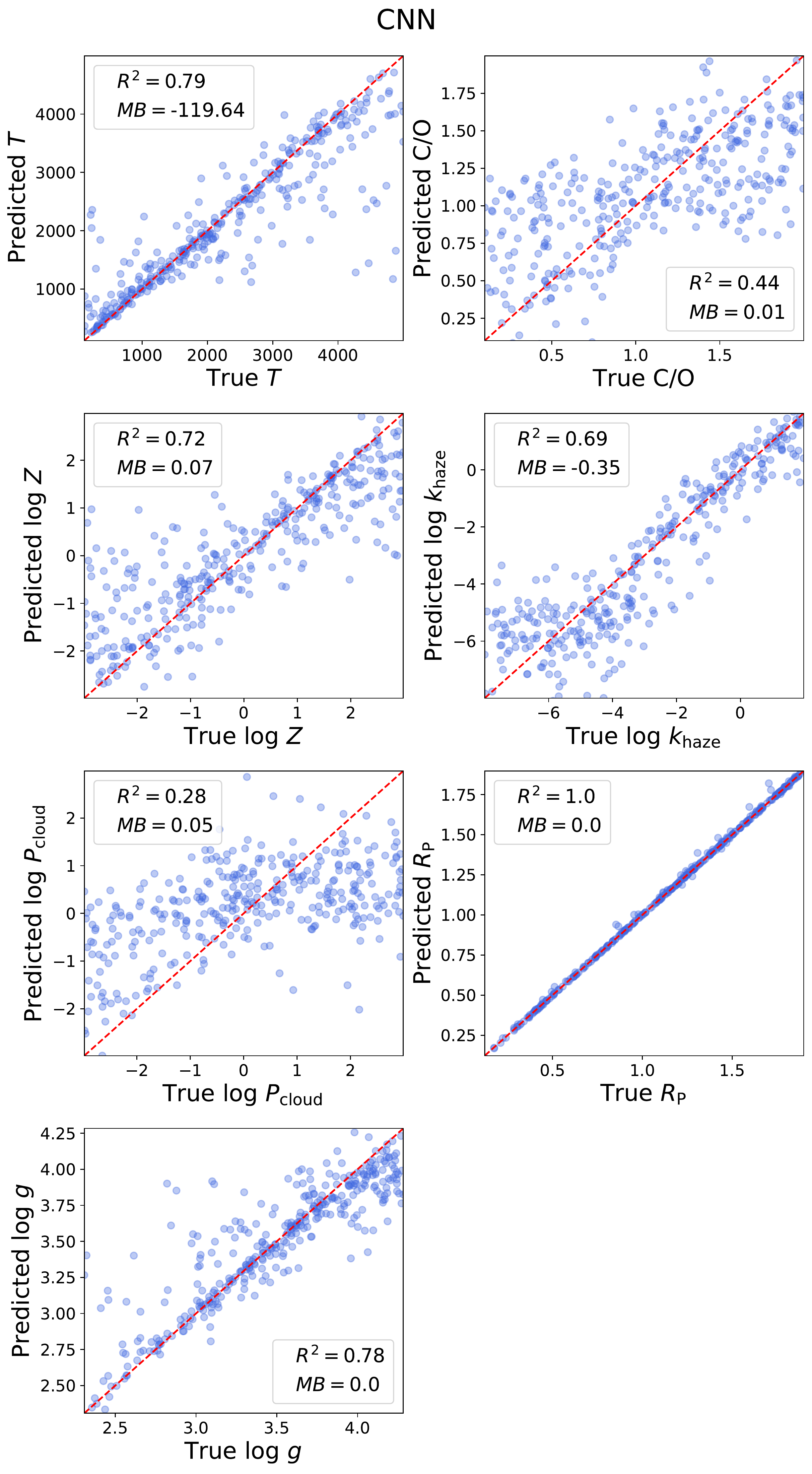}
    \caption{Predicted vs true values for type 2 retrievals performed for simulated NIRSpec transmission spectra of atmospheres without TiO or VO. {(Left)} Multinest. {(Right)} CNN.}
    \label{fig:-tio-vo_corr}
\end{figure*}

\begin{figure*}
    \centering
    \includegraphics[width=0.49\textwidth]{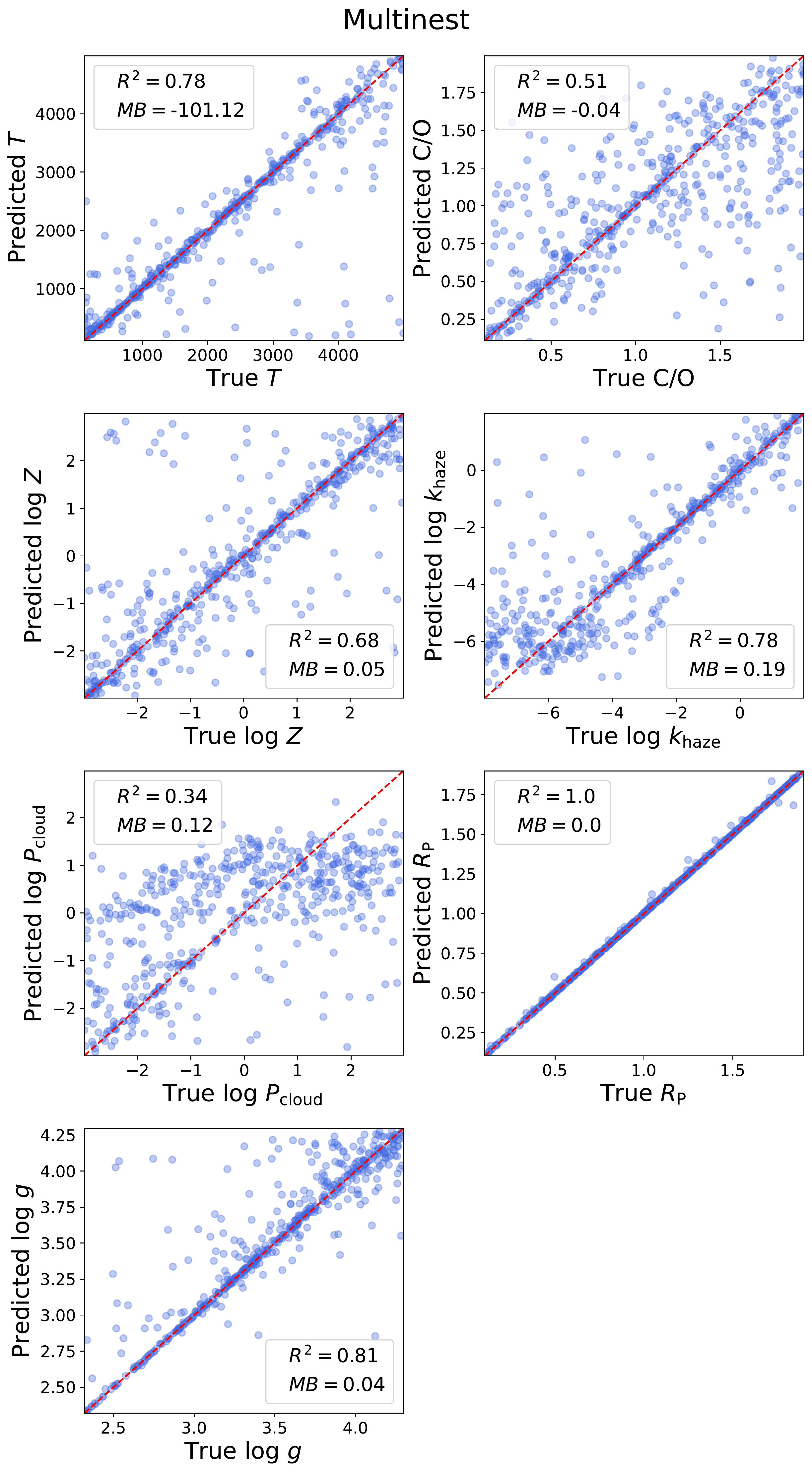}
    \includegraphics[width=0.49\textwidth]{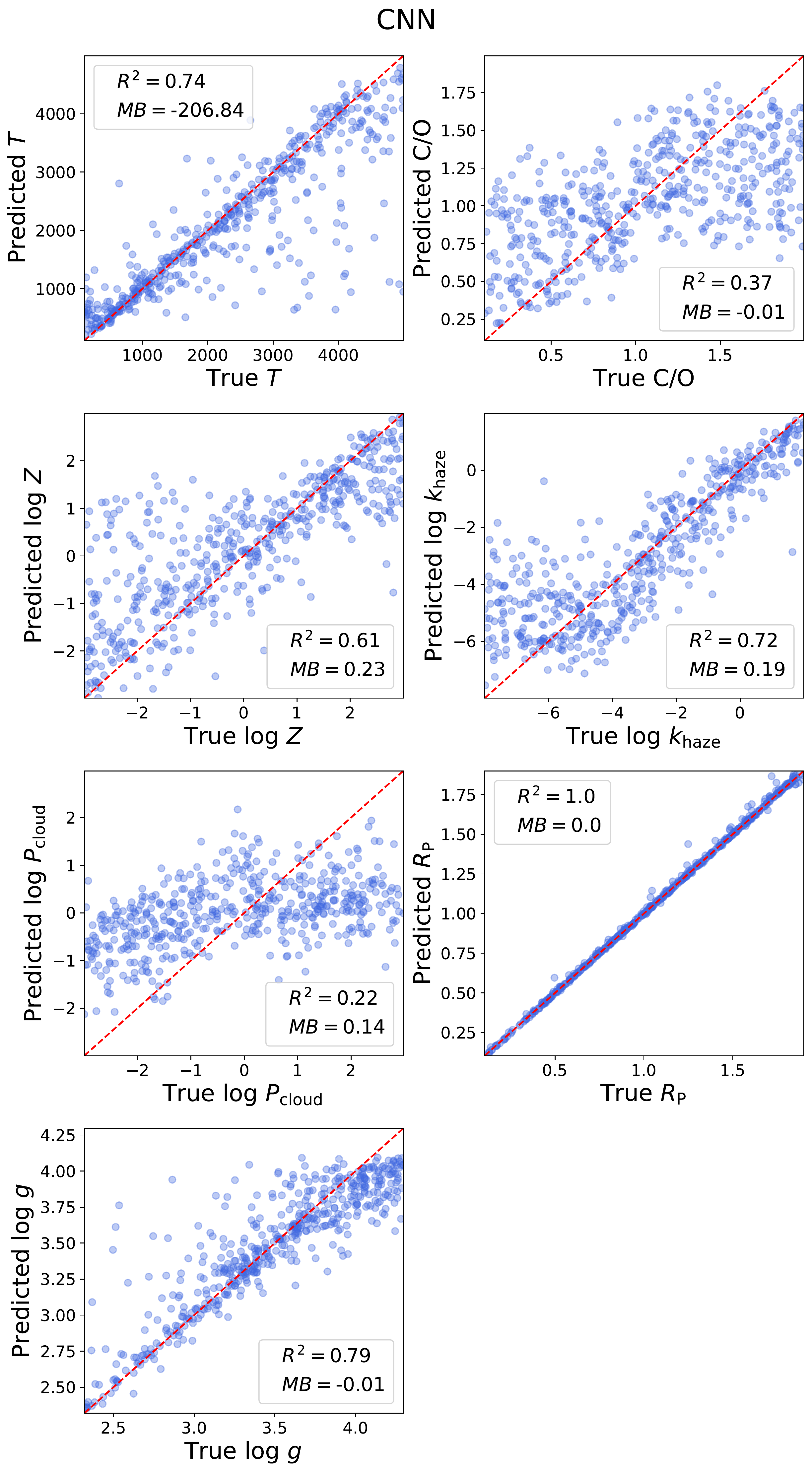}
    \caption{Predicted vs true values for type 2 retrievals performed for simulated NIRSpec transmission spectra of exoplanets orbiting heterogeneously spotted host stars. {(Left)} Multinest. {(Right)} CNN.}
    \label{fig:star-spots_corr}
\end{figure*}

\clearpage
\section{Difference between predictions and truths}\label{app:sigma_diff}
Here we present the distributions of $\sigma_eq$ between the predicted and true values of the parameters that were not shown in the main text. The `ideal case' represents the fraction of predictions we would expect to find in each bin if both methods were correctly estimating the uncertainty.

\begin{figure*}[b]
    \centering
    \includegraphics[width=0.49\textwidth]{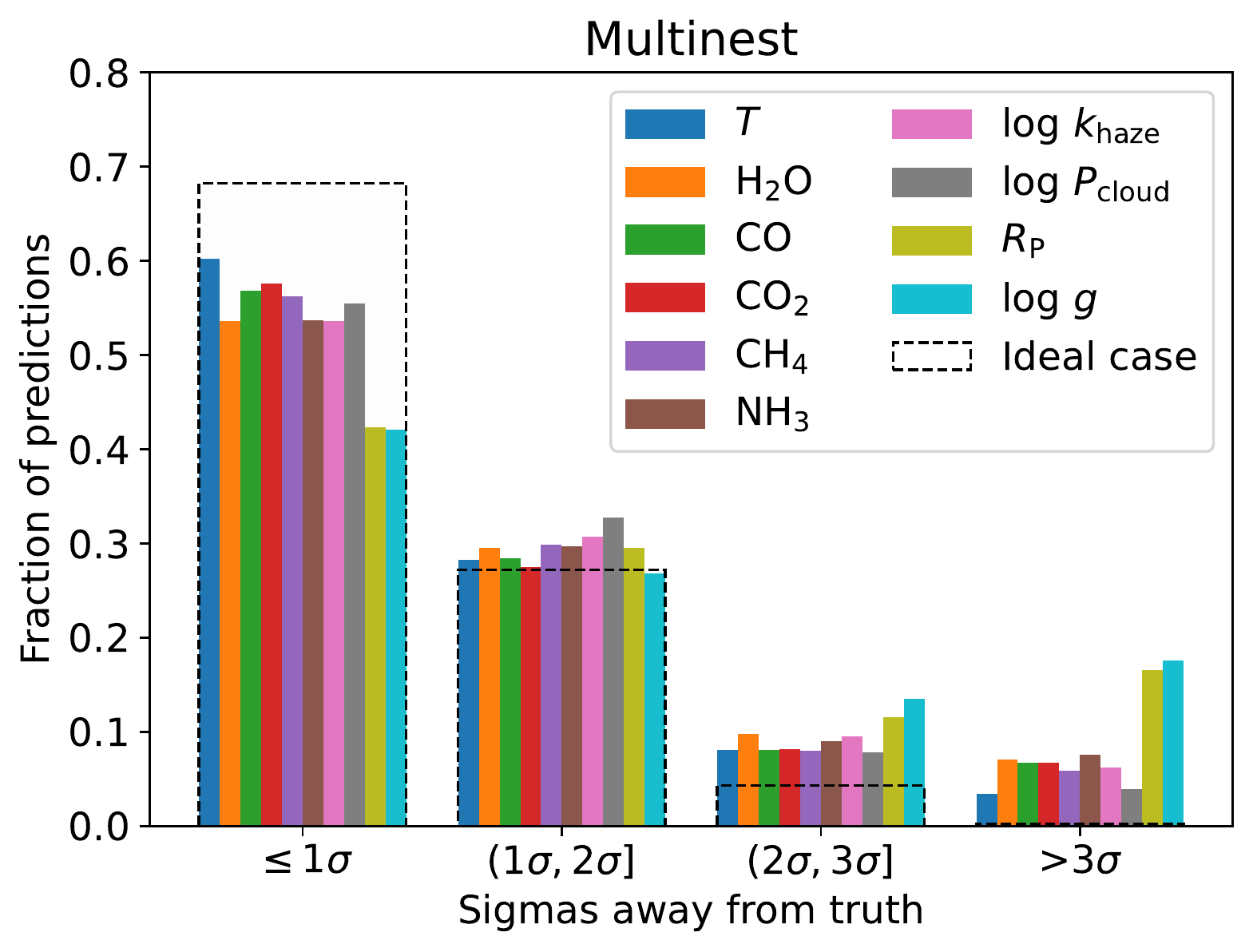}
    \includegraphics[width=0.49\textwidth]{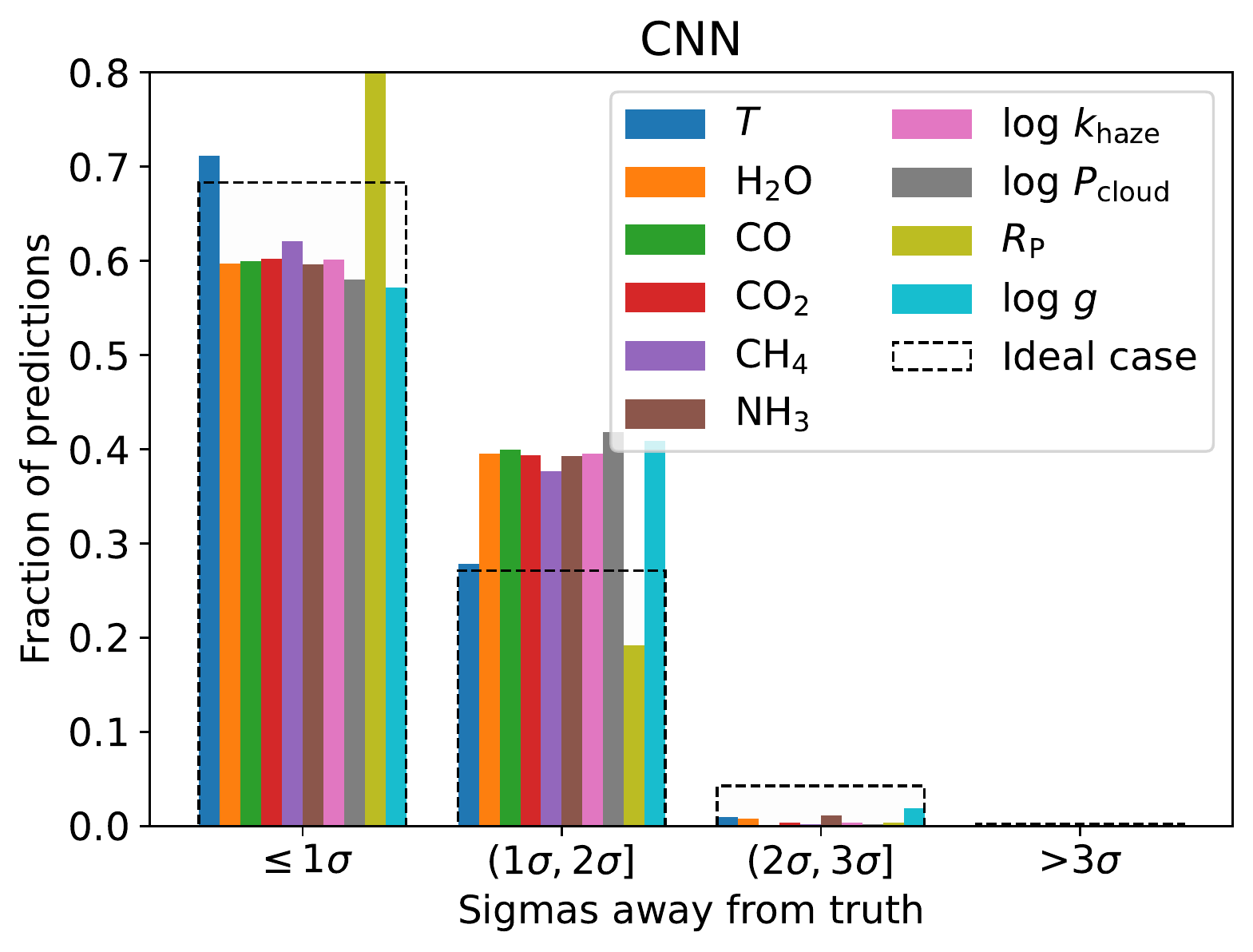}
    \caption{Distance in $\sigma$ between predictions and ground truths for type 1 retrievals performed for simulated WFC3 observations. {(Left)} Multinest. {(Right)} CNN.}
    \label{fig:sigmas_away_wfc3_l1}
\end{figure*}

\begin{figure*}[b]
    \centering
    \includegraphics[width=0.49\textwidth]{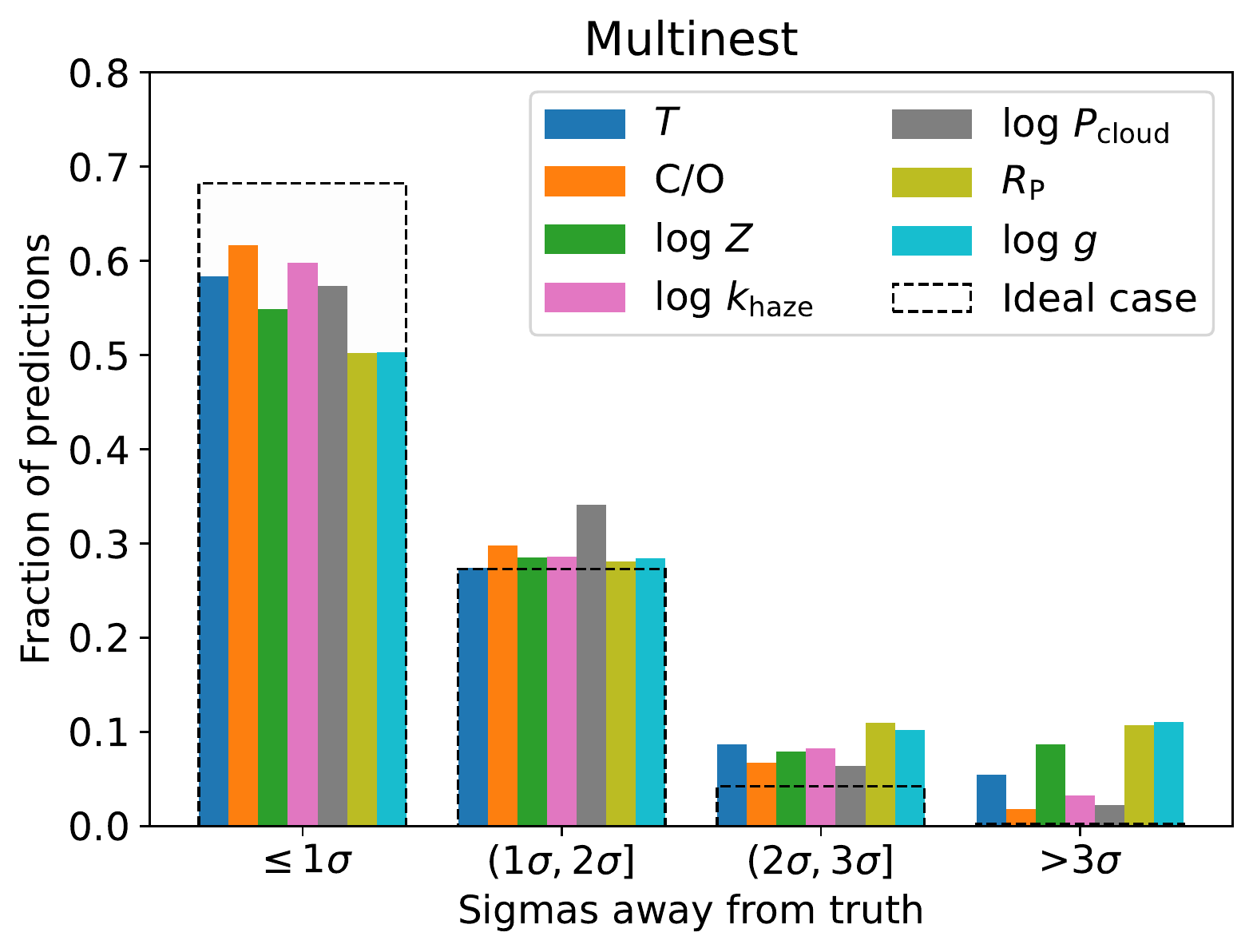}
    \includegraphics[width=0.49\textwidth]{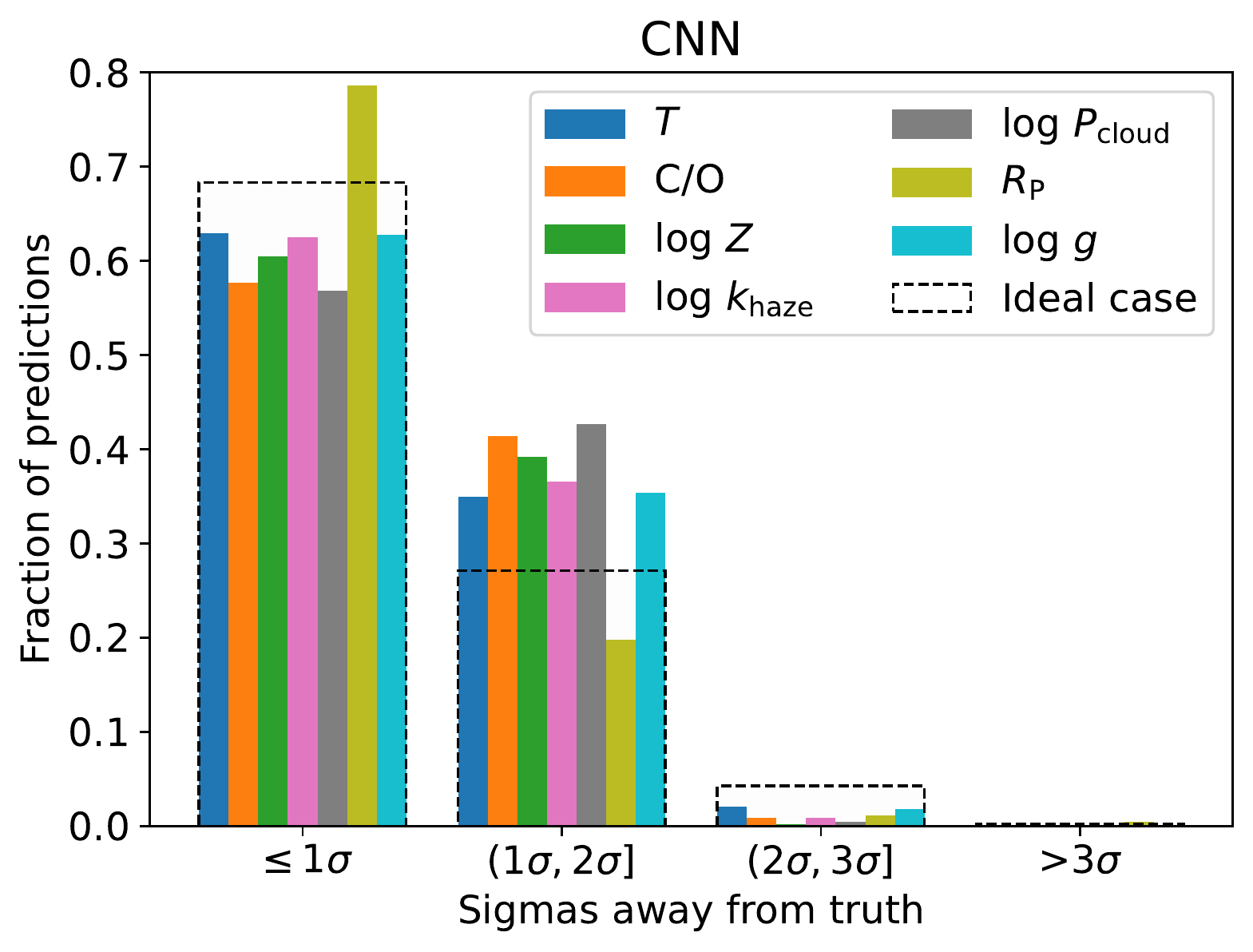}
    \caption{Distance in $\sigma$ between predictions and ground truths for type 2 retrievals performed for simulated WFC3 observations. {(Left)} Multinest. {(Right)} CNN.}
    \label{fig:sigmas_away_wfc3_l2}
\end{figure*}

\begin{figure*}
    \centering
    \includegraphics[width=0.49\textwidth]{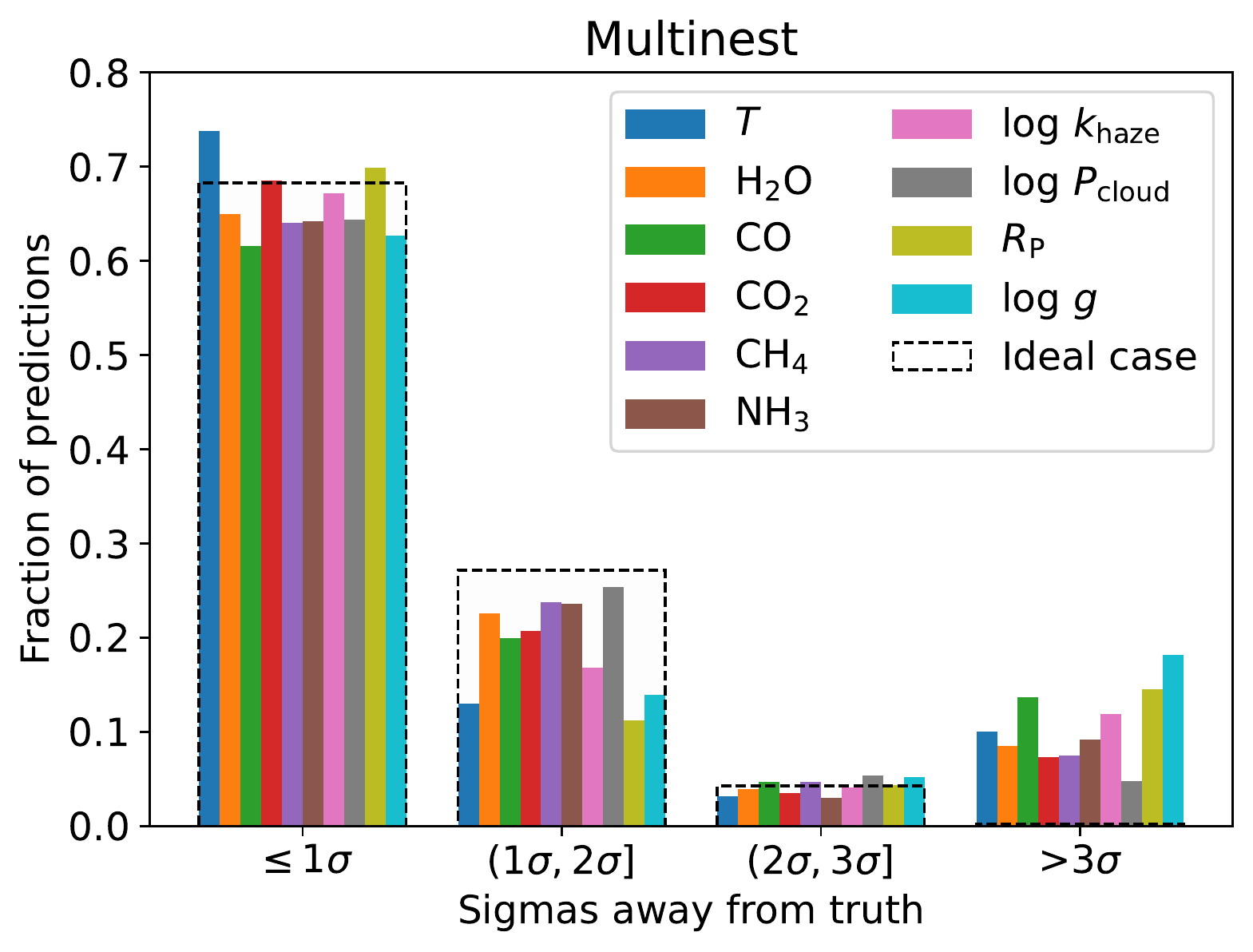}
    \includegraphics[width=0.49\textwidth]{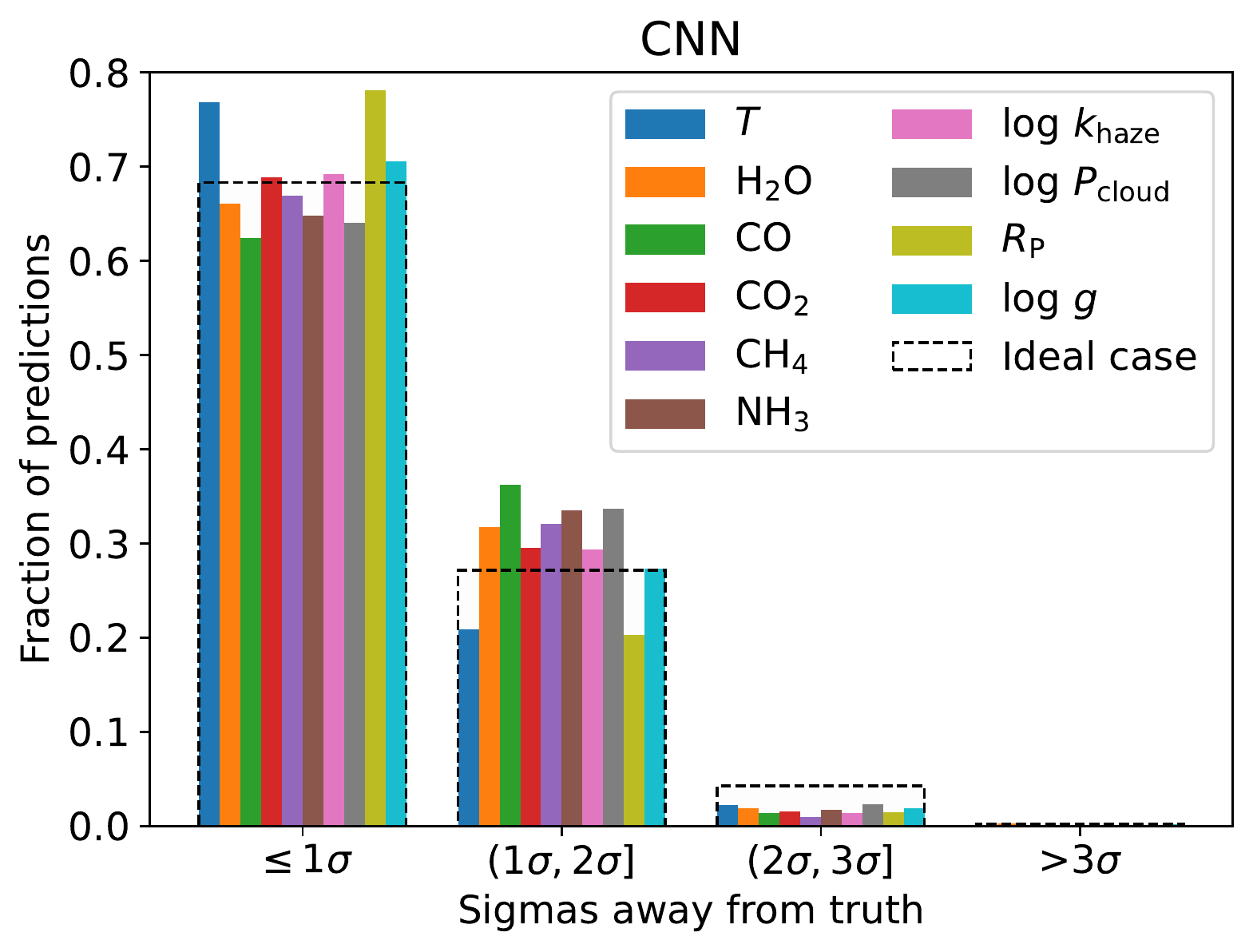}
    \caption{Distance in $\sigma$ between predictions and ground truths for type 1 retrievals performed for simulated NIRSpec observations. {(Left)} Multinest. {(Right)} CNN.}
    \label{fig:sigmas_away_NIRSpec_l1}
\end{figure*}

\clearpage
\section{Corner plots for WASP-12b}\label{app:corner-plots}

Here we present the corner plots of the posterior distributions retrieved for WASP-12b. Notice that both our CNN and Multinest show good agreement for the type 2 retrievals, but type 1 retrievals disagree for $T$ and $\log g$.

\begin{figure*}[b]
    \centering
    \includegraphics[width=\textwidth]{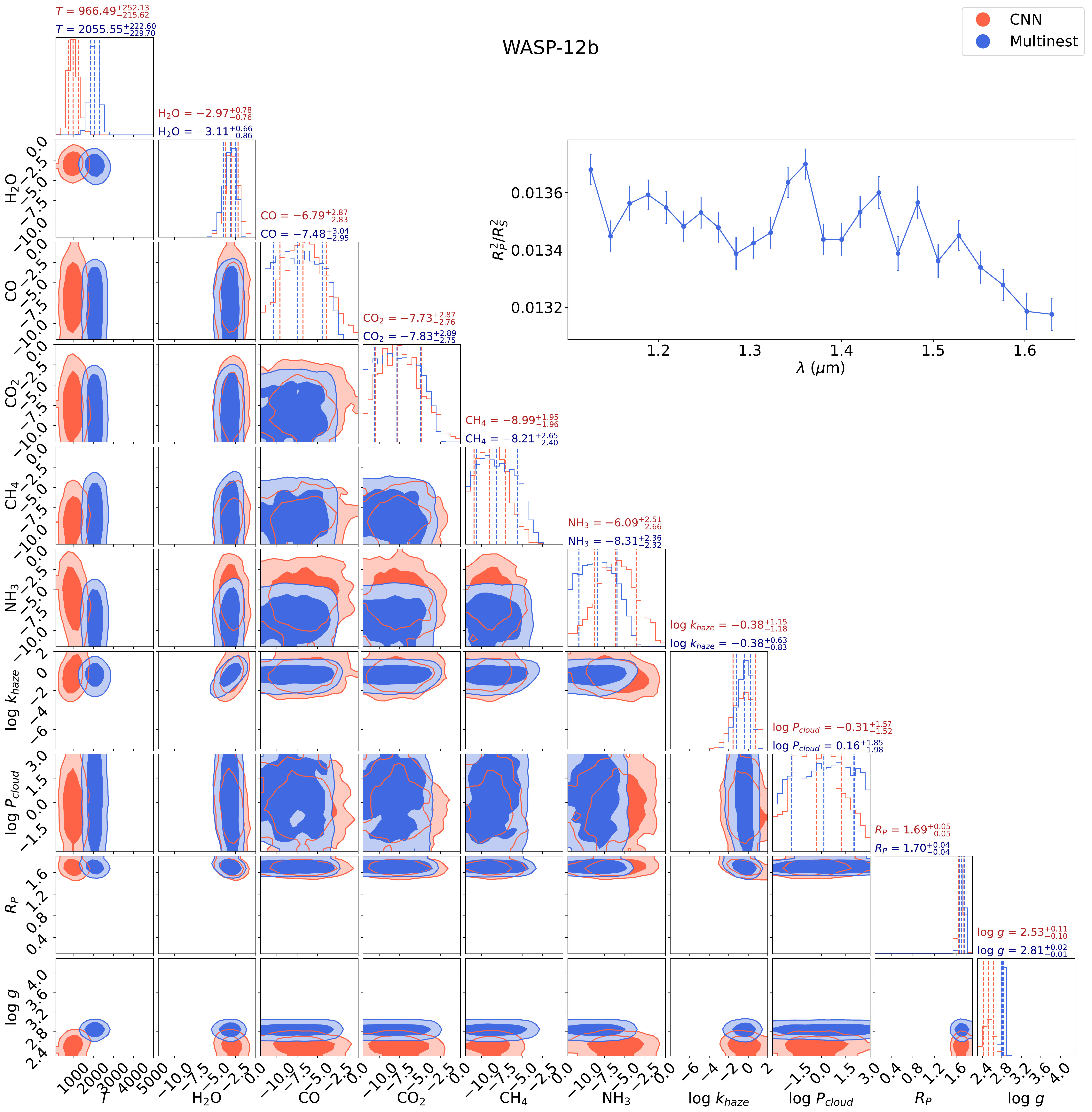}
    \caption{Corner plots for the type 1 retrievals for WASP-12b.}
    \label{fig:corner_l1}
\end{figure*}

\begin{figure*}
    \centering
    \includegraphics[width=\textwidth]{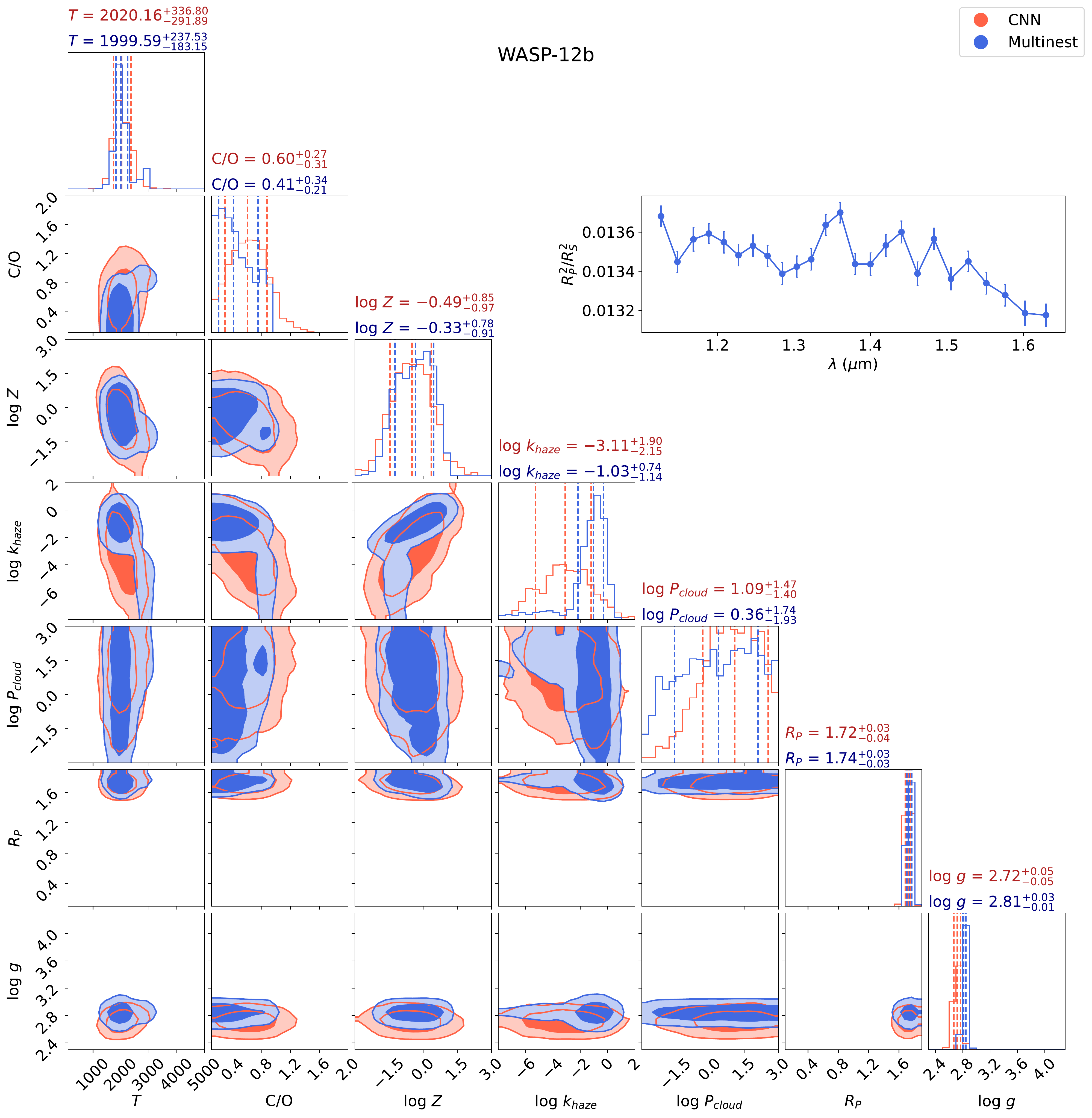}
    \caption{Corner plots for the type 2 retrievals for WASP-12b.}
    \label{fig:corner_l2}
\end{figure*}

\end{appendix}

\end{document}